\documentclass[twocolumn,appendixfloats]{aastex6}

\newcommand{\RE}{R$_{\Earth}$}

\shorttitle{Kepler Follow-Up Observation Program: High-Resolution Imaging}
\shortauthors{Furlan et al.}
\submitted{Accepted to AJ, December 6, 2016}

\begin{document}

\title{The {\it Kepler} Follow-Up Observation Program. I. A Catalog of Companions to 
{\it Kepler} Stars from High-Resolution Imaging}

\author{E. Furlan\altaffilmark{1}, D. R. Ciardi\altaffilmark{1}, M. E. Everett\altaffilmark{2},
M. Saylors\altaffilmark{1,3}, J. K. Teske\altaffilmark{4,5}, E. P. Horch\altaffilmark{6,7}, 
S. B. Howell\altaffilmark{8}, G. T. van Belle\altaffilmark{7}, L. A. Hirsch\altaffilmark{9}, 
Gautier, T. N. III\altaffilmark{10}, E. R. Adams\altaffilmark{11}, D. Barrado\altaffilmark{12}, 
K. M. S. Cartier\altaffilmark{13}, C. D. Dressing\altaffilmark{14,15}, A. K. Dupree\altaffilmark{16}, 
R. L. Gilliland\altaffilmark{13}, J. Lillo-Box\altaffilmark{17}, P. W. Lucas\altaffilmark{18},
J. Wang\altaffilmark{14}}

\altaffiltext{1}{IPAC, Mail Code 314-6, Caltech, 1200 E. California Blvd., Pasadena, 
CA 91125, USA; furlan@ipac.caltech.edu}
\altaffiltext{2}{National Optical Astronomy Observatory, 950 N. Cherry Ave., 
Tucson, AZ 85719, USA}
\altaffiltext{3}{College of the Canyons, 26455 Rockwell Canyon Rd., Santa
Clarita, CA 91355, USA}
\altaffiltext{4}{Carnegie DTM, 5241 Broad Branch Rd., NW, Washington, 
DC 20015, USA}
\altaffiltext{5}{Carnegie Origins Fellow, jointly appointed by Carnegie DTM 
and Carnegie Observatories}
\altaffiltext{6}{Department of Physics, Southern Connecticut State University,
501 Crescent Street, New Haven, CT 06515, USA}
\altaffiltext{7}{Lowell Observatory, 1400 W. Mars Hill Rd., Flagstaff, AZ 86001, USA}
\altaffiltext{8}{NASA Ames Research Center, Moffett Field, CA 94035, USA}
\altaffiltext{9}{Astronomy Department, University of California at Berkeley, 
Berkeley, CA 94720, USA}
\altaffiltext{10}{Jet Propulsion Laboratory/California Institute of Technology,
Pasadena, CA 91109, USA}
\altaffiltext{11}{Planetary Science Institute, Tucson, AZ 85719, USA}
\altaffiltext{12}{Depto. de Astrof\'isica, Centro de Astrobiolog\'ia (CSIC-INTA), ESAC,
Villanueva de la Ca\~nada (Madrid), Spain}
\altaffiltext{13}{Department of Astronomy \& Astrophysics, and Center for Exoplanets
and Habitable Worlds, The Pennsylvania State University, University Park, PA 16802, USA}
\altaffiltext{14}{California Institute of Technology, Pasadena, CA 91125, USA}
\altaffiltext{15}{NASA Sagan Fellow}
\altaffiltext{16}{Harvard-Smithsonian Center for Astrophysics, Cambridge, MA 02138, USA}
\altaffiltext{17}{European Southern Observatory (ESO), Santiago de Chile, Chile}
\altaffiltext{18}{Centre for Astrophysics Research, University of Hertfordshire,
Hatfield AL10 9AB, UK}

\begin{abstract}
We present results from high-resolution, optical to near-IR imaging of host stars of 
{\it Kepler} Objects of Interest (KOIs), identified in the original {\it Kepler} field. Part
of the data were obtained under the {\it Kepler} imaging follow-up observation 
program over seven years (2009 -- 2015). Almost 90\% of stars that are 
hosts to planet candidates or confirmed planets were observed. We combine 
measurements of companions to KOI host stars from different bands to create 
a comprehensive catalog of projected separations, position angles, and magnitude 
differences for all detected companion stars (some of which may not be bound). 
Our compilation includes 2297 companions around 1903 primary stars. 
From high-resolution imaging, we find that $\sim$~10\% ($\sim$~30\%) of 
the observed stars have at least one companion detected within 1\arcsec\ 
(4\arcsec). The true fraction of systems with close ($\lesssim$~4\arcsec) 
companions is larger than the observed one due to the limited sensitivities 
of the imaging data. We derive correction factors for planet radii caused 
by the dilution of the transit depth: assuming that planets orbit the primary stars
or the brightest companion stars, the average correction factors are 1.06
and 3.09, respectively. The true effect of transit dilution lies in between 
these two cases and varies with each system. Applying these factors to 
planet radii decreases the number of KOI planets with radii smaller than 2 \RE\ 
by $\sim$2-23\% and thus affects planet occurrence rates. This effect 
will also be important for the yield of small planets from future transit missions 
such as TESS.
\end{abstract}


\section{Introduction}
\label{intro}

In the last few years our knowledge of extrasolar planetary systems has increased
dramatically, to a large extent due to results from the {\it Kepler} mission 
\citep{borucki10}, which discovered several thousand planet candidates over 
its four years of operation observing more than 150,000 stars in the constellation
of Cygnus-Lyra \citep{borucki11a,borucki11b, batalha13,burke14,
rowe15,seader15,mullally15, coughlin16}. {\it Kepler} measured transit signals, 
which are periodic decreases in the brightness of the star as another object 
passes in front of it. Based on {\it Kepler} data alone, transit events are identified, 
then vetted, and the resulting {\it Kepler} Objects of Interest (KOIs) are categorized 
as planet candidates or false positives. Sorting out false positives is a complex 
process addressed in many publications \citep{fressin13, coughlin14, seader15, 
mullally15, desert15, mccauliff15, santerne16, morton16}, with current estimates 
for false positive rates ranging from $\sim$~10\% for small planets \citep{fressin13} 
to as high as 55\% for giant planets \citep{santerne16,morton16}.
It is essential to identify false positives in order to derive a reliable list of planet 
candidates, which can then be used to study planet occurrence rates.  

Follow-up observations of KOIs play an important role in determining whether a
transit signal is due to a planet or a different astrophysical phenomenon or source, 
such as an eclipsing binary. In addition, these observations can provide further 
constraints on a planet's properties. In particular, high-resolution imaging can 
reveal whether a close companion was included in the photometric aperture, 
given that the {\it Kepler} detector has 4\arcsec\ wide pixels, and photometry 
was typically extracted from areas a few pixels in size. The current list of {\it Kepler} 
planet candidates does not account for any stellar companions within 
$\sim$ 1\arcsec\--2\arcsec of the primary, since these companions are not 
resolved by the Kepler Input Catalog \citep[e.g.,][]{mullally15}; however, if a 
close companion is present, an adjustment to the transit properties, mainly 
the transit depth and thus the planet radius, is necessary. Even if a companion is 
actually a background star and not bound to the planet host star, the transit depth 
would still be diluted by the light of the companion and thus require a correction.
As shown by \citet{ciardi15}, planet radii are underestimated by an average factor 
of 1.5 if all KOI host stars are incorrectly assumed to be single stars. As a result, 
the fraction of Earth-sized planets is overestimated, having implications on the 
occurrence rate of rocky and volatile-rich exoplanets \citep[e.g.,][]{rogers15}. 

In the solar neighborhood, about 56\% of stars are single, while the rest have one 
or more stellar or brown dwarf companions \citep{raghavan10}. High-resolution 
imaging of {\it Kepler} planet candidate host stars found that about one third of 
these stars have companions within several arcseconds \citep{adams12,adams13,
dressing14,lillo-box14}. Given that not all companion stars are detected, the true 
fraction of KOI host stars with companions is larger; \citet{horch14} derived that 
fraction to be 40\%-50\%, consistent with the findings of \citet{raghavan10}.

Since the beginning of the {\it Kepler} mission in March 2009 and beyond its end
in May 2013, high-resolution imaging of KOI host stars has been carried out as 
part of the {\it Kepler} Follow-Up Observation Program (KFOP). In addition, 
several observing teams that were not part of KFOP carried out imaging surveys 
of KOI host stars. Besides imaging, spectroscopic observations were obtained 
by KFOP and other teams both to constrain stellar parameters and to measure 
the planets' radial velocity signals. All these observations focused on targets of 
the original {\it Kepler} mission and not its successor, {\it K2}. Most of the 
results have been posted on the {\it Kepler} Community Follow-Up Observation 
Program (CFOP) website\footnote{https://exofop.ipac.caltech.edu/cfop.php},
which is meant to facilitate information exchange among observers.

In this work we present in detail the follow-up observations by our KFOP team
using adaptive optics in the near-infrared with instrumentation on the Keck II, 
Palomar 5-m, and Lick 3-m telescopes, as well as results from our optical 
imaging using speckle interferometry at the Gemini-North telescope, the 
Wisconsin-Indiana-Yale-NOAO telescope, and the Discovery Channel Telescope. 
We and additional, independent teams have already published other high-resolution 
imaging observations of KOI host stars using some of these telescopes, as well as 
the Calar Alto 2.2-m telescope, Multiple Mirror Telescope, 
Palomar 1.5-m telescope, and Hubble Space Telescope \citep{howell11, adams12, 
lillo-box12, adams13, law14, dressing14, lillo-box14, wang14, gilliland15, everett15, 
cartier15, wang15a, wang15b, kraus16, baranec16, ziegler16}. In particular, the 
Robo-AO {\it Kepler} Planetary Candidate Survey observed almost all KOI host 
stars with planet candidates using automated laser guide star adaptive optics 
imaging at the Palomar 1.5-m telescope \citep{baranec14, law14, baranec16, 
ziegler16}.

We combine the data presented in this work with additional information on multiplicity 
of KOI host stars already published in the literature to create a comprehensive catalog 
of KOI host star multiplicity. As mentioned above and shown by \citet{horch14}, not 
all companions in these ``multiple'' systems are bound (especially if their projected 
separation on the sky to the primary star is larger than about 1\arcsec); however, 
their presence still has to be taken into account for a correct derivation of transit 
depths. The high-resolution imaging observations typically resolve companions 
down to $\sim$~0.1\arcsec, and we list companion stars out to 4\arcsec. We also 
include companions detected in the UKIRT survey of the {\it Kepler} field; the images, 
which are publicly available on the CFOP website, were taken in the $J$-band 
and typically have spatial resolutions of 0.8\arcsec-0.9\arcsec. 
We introduce our sample in section \ref{sample}, present the imaging observations 
in section~\ref{imaging} and our main results in section~\ref{res}. We discuss our 
results in section~\ref{disc} and summarize them in section~\ref{conclude}.

\section{The Sample}
\label{sample}

\begin{figure}[!t]
\centering
\includegraphics[angle=90, scale=0.39]{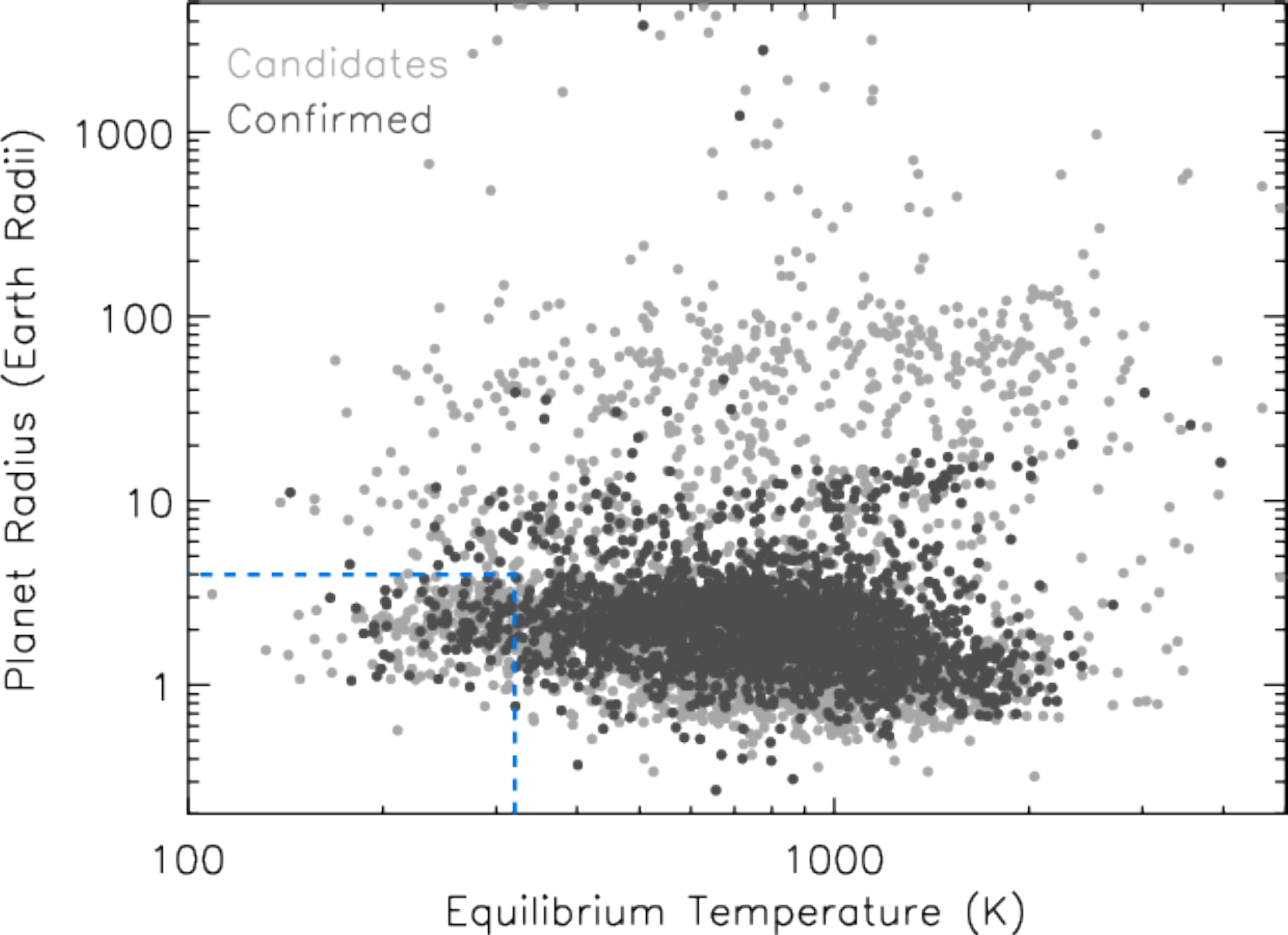}
\caption{The planet radius versus the equilibrium temperature for the 4706
confirmed planets and planet candidates from the latest KOI cumulative
table (note that some candidates with extreme values in these two 
parameters are not shown). The blue dashed lines delineate the region 
of parameter space prioritized in most of the follow-up observations 
($R_p < $ 4 \RE\ and $T_{eq} <$ 320 K). \label{KOIs_Rp_Teq}}
\end{figure}

\begin{figure*}[!t]
\centering
\includegraphics[angle=90, scale=0.61]{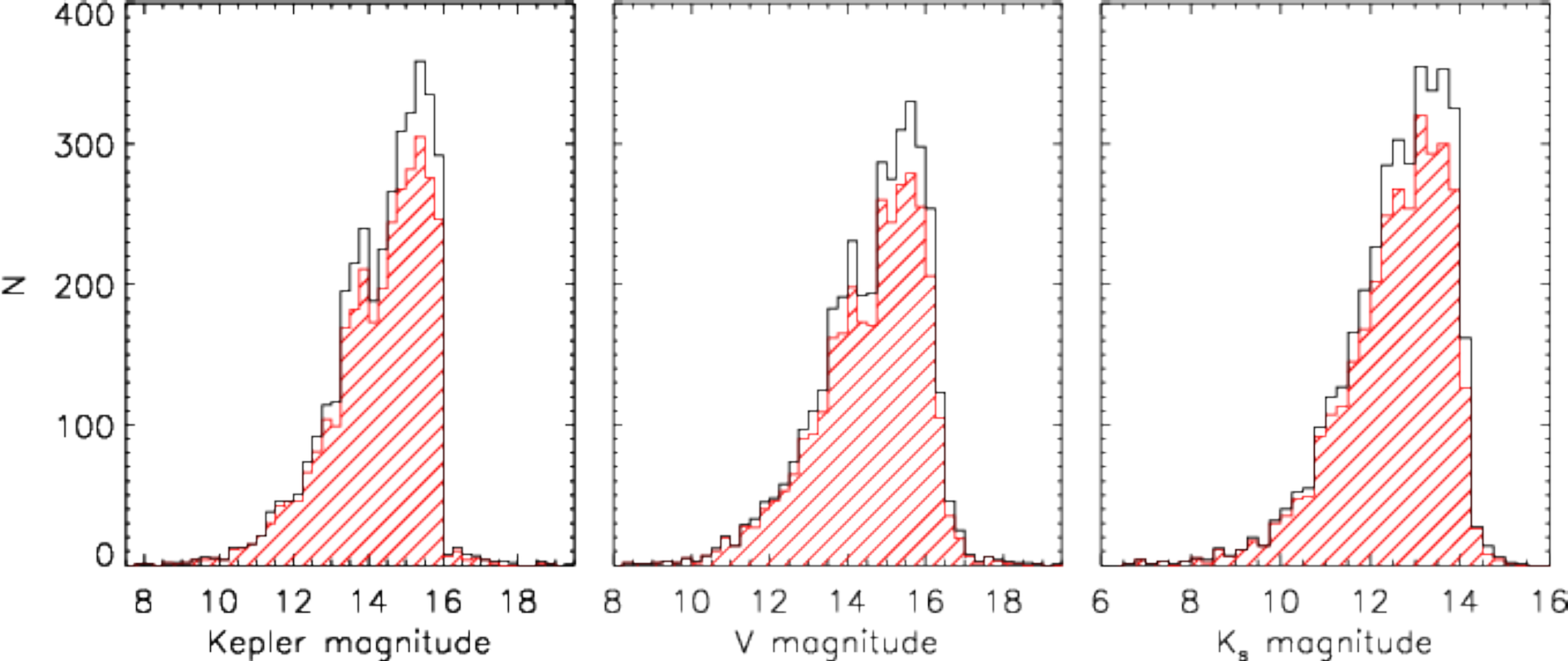}
\caption{Histograms of the magnitudes of all KOI planet host stars (for both
confirmed and candidate planets; black lines) in the {\it Kepler} bandpass 
({\it left}), in the $V$-band ({\it center}), and in the 2MASS $K_s$ band 
({\it right}). The red shaded histograms show the magnitude distributions 
of targets observed with high-resolution imaging.
\label{KOIs_mag_histo}}
\end{figure*}

Over the course of the {\it Kepler} mission, several KOI tables\footnote{All KOI tables can
be accessed at the NASA Exoplanet Archive at http://exoplanetarchive.ipac.caltech.edu.} 
have been released, starting with the Q1-Q6 KOI table in February 2013 and 
ending with the Q1-Q17 DR24 table, which was delivered by the {\it Kepler} project in 
April 2015 and closed to further changes in September 2015. 
The Q1-Q6 table contained 2375 stars (with 2935 KOIs), while the latest KOI table 
includes 6395 stars (with 7470 KOIs). With each new KOI table, some KOIs were
added, others removed, and for some the disposition (planet candidate, false 
positive) or planet parameters changed. The latest KOI cumulative table, which 
mainly incorporated objects from the latest KOI delivery (Q1-Q17 DR24), but also 
has KOIs from previous deliveries, contains a total of 7557 stars (with 8826 KOIs); 
of these, 3665 stars host at least one candidate or confirmed\footnote{Some 
planets were not confirmed with ancillary observations, but rather validated 
using statistical methods \citep[see, e.g.,][]{rowe14,morton16}; in this work 
we we refer to both validated and confirmed planets as confirmed planets.}
planet (we call these stars ``planet host stars'').
As of December 1, 2016, 1627 are stars with confirmed planets (2290 planets), 
2244 are stars with planetary candidates (2416 possible planets), and 4014 are stars 
with transit events classified as false positives. Some stars have both a confirmed 
planet and planet candidate, or a planet (candidate) and a false positive. 
While the cumulative table does not represent a uniform data set, it is the most 
comprehensive list of KOIs with the most accurate dispositions and consistent stellar 
and planetary parameters\footnote{Note that there are a few dozen additional 
confirmed planets in the {\it Kepler} data set that were not identified as KOIs 
by the {\it Kepler} pipeline and are therefore not included in the numbers quoted 
here (but they were assigned {\it Kepler} planet numbers). They can be found in 
the holdings of the NASA Exoplanet Archive.}.

For the {\it Kepler} Follow-Up Observation Program, targets were selected from 
the most recent cumulative KOI list available during each observing season. 
The follow-up program, as well as observing programs by other teams, focused 
almost entirely on planet candidates, and usually prioritized observations based 
on planet radii and equilibrium temperatures, giving higher priority to small 
($\lesssim$ 4 \RE) and cool ($T_{eq} <$ 320 K) planets. A few KOI targets 
were selected based on interesting properties, for example stars with multiple 
planets. These selection criteria narrowed down the original target list of 3665 
planet host stars, but even the high-priority target list contained hundreds of 
{\it Kepler} stars. 
Figure \ref{KOIs_Rp_Teq} shows the distribution of KOI planets (confirmed ones 
and candidates) among different values of equilibrium temperature and planet 
radius. The majority of planets (80\%) have radii less than 4 \RE; 49\% have 
radii less than 2 \RE. 

Obtaining comprehensive imaging and spectroscopic data for the full KOI 
sample is challenging not just due to the large number of targets, but also 
because of the faintness of the sample: 88\% of KOI planet host stars are 
fainter than $V$=13, and 71\% are fainter than $V$=14 (see Figure 
\ref{KOIs_mag_histo}). Many of these faint stars are hosts to Earth-sized 
planet candidates and are thus high-priority targets (see \citealt{everett13} 
and \citealt{howell16} for some recent results). Given the faintness of most 
{\it Kepler} stars, large telescopes are needed to obtain deep limits on the 
presence of nearby companions. In addition, high-resolution imaging
with adaptive optics requires a guide star for wavefront sensing; beyond
$V$ magnitudes of 14-15, the star is often too faint to be used as the guide
star, and a laser guide star has to be used instead.

Different groups observed {\it Kepler} stars with high-resolution imaging techniques 
(adaptive optics, speckle interferometry, lucky imaging, and some space-based 
observations); the magnitude distributions of observed targets are shown in
Figure \ref{KOIs_mag_histo} (red shaded histograms). The Robo-AO imaging 
at the Palomar 1.5-m telescope \citep{baranec14} contributed most of the 
observations: of the 3665 
stars that host at least one KOI planet candidate or confirmed planet, Robo-AO 
observed 3093. In total, 3183 planet host stars (or 87\%) have high-resolution 
images. When considering the 4706 confirmed planets and planet candidates 
from the latest KOI cumulative table (the number is larger than the number of 
stars, since some stars have more than one planet), 90\% (or 4213 planets) 
have been covered by high-resolution imaging. 
It should be noted that the imaging data are not only important for detecting
companion stars, but were also useful in confirming many of the planet
candidates \citep[e.g.,][]{batalha11}.

\begin{figure}[!t]
\centering
\includegraphics[angle=90, scale=0.4]{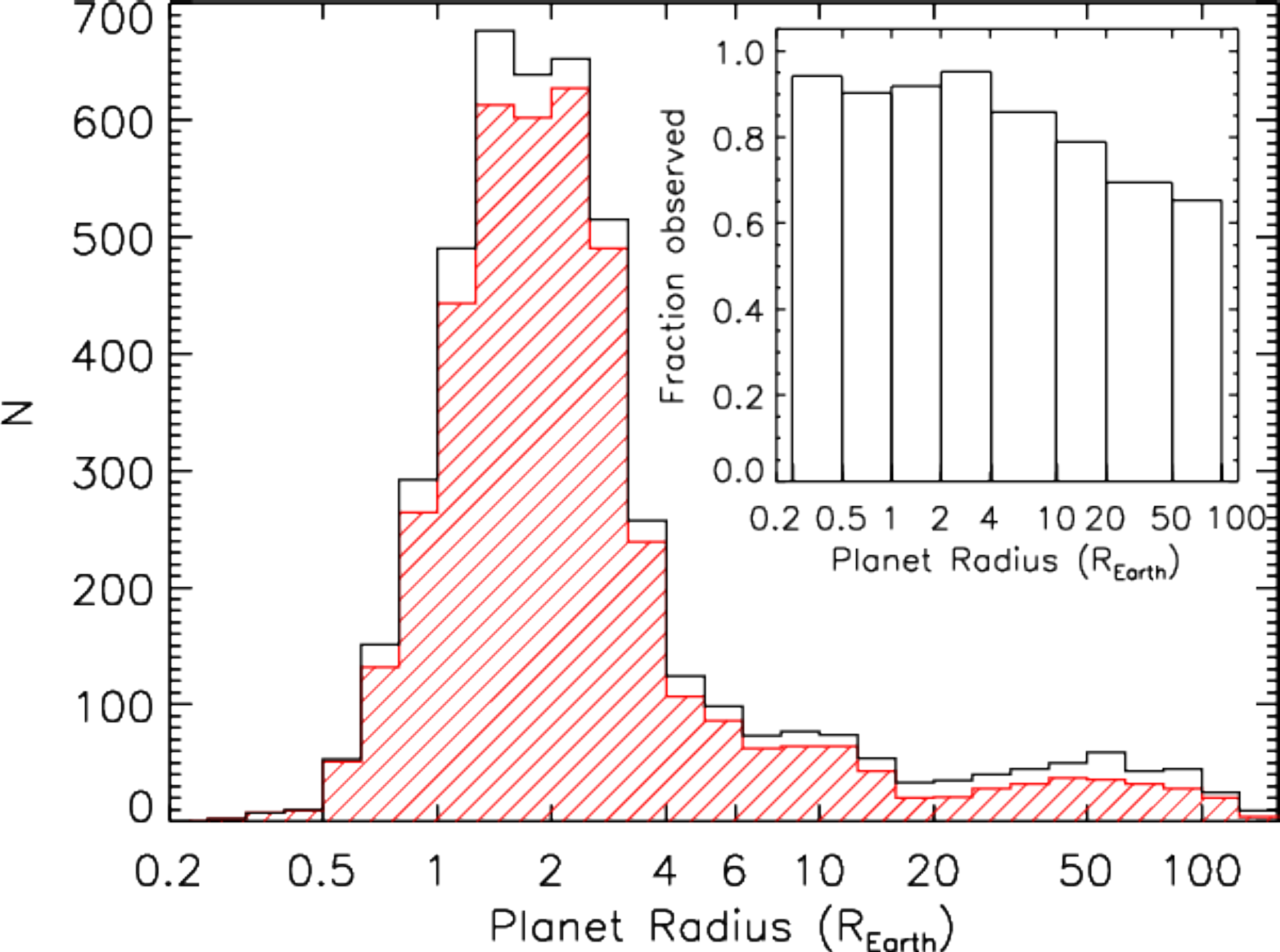}
\caption{Histogram of planet radii of all KOI planet candidates and 
confirmed planets from the latest KOI cumulative table ({\it black})
and for those targeted by high-resolution imaging ({\it red}). The 
insert shows the fraction of planets observed for different bins of
planet radii (0.25-0.5, 0.5-1, 1-2, 2-4, 4-10, 10-20, 20-50, and 
50-100 \RE).
\label{KOIs_Rp_histo}}
\end{figure}

The distribution of planet radii (taken from the latest KOI cumulative table) 
for the whole sample and the high-resolution imaging sample of KOI planets 
can be seen in Figure \ref{KOIs_Rp_histo}. 
About 93\% of planets with radii less than 4 \RE\ were observed, while this 
fraction is about 76\% for larger planets ($R_p > $ 4 \RE); about two thirds
of KOI planets larger than 20 \RE\ have been targeted by high-resolution 
imaging. This just follows from the selection criteria for targets for most of 
the imaging programs, since the smallest planets had the highest priority.
We note that the majority of the very large planets ($R_p > $ 20 \RE) are 
still planet candidates (they also constitute just $\sim$ 9\% of the planet 
sample). It is likely that most of them will not be confirmed as planets,
but instead as brown dwarfs and eclipsing binaries (\citealt{santerne16} 
determined a false-positive rate of 55\% for giant planets with periods 
$<$ 400 d; \citealt{morton16} found a mean false positive probability of
84\% for planet candidates with radii $>$ 15 \RE). Others likely have 
highly inaccurate planet radii due to very uncertain stellar radii or 
unreliable transit fits by the {\it Kepler} pipeline (see, e.g., KOI 1298.02 
and KOI 2092.03 with $R_p$ of 39 and 30 \RE, respectively, from the 
KOI table, which were validated as planets with radii of 1.82 \RE\ and 
4.01\RE, respectively; \citealt{rowe14}).

\section{Observations}
\label{imaging}

\begin{deluxetable*}{llcccl}[h]
\rotate
\tablewidth{0.95\linewidth}
\tablecaption{High-Resolution Imaging Observations of KOI Host Stars 
\label{obs_list}}
\tablehead{
\colhead{Telescope} & \colhead{Instrument} & \colhead{Band} &
\colhead{PSF} & \colhead{N} & \colhead{References} \\
\colhead{(1)} & \colhead{(2)} & \colhead{(3)} & \colhead{(4)} &
\colhead{(5)} & \colhead{(6)}}
\startdata
Calar Alto (2.2 m) & AstraLux & $i', z'$ & 0.21\arcsec & 234  & \citet{lillo-box12, lillo-box14} \\
DCT (4 m) & DSSI & 692, 880 nm & 0.04\arcsec & 75  & this work \\
Gemini North (8 m) & DSSI & 562, 692, 880 nm & 0.02\arcsec & 158  & \citet{horch12,horch14}, 
\citet{everett15}, this work \\
HST (2.4 m) & WFC3 & $F555W, F775W$ & 0.08\arcsec & 34  & \citet{gilliland15, cartier15} \\
Keck II (10 m) & NIRC2 & $J, H, K$ & 0.05\arcsec & 667  & 
                           \citet{wang15a,wang15b,kraus16,baranec16,ziegler16}, this work \\
LBT (8 m) & LMIRCam & $K_s$ & \nodata & 24  & unpublished \\
Lick (3 m) & IRCAL & $J, H$ & 0.35\arcsec & 324  & this work \\
MMT (6.5 m) & ARIES & $J, K_s$ & 0.15\arcsec & 128  & \citet{adams12, adams13, dressing14} \\
Palomar (1.5 m) & Robo-AO & $i', LP600$ & 0.15\arcsec & 3320 & \citet{law14,baranec16,ziegler16} \\
Palomar (5 m) & PHARO & $J, H, K$ & 0.12\arcsec & 449  & \citet{adams12,wang15a,wang15b}, 
this work \\
WIYN (3.5 m) & DSSI & 562, 692, 880 nm & 0.05\arcsec & 681  & \citet{howell11}, \citet{horch14}, 
this work \\
\enddata
\tablecomments{Column (1) lists the telescope and the mirror size (in parentheses), 
column (2) the instrument used, column (3) the various bands/filters of the observations,
column (4) the typical width of the point spread function in arcseconds, column (5) the 
number of KOI host stars observed at each facility, and column (6) the references where the
data are published.}
\end{deluxetable*}

\begin{deluxetable*}{llccccccccccl}\scriptsize
\rotate
\tablewidth{0.95\linewidth}
\tablecaption{Summary of KOI Host Stars Observed with High-Resolution Imaging 
Techniques  \label{KOI_obs_summary}}
\tablehead{
\colhead{KOI} & \colhead{KICID} & \colhead{CP} & \colhead{PC} & \colhead{FP} & 
\colhead{$R_{p,min}$} & \colhead{KOI($R_{p,min}$)} & \colhead{$T_{eq,min}$} & 
\colhead{KOI($T_{eq,min}$)} & \colhead{$Kp$} & \colhead{$V$} & \colhead{$K_s$} & 
\colhead{Observatories}  \\
\colhead{(1)} & \colhead{(2)} & \colhead{(3)} & \colhead{(4)} &
\colhead{(5)} & \colhead{(6)} & \colhead{(7)} & \colhead{(8)} &
\colhead{(9)} & \colhead{(10)} & \colhead{(11)} & \colhead{(12)} &
\colhead{(13)}} 
\startdata
   1 &  11446443 &  1 &  0 &  0 &   12.9 &   1.01 &  1344 &    1.01 &  11.34 &  11.46 &   9.85 & Keck,Pal1.5,WIYN \\
   2 &  10666592 &  1 &  0 &  0 &   16.4 &   2.01 &  2025 &    2.01 &  10.46 &  10.52 &   9.33 & Keck,Pal1.5,WIYN \\
   3 &  10748390 &  1 &  0 &  0 &    4.8 &   3.01 &   801 &    3.01 &   9.17 &   9.48 &   7.01 & Keck,MMT,Pal1.5,WIYN \\
   4 &   3861595 &  0 &  1 &  0 &   13.1 &   4.01 &  2035 &    4.01 &  11.43 &  11.59 &  10.19 & Pal1.5,WIYN \\
   5 &   8554498 &  0 &  2 &  0 &    0.7 &   5.02 &  1124 &    5.02 &  11.66 &  11.78 &  10.21 & Keck,Pal1.5,Pal5,WIYN \\
   6 &   3248033 &  0 &  0 &  1 &   50.7 &   6.01 &  2166 &    6.01 &  12.16 &  12.33 &  10.99 & CAHA \\
   7 &  11853905 &  1 &  0 &  0 &    4.1 &   7.01 &  1507 &    7.01 &  12.21 &  12.39 &  10.81 & Pal1.5,Pal5,WIYN \\
   8 &   5903312 &  0 &  0 &  1 &    2.0 &   8.01 &  1752 &    8.01 &  12.45 &  12.62 &  11.04 & Pal5 \\
  10 &   6922244 &  1 &  0 &  0 &   14.8 &  10.01 &  1521 &   10.01 &  13.56 &  13.71 &  12.29 & Pal1.5,Pal5,WIYN \\
  11 &  11913073 &  0 &  0 &  1 &   10.5 &  11.01 &  1031 &   11.01 &  13.50 &  13.75 &  11.78 & Pal5 \\
  12 &   5812701 &  1 &  0 &  0 &   14.6 &  12.01 &   942 &   12.01 &  11.35 &  11.39 &  10.23 & CAHA,Keck,Lick,Pal1.5,WIYN \\
  13 &   9941662 &  1 &  0 &  0 &   25.8 &  13.01 &  3560 &   13.01 &   9.96 &   9.87 &   9.43 & DCT,Gem,Keck,MMT,Pal1.5,Pal5,WIYN \\
  14 &   7684873 &  0 &  0 &  1 &    5.9 &  14.01 &  2405 &   14.01 &  10.47 &  10.62 &   9.84 & Pal5 \\
  17 &  10874614 &  1 &  0 &  0 &   13.4 &  17.01 &  1355 &   17.01 &  13.30 &  13.41 &  11.63 & Pal1.5,Pal5 \\
  18 &   8191672 &  1 &  0 &  0 &   15.3 &  18.01 &  1640 &   18.01 &  13.37 &  13.47 &  11.77 & Gem,Pal1.5,Pal5 \\
  20 &  11804465 &  1 &  0 &  0 &   18.2 &  20.01 &  1338 &   20.01 &  13.44 &  13.58 &  12.07 & Pal1.5,Pal5,WIYN \\
  22 &   9631995 &  1 &  0 &  0 &   12.2 &  22.01 &  1000 &   22.01 &  13.44 &  13.64 &  12.04 & Pal1.5,Pal5,WIYN \\
  28 &   4247791 &  0 &  0 &  1 &   83.1 &  28.01 &  1412 &   28.01 &  11.26 &  11.79 &  10.29 & Pal5,WIYN \\
  31 &   6956014 &  0 &  0 &  1 &   45.3 &  31.01 &  6642 &   31.01 &  10.80 &  11.92 &   7.94 & Pal5 \\
  33 &   5725087 &  0 &  0 &  1 &   63.1 &  33.01 &  9970 &   33.01 &  11.06 &  11.10 &   7.59 & Pal5 \\
  41 &   6521045 &  3 &  0 &  0 &    1.3 &  41.02 &   674 &   41.03 &  11.20 &  11.36 &   9.77 & CAHA,Keck,MMT,Pal1.5,Pal5,WIYN \\
  42 &   8866102 &  1 &  0 &  0 &    2.5 &  42.01 &   859 &   42.01 &   9.36 &   9.60 &   8.14 & Keck,MMT,Pal1.5,Pal5,WIYN \\
  44 &   8845026 &  0 &  0 &  1 &   11.9 &  44.01 &   462 &   44.01 &  13.48 &  13.71 &  11.66 & Keck,Lick,Pal1.5,WIYN \\
  46 &  10905239 &  2 &  0 &  0 &    0.9 &  46.02 &  1075 &   46.02 &  13.77 &  13.80 &  12.01 & Keck,Pal1.5,WIYN \\
  49 &   9527334 &  1 &  0 &  0 &    2.7 &  49.01 &   886 &   49.01 &  13.70 &  13.56 &  11.92 & CAHA,Pal1.5,WIYN \\
  51 &   6056992 &  0 &  1 &  0 &   49.8 &  51.01 &   833 &   51.01 &  13.76 &  14.02 &  14.31 & CAHA,Pal1.5,WIYN \\
\enddata
\tablecomments{The full table is available in a machine-readable form in the online
journal. A portion is shown here for guidance regarding content and form. \\
Column (1) lists the KOI number of the star, column (2) its identifier from the Kepler
Input Catalog (KIC), columns (3) to (5) the number of confirmed planets (CP), planet 
candidates (PC), and false positives (FP), respectively, in the system, column (6) radius 
of the smallest planet in the system (in \RE) and column (7) its KOI number, column (8)
the equilibrium temperature of the coolest planet in the system (in K) and column (9)
its KOI number, columns (10) to (12) the {\it Kepler}, $V$, and $K_s$ magnitudes
of the KOI host stars, and column (13) the observatories where data were taken.
Note that if a system contains both planets and false positives, only the planets are
used to determine the smallest planet radius and lowest equilibrium temperature.
The abbreviations in column (13) have the following meaning: CAHA -- Calar Alto,
DCT -- Discovery Channel Telescope, Gem -- Gemini N, HST -- Hubble Space
Telescope, Keck -- Keck II, LBT - Large Binocular Telescope, Lick -- Lick-3m,
MMT -- Multiple Mirror Telescope, Pal1.5 -- Palomar-1.5m, Pal5 -- Palomar-5m,
WIYN -- Wisconsin-Indiana-Yale-NOAO telescope.}
\end{deluxetable*}

\begin{deluxetable*}{llccccccl}[h] \scriptsize
\tablewidth{0.95\linewidth}
\tablecaption{Summary of High-Resolution Imaging Observations of KOI Host Stars 
\label{KOI_imaging_properties}}
\tablehead{
\colhead{KOI} & \colhead{KICID} & \colhead{Telescope} & \colhead{Instrument} & 
\colhead{Filter/Band} & \colhead{PSF (\arcsec)} & \colhead{$\Delta m$} & 
\colhead{$d_{\Delta m}$ (\arcsec)} & \colhead{Obs. Date}  \\
\colhead{(1)} & \colhead{(2)} & \colhead{(3)} & \colhead{(4)} &
\colhead{(5)} & \colhead{(6)} & \colhead{(7)} & \colhead{(8)} &
\colhead{(9)}} 
\startdata
   1 &  11446443 &   Keck &    NIRC2 &   $K'$ & \nodata &   7.60 & 0.50 &  2012-07-06 \\
   1 &  11446443 &   Keck &    NIRC2 &   $K'$ & \nodata &   4.11 & 0.03 &  2012-07-06 \\
   1 &  11446443 &   Keck &    NIRC2 &   $K'$ & \nodata &   5.90 & 0.50 &  2012-07-06 \\
   1 &  11446443 &   Keck &    NIRC2 &    $J$ &  0.04 &   5.61 &  0.5 &  2014-07-17 \\
   1 &  11446443 &   Keck &    NIRC2 &    $H$ &  0.04 &   6.07 &  0.5 &  2014-07-17 \\
   1 &  11446443 &   Keck &    NIRC2 &   $Ks$ &  0.04 &   4.93 &  0.5 &  2014-07-17 \\
   1 &  11446443 & Pal1.5 &  Robo-AO &   $i'$ &  0.12 &   5.40 &  0.5 &  2012-07-16 \\
   1 &  11446443 &   WIYN &     DSSI & 880 nm &  0.05 &   2.73 &  0.2 &  2011-06-13 \\
   1 &  11446443 &   WIYN &     DSSI & 692 nm &  0.05 &   3.28 &  0.2 &  2011-06-13 \\
   1 &  11446443 &   WIYN &     DSSI & 880 nm &  0.05 &   2.67 &  0.2 &  2013-09-21 \\
   1 &  11446443 &   WIYN &     DSSI & 692 nm &  0.05 &   3.50 &  0.2 &  2013-09-21 \\
   1 &  11446443 &   WIYN &     DSSI & 880 nm &  0.05 &   2.84 &  0.2 &  2013-09-23 \\
   1 &  11446443 &   WIYN &     DSSI & 692 nm &  0.05 &   2.82 &  0.2 &  2013-09-23 \\
   2 &  10666592 &   Keck &    NIRC2 &   $K'$ & \nodata &   7.20 & 0.50 &  2012-08-14 \\
   2 &  10666592 &   Keck &    NIRC2 &   $K'$ & \nodata &   5.80 & 0.50 &  2012-08-14 \\
   2 &  10666592 &   Keck &    NIRC2 &   $K'$ & \nodata &   4.56 & 0.03 &  2014-08-13 \\
   2 &  10666592 & Pal1.5 &  Robo-AO &   $i'$ &  0.12 &   4.60 &  0.2 &  2012-07-16 \\
   2 &  10666592 &   WIYN &     DSSI & 880 nm &  0.05 &   2.78 &  0.2 &  2011-06-13 \\
   2 &  10666592 &   WIYN &     DSSI & 692 nm &  0.05 &   4.01 &  0.2 &  2011-06-13 \\
   3 &  10748390 &   Keck &    NIRC2 &   $K'$ & \nodata &   7.70 & 0.50 &  2012-07-05 \\
   3 &  10748390 &   Keck &    NIRC2 &   $K'$ & \nodata &   4.17 & 0.03 &  2012-07-05 \\
   3 &  10748390 &    MMT &    ARIES &   $Ks$ &  0.15 &   8.00 &  1.0 &  2012-10-02 \\
   3 &  10748390 &    MMT &    ARIES &    $J$ &  0.20 &   8.00 &  1.0 &  2012-10-02 \\
   3 &  10748390 & Pal1.5 &  Robo-AO &   $i'$ &  0.12 &   4.60 &  0.2 &  2012-07-16 \\
   3 &  10748390 &   WIYN &     DSSI & 880 nm &  0.05 &   3.45 &  0.2 &  2011-06-13 \\
   3 &  10748390 &   WIYN &     DSSI & 692 nm &  0.05 &   3.76 &  0.2 &  2011-06-13 \\
   4 &   3861595 & Pal1.5 &  Robo-AO &   $i'$ &  0.12 &  \nodata & \nodata &  2012-07-16 \\
   4 &   3861595 &   WIYN &     DSSI & 692 nm &  0.05 &   3.06 &  0.2 &  2010-09-17 \\
   4 &   3861595 &   WIYN &     DSSI & 562 nm &  0.05 &   3.46 &  0.2 &  2010-09-17 \\
   4 &   3861595 &   WIYN &     DSSI & 692 nm &  0.05 &   3.58 &  0.2 &  2010-09-18 \\
   4 &   3861595 &   WIYN &     DSSI & 562 nm &  0.05 &   4.01 &  0.2 &  2010-09-18 \\
   5 &   8554498 &   Keck &    NIRC2 &   $K'$ & \nodata &   6.70 & 0.50 &  2012-08-14 \\
   5 &   8554498 &   Keck &    NIRC2 &   $K'$ & \nodata &   1.12 & 0.03 &  2012-08-14 \\
   5 &   8554498 &   Keck &    NIRC2 &   $K'$ &  0.05 &   8.00 &  0.5 &  2013-08-20 \\
   5 &   8554498 & Pal1.5 &  Robo-AO &   $i'$ &  0.12 &   4.60 &  0.2 &  2012-07-16 \\
   5 &   8554498 &   Pal5 &    PHARO &    $J$ &  0.24 &   5.08 &  0.5 &  2009-09-10 \\
   5 &   8554498 &   WIYN &     DSSI & 692 nm &  0.05 &   3.02 &  0.2 &  2010-09-17 \\
   5 &   8554498 &   WIYN &     DSSI & 562 nm &  0.05 &   3.44 &  0.2 &  2010-09-17 \\
   5 &   8554498 &   WIYN &     DSSI & 692 nm &  0.05 &   3.13 &  0.2 &  2010-09-18 \\
   5 &   8554498 &   WIYN &     DSSI & 562 nm &  0.05 &   3.50 &  0.2 &  2010-09-18 \\
   5 &   8554498 &   WIYN &     DSSI & 692 nm &  0.05 &   3.12 &  0.2 &  2010-09-21 \\
   5 &   8554498 &   WIYN &     DSSI & 880 nm &  0.05 &   2.38 &  0.2 &  2010-09-21 \\
   6 &   3248033 &   CAHA & AstraLux &   $i'$ &  0.16 &   3.24 &  0.5 &  2013-06-23 \\
\enddata
\tablecomments{The full table is available in a machine-readable form in the online
journal. A portion is shown here for guidance regarding content and form. \\
Column (1) lists the KOI number of the star, column (2) its identifier from the Kepler
Input Catalog (KIC), column (3) the telescope where the images were taken (see
the notes of Table \ref{KOI_obs_summary} for an explanation of the abbreviations),
column (4) the instrument used, column (5) the filter/band of the observation, column (6)
the typical width of the stellar PSF in arcseconds, column (7) the typical sensitivity 
$\Delta m$ (usually 5$\sigma$) at a certain separation (in arcseconds) from the primary 
star, column (8) the separation for the $\Delta m$ value from column (7), and column 
(9) the date of the observation (in year-month-day format). Sensitivity curves with 
$\Delta m$ values measured at a range of separations are available on the CFOP 
website at  https://exofop.ipac.caltech.edu/cfop.php.}
\end{deluxetable*}

Several observing facilities were used to obtain high-resolution images of
KOI host stars. Table \ref{obs_list} lists the various telescopes, instruments 
used, filter bandpasses, typical PSF widths,
number of targets observed, and main references for the published results.
The four main observing techniques employed are adaptive optics 
(Keck, Palomar, Lick, MMT), speckle interferometry (Gemini North, WIYN, 
DCT), lucky imaging (Calar Alto), and imaging from space with HST. 
A total of 3557 KOI host stars were observed at 11 facilities with 9 different
instruments, using filters from the optical to the near-infrared. 
In addition, 10 of these stars were also observed at the 8-m Gemini North 
telescope by \citet{ziegler16} using laser-guide-star adaptive optics. The 
largest number of KOI host stars (3320) were observed using Robo-AO at 
the Palomar 1.5-m telescope \citep{baranec14,law14,baranec16,ziegler16}.

Table \ref{KOI_obs_summary} lists the KOI host stars that were observed
with high-resolution imaging, together with the observatories that were used and 
some of the planet parameters and stellar magnitudes. Some KOIs that are 
currently dispositioned as false positives (i.e., there is no planet, candidate or 
confirmed, orbiting the star) were observed, too, since at the time their 
observations were carried out the disposition was either set to planet 
candidate or was not set. 
Of the 3557 observed stars, almost two thirds (61\% or 2187 stars) were 
observed at only one telescope facility with one instrument, usually just using 
one filter; 696 stars were observed at two telescopes, while the remaining 674 
stars were observed at two or more facilities. Combining the data from all 
telescopes, 1431 stars were observed with two or more filters.

In Table \ref{KOI_imaging_properties} we provide a more detailed summary
of the high-resolution observations, including the dates of the observations, the
telescopes, instruments, and filters used, and, for most observations, the typical 
PSF width and sensitivity (given as $\Delta m$ -- typically a 5$\sigma$ measurement -- 
at a certain separation from the primary star). A total of 8332 observations were 
carried out from 2009 September to 2015 October covering 3557 stars. 
The median and mean PSF widths of all the high-resolution imaging observations 
where this parameter was reported are both 0.12\arcsec; $\sim$ 90\% of the 
observations have PSF widths smaller than 0.16\arcsec. 
For the image sensitivities, the majority of $\Delta m$ values are given at a projected 
separation of 0.5\arcsec\ (for most AO observations and lucky imaging) or 0.2\arcsec\ 
(for speckle observations) from the primary star. Median values for $\Delta m$ at 
0.2\arcsec\ and 0.5\arcsec\ are 3.0 and 6.0, respectively. The $\Delta m$ 
values at 0.03\arcsec\ from the primary star are measurements from images 
using non-redundant aperture masking at the Keck telescope \citep{kraus16}; 
this technique enables binaries to be resolved at projected separations of just 
a few tenths of an arcsecond \citep[see][]{kraus16}. The median $\Delta m$ 
value at 0.03\arcsec\ is 3.94.

For this work, we reduced and analyzed our (for the most part not yet published)
AO observations at Keck, Palomar, and Lick (see section \ref{our_AO}), and our 
speckle imaging observations from Gemini North, WIYN, and DCT (see section 
\ref{our_speckle}). We also gathered results from all {\it Kepler} follow-up imaging 
observations, carried out by KFOP and other observing teams, from the literature 
and a few unpublished results from CFOP. These observations will be briefly 
introduced in section \ref{other_highres}.

\vspace{1ex}

\subsection{Adaptive Optics at Keck, Palomar, and Lick}
\label{our_AO}

\begin{deluxetable*}{ll}
\tablewidth{0.95\linewidth}
%\tabletypesize{\small}     %{\scriptsize}   
\tablecaption{Observing Log for our Adaptive Optics Runs at Keck, Palomar, and Lick
\label{obs_table_AO}}
\tablehead{
 \colhead{Telescope and Instrument} & \colhead{UT Dates (YYYYMMDD)}}
\startdata
KeckII, NIRC2 & 20120505, 20120606, 20120704, 20120825, 20130615, 20130706, 20130723, \\
 & 20130808, 20130819, 20140612, 20140613, 20140702, 20140717, 20140718, \\
 & 20140811, 20140812, 20140817, 20140904, 20140905, 20150714, 20150731, \\
 & 20150804, 20150806, 20150807 \\
Palomar Hale, PHARO & 20090907, 20090908, 20090909, 20090910, 20100630, 20100701, 20100702, \\
 & 20120907, 20120908, 20130624, 20140710, 20140711, 20140712, 20140713, \\
 & 20140714, 20140716, 20140717, 20140807, 20140808, 20140810, 20140813, \\
 & 20150527, 20150528, 20150529, 20150827, 20150828, 20150829, 20150830, \\
 & 20150831 \\
Lick Shane, IRCAL & 20110908, 20110909, 20110910, 20110911, 20110912, 20120706, 20120707, \\
 & 20120708, 20120709, 20120710, 20120805, 20120806, 20120901, 20120902, \\
 & 20120903, 20130715, 20130716, 20130717, 20130718, 20130916, 20130918 \\
\enddata
\end{deluxetable*}

We carried out observations at the Keck, Palomar, and Lick Observatory using
the facility adaptive optics systems and near-infrared cameras from 2009 to 2015. 
Table \ref{obs_table_AO} lists the various observing runs whose results are
presented here. At Palomar and Lick, we used the targets themselves as natural 
guide stars (NGS) for the adaptive optics system, while at Keck we used our targets 
as natural guide stars when they were sufficiently bright, and the laser guide star 
(LGS) for the fainter targets (roughly $Kp$ $>$ 14.5). The majority of our nights 
at Keck employed NGS.   

At Keck, we observed with the 10-m Keck II telescope and NIRC2
\citep{wizinowich04}. The pixel scale of NIRC2 was 0.01\arcsec/pixel, 
resulting in a field of view of about 10\arcsec $\times$ 10\arcsec. 
We observed our targets in a narrow $K$-band filter, Br$\gamma$,
which has a central wavelength of 2.1686 $\mu$m. In most cases, when
a companion was detected, we also observed the target in a narrow-band
$J$ filter, $Jcont$, which is centered at 1.2132 $\mu$m. We dithered the
target in a 3-point pattern to place it in all quadrants of the array except 
for the lower left one (which has somewhat larger noise levels).  

At Palomar, we used the 5-m Hale telescope with PHARO \citep{hayward01}. 
We used the 0.025\arcsec/pixel scale, which yielded a field of view of about 
25\arcsec $\times$ 25\arcsec. As at Keck, we typically used a narrow-band 
filter in the $K$-band, Br$\gamma$ centered at 2.18 $\mu$m, to 
observe our targets. When a companion was detected, we usually also 
observed our targets in the $J$ filter (centered at 1.246 $\mu$m). We dithered
each target in a 5-point quincunx pattern to place it in all four quadrants of
the array and at the center. 

At Lick, we used the 3-m Shane telescope and IRCAL \citep{lloyd00}. 
With its 0.075\arcsec/pixel scale, it offered a field of view of about 19\arcsec $\times$ 
19\arcsec. We observed our targets with the $J$ filter (centered at 1.238 $\mu$m) 
or the $H$ filter (centered at 1.656 $\mu$m). Each target was dithered on the 
array in a 5-point pattern.

At all three telescopes, the integration time for each target varied, depending
on its brightness. It was typically between 5 and 60 sec per frame, for a total
exposure time of 10-15 minutes. Some of the fainter targets required longer
exposures, but, in order to cover a reasonable number of targets on any 
given night, we tried to limit the time spent on any target to about half an hour.
Over all observing runs at Keck, Palomar, and Lick, we observed 253, 317, 
and 310 unique KOI host stars, respectively. Some were observed in more 
than one filter, and some were observed at more than one telescope. Overall, 
we covered 770 unique KOI host stars with our adaptive optics imaging. 

To reduce the images, we first created nightly flatfields, and for each target
we constructed a sky image by median-filtering and coadding the dithered
frames. Each frame was then flatfielded and sky-subtracted, and the dithered 
frames combined.
The final, co-added images obtained at Palomar are typically 14\arcsec $\times$ 
14\arcsec\ in size, but there is a spread ranging from 10\arcsec\ to 34\arcsec.
The final images from Keck are usually 4\arcsec $\times$ 4\arcsec\ in size, with
some up to 16\arcsec $\times$ 16\arcsec. Finally, the reduced Lick images
are $\sim$ 23\arcsec $\times$ 23\arcsec\ in size.

We used aperture photometry to measure the relative brightness of the stars
in each reduced frame. We used an aperture radius equal to the FWHM of
the primary star and a sky annulus between about 3 and 5 times the FWHM.
For close companions, we reduced the FWHM to minimize contamination,
and we adjusted the sky annulus to exclude emission from the sources.
The FWHM values varied depending on the observing conditions; at
Palomar, the mean and median FWHM values were 6.6 and 5.4 pixels
(or 0.165\arcsec\ and 0.135\arcsec), respectively, at Keck 5.7 and 5.3 pixels
(0.057\arcsec\ and 0.053\arcsec), and at Lick 5.0 and 4.6 pixels (0.375\arcsec\
and 0.345\arcsec).  
The $J$-, $H$-, and $K$-band measurements were converted from data
numbers to magnitudes using the magnitudes of the primary source from
the Two Micron All Sky Survey \citep[2MASS;][]{skrutskie06}. 

We also measured image sensitivities for each target by calculating the standard 
deviation of the background ($\sigma$) in concentric annuli around the main star; 
the radii of the annuli were set to multiples of the FWHM of the primary star.  
As can be seen from Table \ref{obs_list}, the typical FWHM of the stellar PSF 
was 0.05\arcsec\ at Keck, 0.12\arcsec\ at Palomar, and 0.2\arcsec\ at Lick. 
Within each ring, we determined 5$\sigma$ limits.

Some of the AO data presented here (mostly from Palomar and Keck) have 
already been published in the literature \citep{torres11, batalha11, fortney11, 
ballard11, ballard13, borucki12, borucki13, gautier12, adams12, marcy14, 
everett15, torres15, teske15}; they were typically used to confirm {\it Kepler} 
planet candidates.

\subsection{Speckle Interferometry}
\label{our_speckle}

\begin{deluxetable*}{ll}
\tablewidth{0.99\linewidth}
%\tabletypesize{\small}     %{\scriptsize}   
\tablecaption{Observing Log for our Speckle Interferometry Runs at WIYN, Gemini North,
and DCT
\label{obs_table_speckle}}
\tablehead{
 \colhead{Telescope} & \colhead{UT Dates (YYYYMMDD)}}
\startdata
WIYN  & 20100618, 20100619, 20100620, 20100621, 20100622, 20100624,  20100917, \\ 
           & 20100918, 20100919, 20100920, 20100921, 20101023, 20101024, 20101025, \\ 
           & 20110611, 20110612, 20110613, 20110614, 20110615, 20110616, 20110907, \\
           & 20110908, 20110909, 20110910, 20110911, 20120927, 20120929, 20120930, \\
           & 20121001, 20121003, 20121004, 20121005, 20130525, 20130526, 20130527, \\
           & 20130528, 20130921, 20130922, 20130923, 20130924, 20130925, 20150927, \\
           & 20150928, 20150929, 20150930, 20151002, 20151003, 20151004, 20151023, \\
           & 20151024, 20151027 \\
Gemini North & 20120727, 20120728, 20130725, 20130726, 20130727, 20130728, 20130729, \\
           & 20130731, 20140719, 20140722, 20140723, 20140724, 20140725, 20150711, \\
           & 20150712, 20150714, 20150715, 20150718, 20150719, 20150720 \\
DCT & 20140321, 20140323, 20140617, 20140618, 20141001, 20141002 \\
\enddata
\end{deluxetable*}

Our team also carried out speckle imaging using the Differential Speckle Survey 
Instrument (DSSI; \citealt{horch09,horch10}) at Gemini North, the 
Wisconsin-Indiana-Yale-NOAO (WIYN) telescope, and at the Discovery Channel 
Telescope (DCT) from 2010 to 2015. Table \ref{obs_table_speckle} lists the various 
observing dates at the three telescopes. 
At the 8-m Gemini North telescope, 158 unique KOI host stars were observed, while at 
the 3.5-m WIYN telescope, 681 stars were targeted. The more recent observing 
runs at the 4-m DCT telescope covered 75 stars. Overall, at all three telescopes
the observations were directed at 828 unique KOI host stars. 

Targets were observed simultaneously in two bands, centered at 562 nm and 692 nm 
(both with a band width of 40 nm), or at 692 nm and 880 nm (the latter with a band 
width of 50 nm). Some targets have data in all three bands.
The field of view of the speckle images is smaller than that of the AO images,
about 3\arcsec\ on each side, but the PSF widths are narrower (0.02\arcsec-0.05\arcsec),
resulting in better spatial resolution. Some of the results on DSSI observations
of KOIs can be found in \citet{howell11}, \citet{horch12, horch14}, \citet{everett15}, 
and \citet{teske15}. 

A description of typical observing sequences done for KOI host stars using DSSI 
and a detailed explanation of the data reduction methods can be found in 
\citet{horch11} and \citet{howell11}. In addition, M. E. Everett et al. (2017, in 
preparation) will document all of the speckle imaging in more detail.
Here we briefly outline the reduction and analysis of the speckle data.
The reduction of speckle observations takes place in both image and Fourier 
space. First, the autocorrelation function and triple correlation function are 
calculated for each frame of an image set centered on the target star's speckle 
pattern. These functions are averaged over all frames and then converted 
through Fourier transforms into a power spectrum and bispectrum. The same 
procedure is applied to the speckle observations of single (point source) 
calibrator stars. For each target, the power spectrum is divided by the power 
spectrum of the point source calibrator to yield a fringe pattern, which contains 
information on the separation, relative position angle, and brightness of any 
pair of stars, or a pattern containing no significant fringes in the case of a 
single star. Using the methods described by \citet{meng90}, a reconstructed 
image of the target star and its surroundings is made from the power spectrum 
and bispectrum (the bispectrum contains the phase information to properly
orient the position angle).

We fit a model fringe pattern to the observations to determine the separation 
and position angle of any detected companion relative to the primary, as well as 
the magnitude difference between companion and primary encoded in the amplitude 
of the fringes. Besides the relative positions and magnitudes, we also derived 
background sensitivities in the speckle fields using the reconstructed images. 
We used the fluxes relative to the primary star of all local maxima and minima 
noise features in the background of the reconstructed image to derive average 
$\Delta m$ values and their standard deviation within certain bins of separation 
from the primary star. From this, we adopted a contrast curve 5$\sigma$ brighter 
than the average $\Delta m$ values to represent the detection limits for any
given image.

Given this reduction and analysis method, it is difficult to determine
individual uncertainties for the $\Delta m$ measurements of detected
companions (the most challenging of the measurements we made). We took 
a conservative approach and adopted an uncertainty of 0.15 mag for all
measurements (roughly twice the uncertainty determined empirically in,
e.g., \citealt{horch11}, as KOI host stars are almost all fainter stars). When
comparing targets observed in the same band multiple times, we note
just a few outliers that are likely affected by poor fits between the model 
and observed power spectrum, or a poor match between the science
target and point source calibrator. In addition, the photometric accuracy 
of speckle observations degrades with a combination of poor seeing and 
large angular separations, as well as with fainter targets.

\subsection{Other High-Resolution Imaging}
\label{other_highres}

\citet{wang15a,wang15b} used the adaptive optics systems at Keck and Palomar 
with NIRC2 and PHARO, respectively, and typically observed each target with 
the $J$-, $H$-, and $K$-band filters.

\citet{adams12,adams13} and \citet{dressing14} mainly used the $K_s$ filter 
in their AO observations at the MMT; in addition, they often used the $J$-band 
filter when a companion was detected in the $K_s$ image. The field of view of 
the ARIES instrument on the MMT was 20\arcsec $\times$ 20\arcsec, somewhat 
smaller than that of PHARO at Palomar, and the FWHM of the stellar images 
varied between about 0.1\arcsec\ and 0.6\arcsec.

\citet{kraus16} employed adaptive optics imaging and also nonredundant 
aperture-mask interferometry at Keck with the NIRC2 instrument; the latter 
technique is limited only by the diffraction limit of the 10-m Keck telescope. 
They used the $K'$ filter for their observations.

\citet{baranec16} and \citet{ziegler16} observed a sample of KOI host stars at 
Keck using mostly the $K'$ filter on NIRC2.
With the Robo-AO imaging at the Palomar 1.5-m telescope, \citet{law14},
\citet{baranec16}, and \citet{ziegler16} covered a total of 3320 KOI host stars. 
For most observations, they used a long-pass filter whose window starts at 
600 nm ($LP600$), which is similar to the {\it Kepler} bandpass; they also took 
data for some stars in the Sloan $i$-band filter and, more rarely, Sloan $r$ 
and $z$ filters. Typical FWHM of the observed stellar PSF amounted to 
0.12\arcsec-0.15\arcsec; the images covered a field of view of 44\arcsec\
$\times$ 44\arcsec.

\citet {lillo-box12, lillo-box14} used the 2.2-m Calar Alto telescope with
the AstraLux instrument to obtain diffraction-limited imaging with the
lucky imaging technique, typically observing in the $i$- and $z$-band filters. 
This technique involves taking a very large number of short exposures and 
then combining only those images with the best quality (i.e., with the highest 
Strehl ratios). The FWHM of the stellar PSF in their 24\arcsec\ $\times$
24\arcsec\ images was typically 0.21\arcsec, which is somewhat larger than 
the value from AO images ($\sim$ 0.15\arcsec).

HST imaging using the WFC3 was carried out in the $F555W$ and $F775W$ 
bands \citep{gilliland15, cartier15}. The images spanned a relatively large
field of view of 40\arcsec $\times$ 40\arcsec, and the typical FWHM of the 
stellar PSF was 0.08\arcsec. 

One additional facility, the 8-m Large Binocular Telescope, was used with 
LMIRCam to observe 24 KOI host stars in the $K_s$ band, but results have 
not yet been published and are not available on CFOP. Except for one of 
these 24 stars (which has only one false positive transit signal), all have 
been observed with one or more other facilities, too.

\begin{figure*}[!t]
\centering
\includegraphics[scale=0.4]{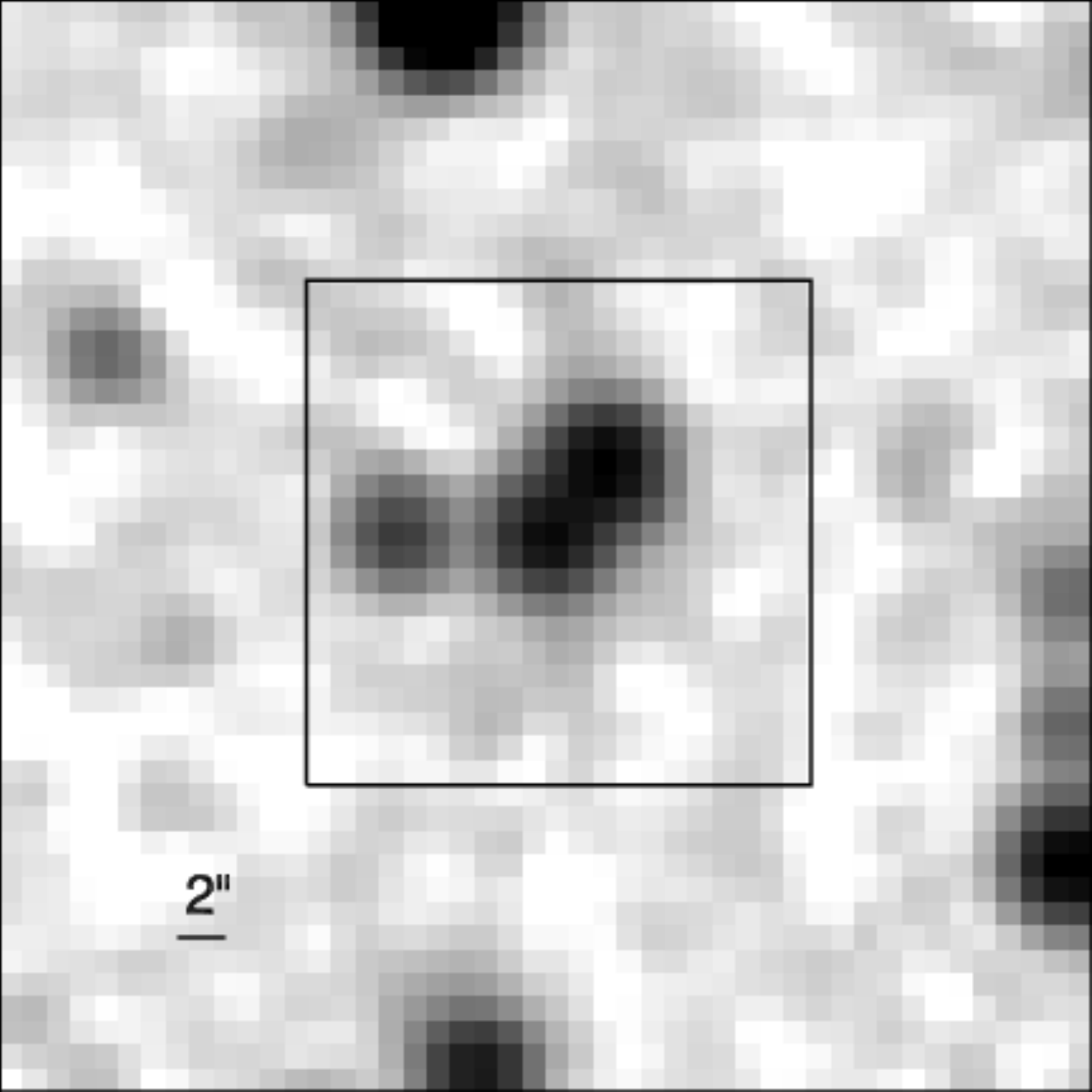}
\includegraphics[scale=0.4]{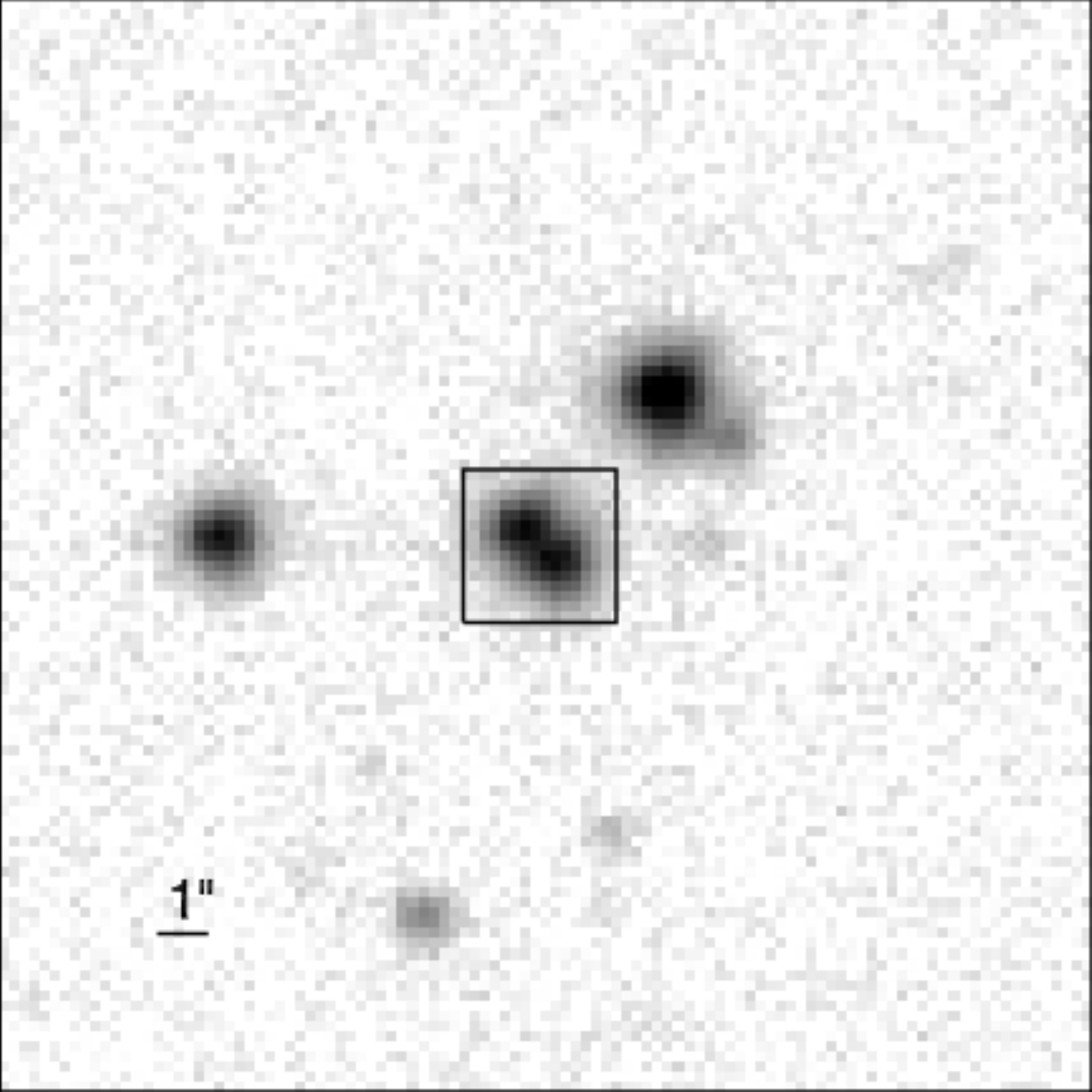}
\caption{Images of KOI 2174 in the $J$-band filter. The target star is at the center
of the images, and north is up and east is to the left. {\it Left:} 2MASS, with 
an image scale of 1\arcsec/pixel. The box shows the size of the UKIRT image 
displayed on the right. {\it Right:} UKIRT, with an image scale of 0.2\arcsec/pixel. 
The box shows the size of the Keck images shown in Figure \ref{KOI2174_highres}.
\label{KOI2174_lowres}}
\end{figure*}

\begin{figure*}[!t]
\centering
\includegraphics[scale=0.4]{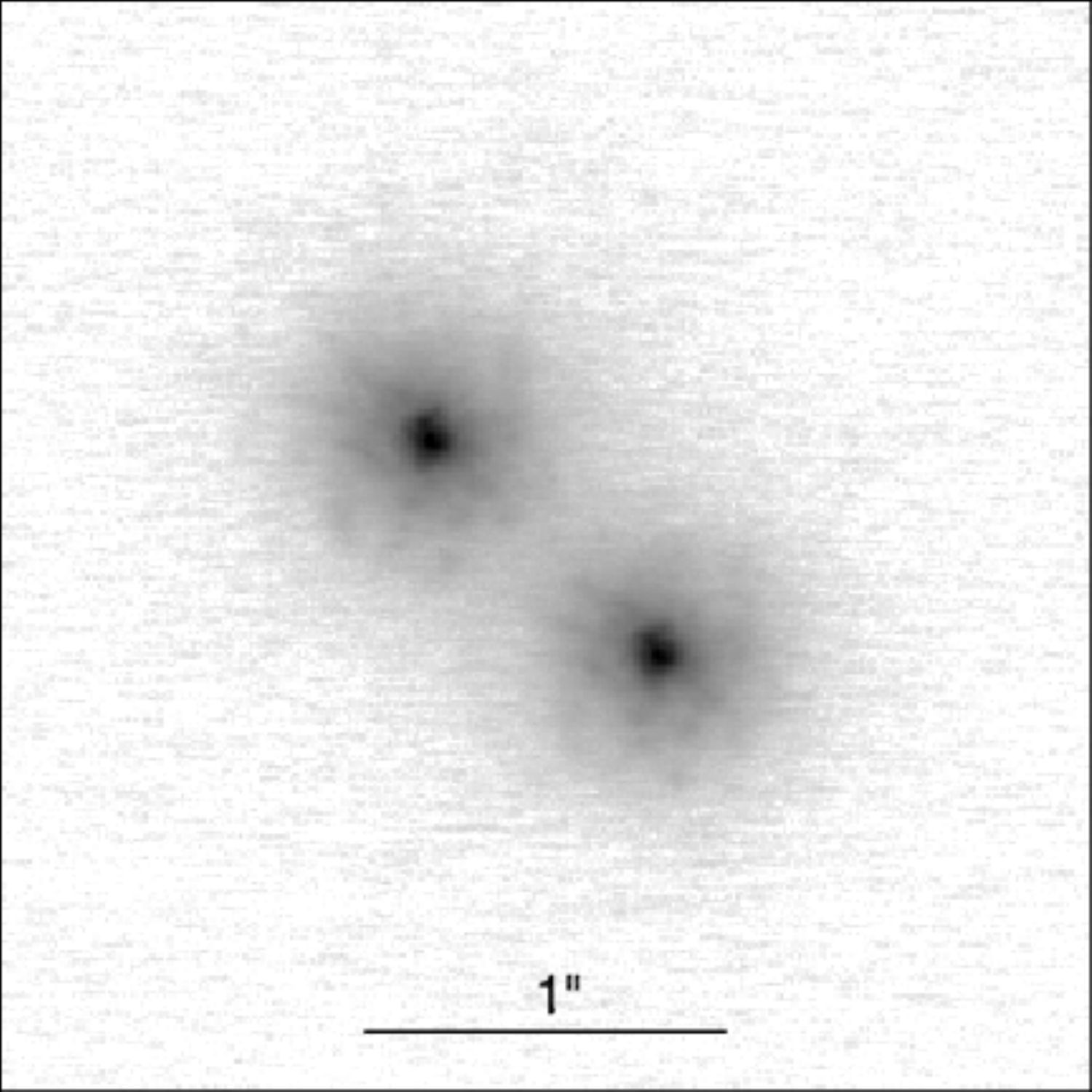}
\includegraphics[scale=0.4]{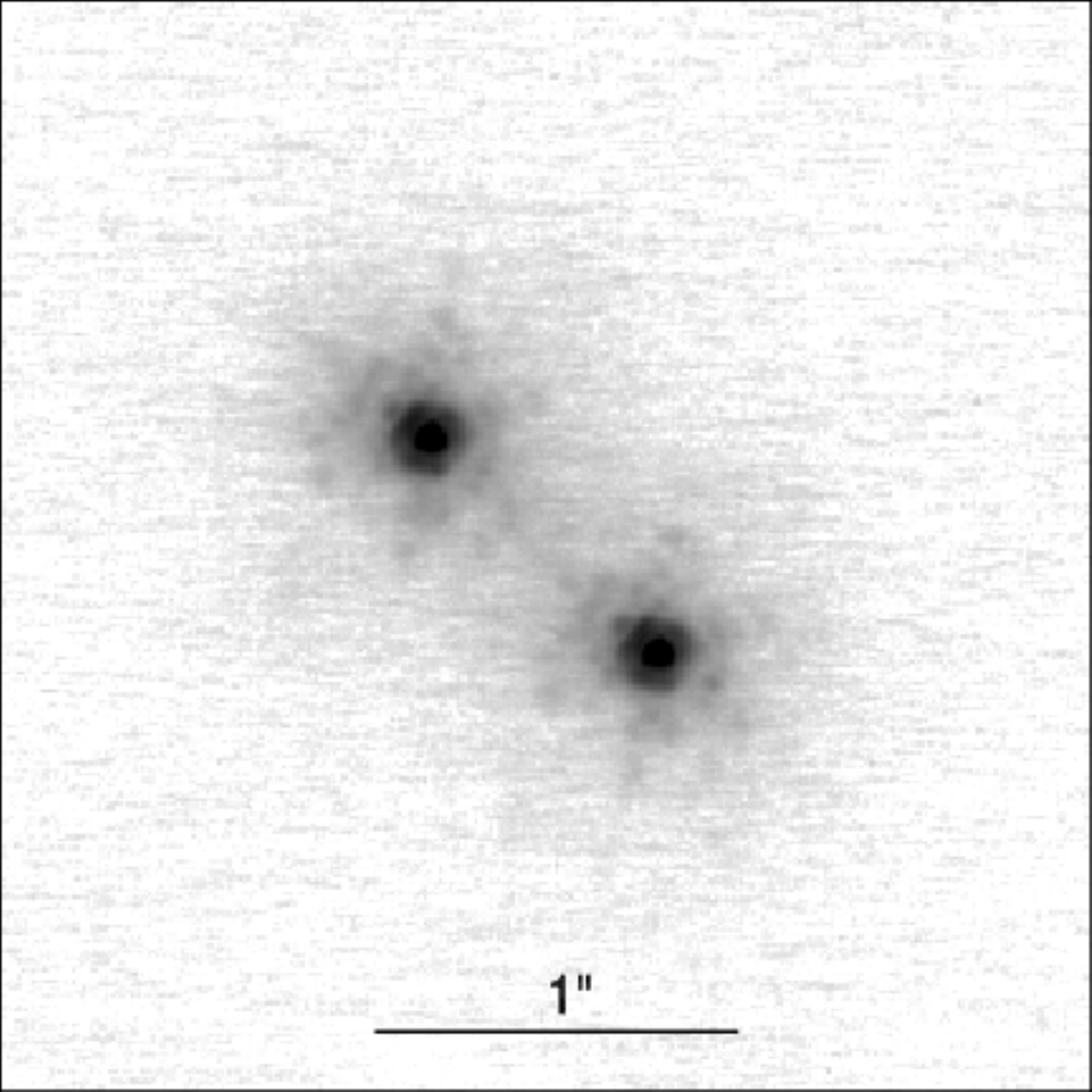}
\caption{Images of KOI 2174 observed with Keck/NIRC2 in the $J$-band filter
({\it left}) and in the $K$-band filter ({\it right}). The image scale is 0.01\arcsec/pixel; 
the images shown are 3\arcsec\ on each side. North is up and east is to the
left.
\label{KOI2174_highres}}
\end{figure*}

\vspace{1ex}

\subsection{Other Imaging}
\label{other_ima}

\subsubsection{UKIRT Survey}

The {\it Kepler} field was observed at the United Kingdom Infrared Telescope (UKIRT)
in 2010 using the UKIRT Wide Field Camera (WFCAM). The images were taken
in the $J$-band and have a typical spatial resolution of 0.8\arcsec-0.9\arcsec.
For each KOI host star, UKIRT image cut-outs and tables with nearby stars are available
on CFOP. We used that information to extract companions located within a radius of 
4\arcsec\ around each KOI host star. We considered all sources listed in the
UKIRT catalog that were not affected by saturation; thus, we also included objects 
with a high ``galaxy probability'' ($>0.9$), which applies to most faint sources, as well
as objects with a larger ``noise probability'' ($>0.1$). We vetted each companion by 
checking the UKIRT images for artifacts or spurious source detections. This vetting 
process led us to identify the following 18 KOIs as galaxies (which are clearly 
resolved in the UKIRT images): 51, 311, 1836, 1926, 2826, 3172, 3174, 3193, 
3206, 5014, 5190, 5238, 5595, 5668, 6817, 6864, 7213, 7612. Most are identified 
as false positives, but KOI 51, 3193, and 3206 are dispositioned as planet candidates. 

We cross-checked the UKIRT detections with 2MASS, which has a much lower
spatial resolution (1\arcsec/pixel for 2MASS, compared to 0.2\arcsec/pixel for 
UKIRT). We found that only companions at separations $\gtrsim$ 3.5\arcsec\ 
are resolved by 2MASS, and only if the companion is not too faint ($\Delta m 
\lesssim 4$) and the region within 4\arcsec-5\arcsec\ from the star is not crowded 
by multiple sources. The search for companions in 2MASS data yielded $H$- and 
$K_s$-band magnitudes for some of the wider companions.

To illustrate the importance of high-resolution imaging, Figures \ref{KOI2174_lowres}
and \ref{KOI2174_highres} show $J$-band images of KOI 2174 (which is a star with
3 planetary candidates with $R_p < $~2~\RE) with increasing spatial resolution. 
The 2MASS images do not resolve the central $\sim$ 0.9\arcsec\ binary; even though 
it is discernible in the UKIRT image, the UKIRT source catalog does not resolve 
the two sources (Fig.\ \ref{KOI2174_lowres}). The only companion within 4\arcsec\ 
resolved by the UKIRT (and also 2MASS) catalog is the star at a separation 
of 3.8\arcsec\ and position angle of $\sim$ 320\degr\ (i.e., to the northwest). 
The small field of view of Keck (Fig.\ \ref{KOI2174_highres}) does not include any 
star beyond about 2.5\arcsec\ from the close binary, but the Keck images clearly 
separate the two components of the 0.9\arcsec\ binary.

\vspace{1ex}

\subsubsection{UBV Survey}

\citet{everett12} carried out a survey of the {\it Kepler} field in 2011 using the
NOAO Mosaic-1.1 Wide Field Imager on the WIYN 0.9-m telescope. They 
observed the field in $UBV$ filters; the FWHM of the stellar PSF due to seeing 
ranged from 1.2\arcsec\ to 2.5\arcsec\ in the $V$-band (with somewhat larger 
values in the $U$- and $B$-band). The source catalog and the images are
available on CFOP.  We searched the catalog to find companions within
4\arcsec\ for each KOI host star. Due to the lower spatial resolution, just 132 KOI
host stars were found to have such a companion; the smallest companion 
separation is 1.4\arcsec. Almost all the companions detected in the $UBV$ 
survey are also found in UKIRT images. In a few cases their positions disagree 
somewhat (up to $\sim$~0.5\arcsec\ in radial separation and 10\degr-15\degr\ in 
position angle relative to the primary star) due to the presence of additional nearby 
stars, which make the positions from the lower-resolution $UBV$ data more uncertain.
In one case (KOI 6256), there are two companion stars detected in the
$UBV$ survey, but only one of them is also resolved in the UKIRT $J$-band image.
In another case (KOI 5928), a companion is detected at a projected separation 
of 3.3\arcsec\ in $UBV$ images, but the primary star is saturated in the UKIRT 
data, and so no reliable position and magnitude for the companion could be 
determined in the $J$-band.

\newpage

\section{Results}
\label{res}

\subsection{Companions and Sensitivity Curves}

\subsubsection{Keck, Palomar, and Lick}

\begin{figure}[!t]
\centering
\includegraphics[angle=90, scale=0.47]{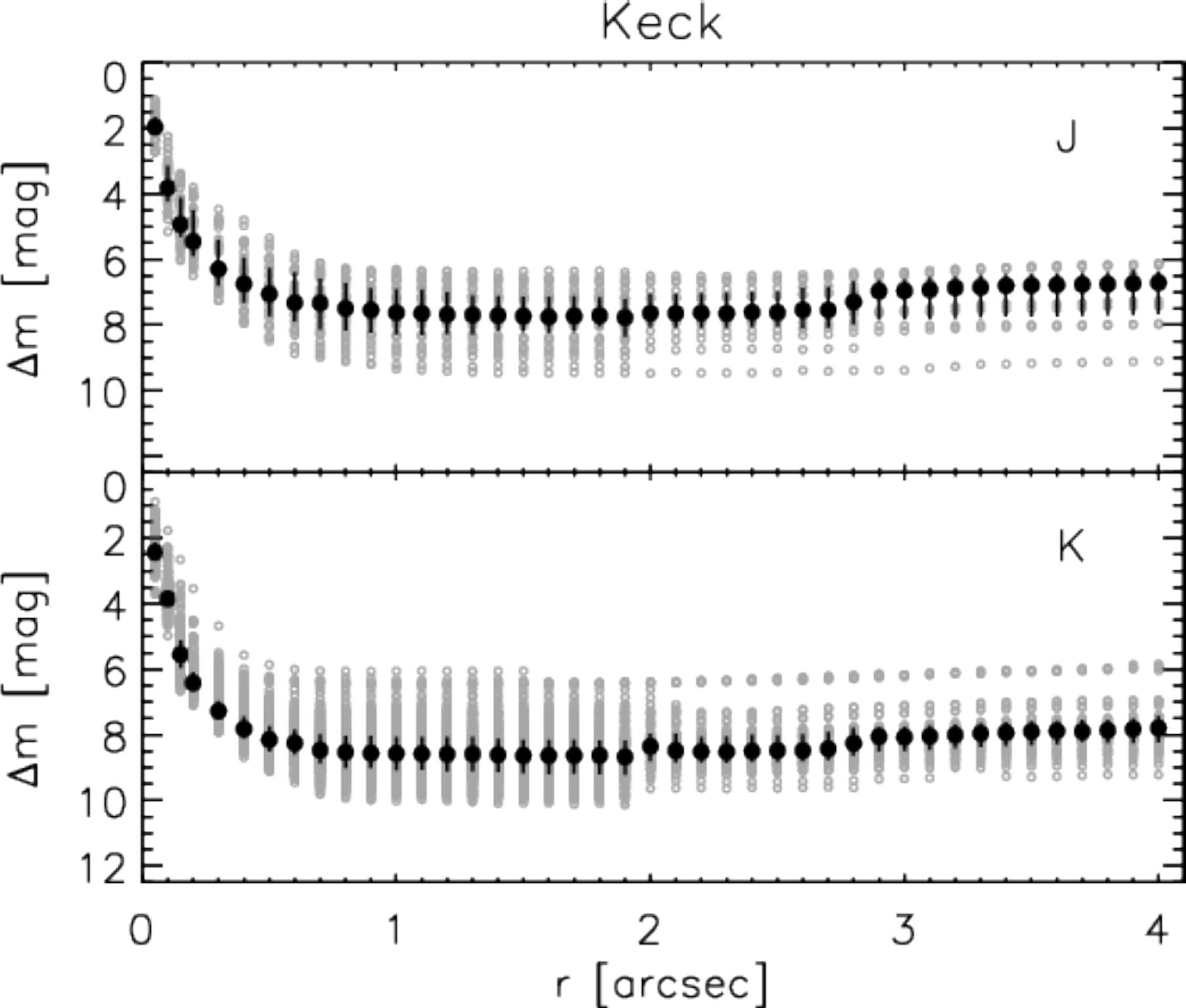}
\caption{Image sensitivities for observations of {\it Kepler} stars with the 
Keck 10-m telescope. The median sensitivities and quartiles are plotted 
with black symbols. The median FWHM of the stellar images was 
0.05\arcsec.
\label{sensitivites_Keck}}
\end{figure}

\begin{figure}[!t]
\centering
\includegraphics[angle=90, scale=0.47]{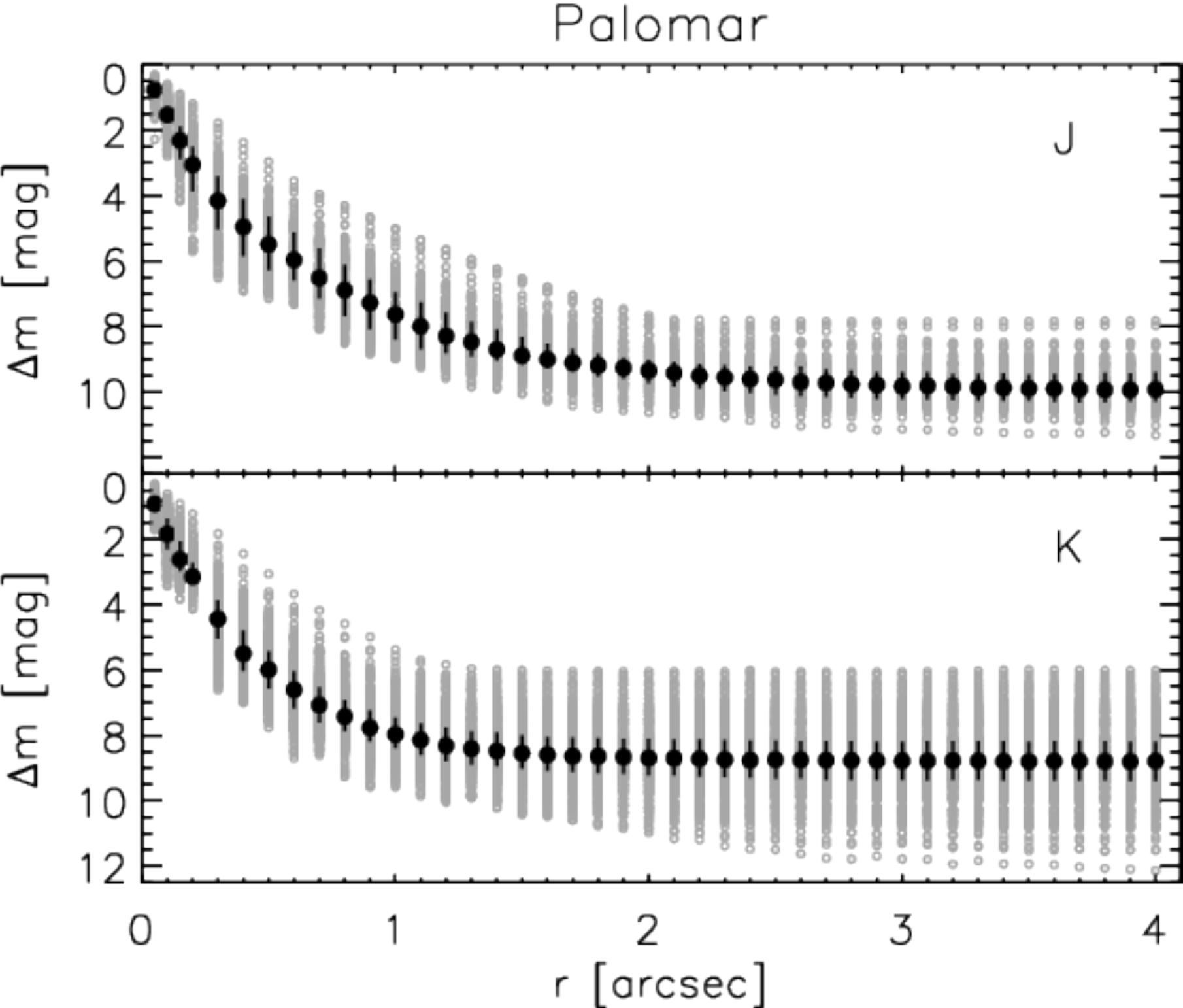}
\caption{Image sensitivities for observations of {\it Kepler} stars with the 
Palomar 5-m telescope. The median sensitivities and quartiles are plotted 
with black symbols. The median FWHM of the stellar images was
0.13\arcsec\ for the $J$-band and 0.12\arcsec\ for the $K$-band.
\label{sensitivites_Palomar}}
\end{figure}

\begin{figure}[!t]
\centering
\includegraphics[angle=90, scale=0.47]{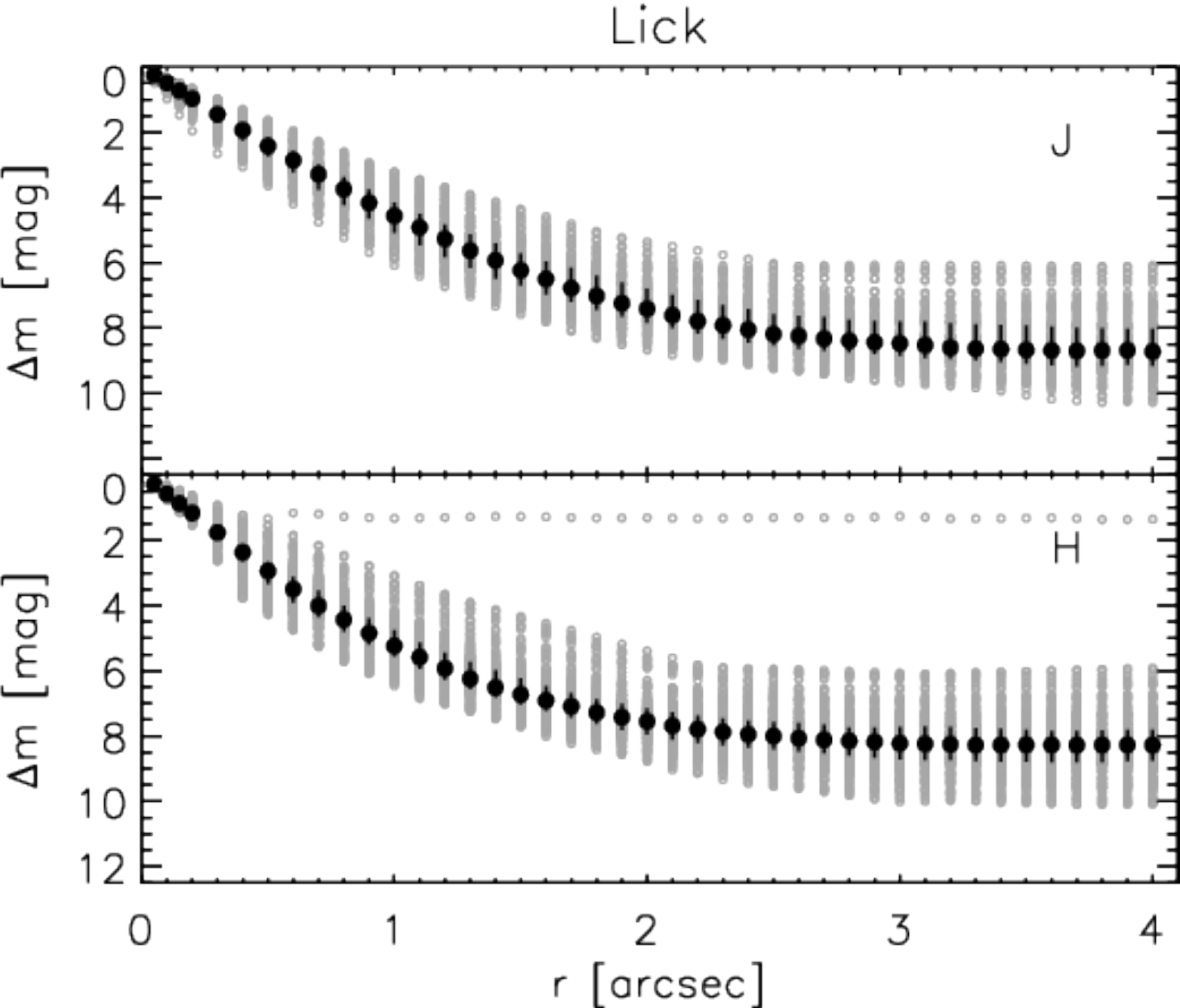}
\caption{Image sensitivities for observations of {\it Kepler} stars with the 
Lick 3-m telescope. The median sensitivities and quartiles are plotted 
with black symbols. The median FWHM of the stellar images was 0.43\arcsec\ 
for the $J$-band and 0.31\arcsec\ for the $H$-band.
\label{sensitivites_Lick}}
\end{figure}

\begin{deluxetable*}{llccc}
\tablewidth{0.99\linewidth}
%\tabletypesize{\small}     %{\scriptsize}   
\tablecaption{Typical FWHM values of the stellar PSFs for the AO and speckle images
\label{FWHM_table}}
\tablehead{
 \colhead{Telescope} & \colhead{Technique} & \colhead{Band} & 
 \colhead{Mean FWHM} & \colhead{Median FWHM}}
\startdata
Keck & AO & $J$ & 0.063\arcsec & 0.053\arcsec \\
   & & $K$ & 0.057\arcsec & 0.053\arcsec \\
Palomar & AO & $J$ & 0.166\arcsec & 0.127\arcsec \\
   & & $K$ & 0.140\arcsec & 0.120\arcsec \\
Lick & AO & $J$ & 0.453\arcsec & 0.431\arcsec \\
   & & $H$ & 0.347\arcsec & 0.314\arcsec \\
Gemini North & speckle & 562, 692, 880 nm & 0.02\arcsec & 0.02\arcsec \\
WIYN & speckle & 562, 692, 880 nm & 0.05\arcsec & 0.05\arcsec \\
DCT & speckle & 692, 880 nm & 0.04\arcsec & 0.04\arcsec \\
\enddata
\end{deluxetable*}

As described in section \ref{our_AO}, we have observed several hundred KOI host 
stars at Keck, Palomar, and Lick. Here we present the results of our measurements
of the image sensitivity and of companions detected in the images. For the former, we 
combined the measurements from each image (5$\sigma$ limits in annuli around the 
main star) to determine the median, lower and upper quartiles for each filter 
at each telescope; the resulting plots are shown in Figures \ref{sensitivites_Keck} 
to \ref{sensitivites_Lick}. Typical FWHM values (mean and median) of the stellar 
PSFs are listed in Table \ref{FWHM_table}; we used the 5$\sigma$ limits measured
at radial separations equal to multiples of the FWHM and interpolated them at
the radial values shown in the plots.
Of the three observing facilities, we reach the highest sensitivity close to the 
primary star with Keck; already at a separation of $\sim$~0.5\arcsec\ we reach 
a median sensitivity of $\Delta m \sim$~8 mag in the $K$-band. At Palomar, 
the median sensitivity reaches $\Delta m \sim$ 8 mag in the $K$-band at 
$\sim$~1\arcsec\ from the primary. We are particularly sensitive to companions 
in the $J$-band; in this band, at a separation of a few arcseconds, we are sensitive 
to companions up to 10 magnitudes fainter than the primary. Finally, at Lick we 
achieve a median sensitivity of $\Delta m \sim$ 8 mag in both the $J$- and 
$H$-band at about 2.5\arcsec\ from the primary.

\begin{figure}[!]
\centering
\includegraphics[angle=90, scale=0.48]{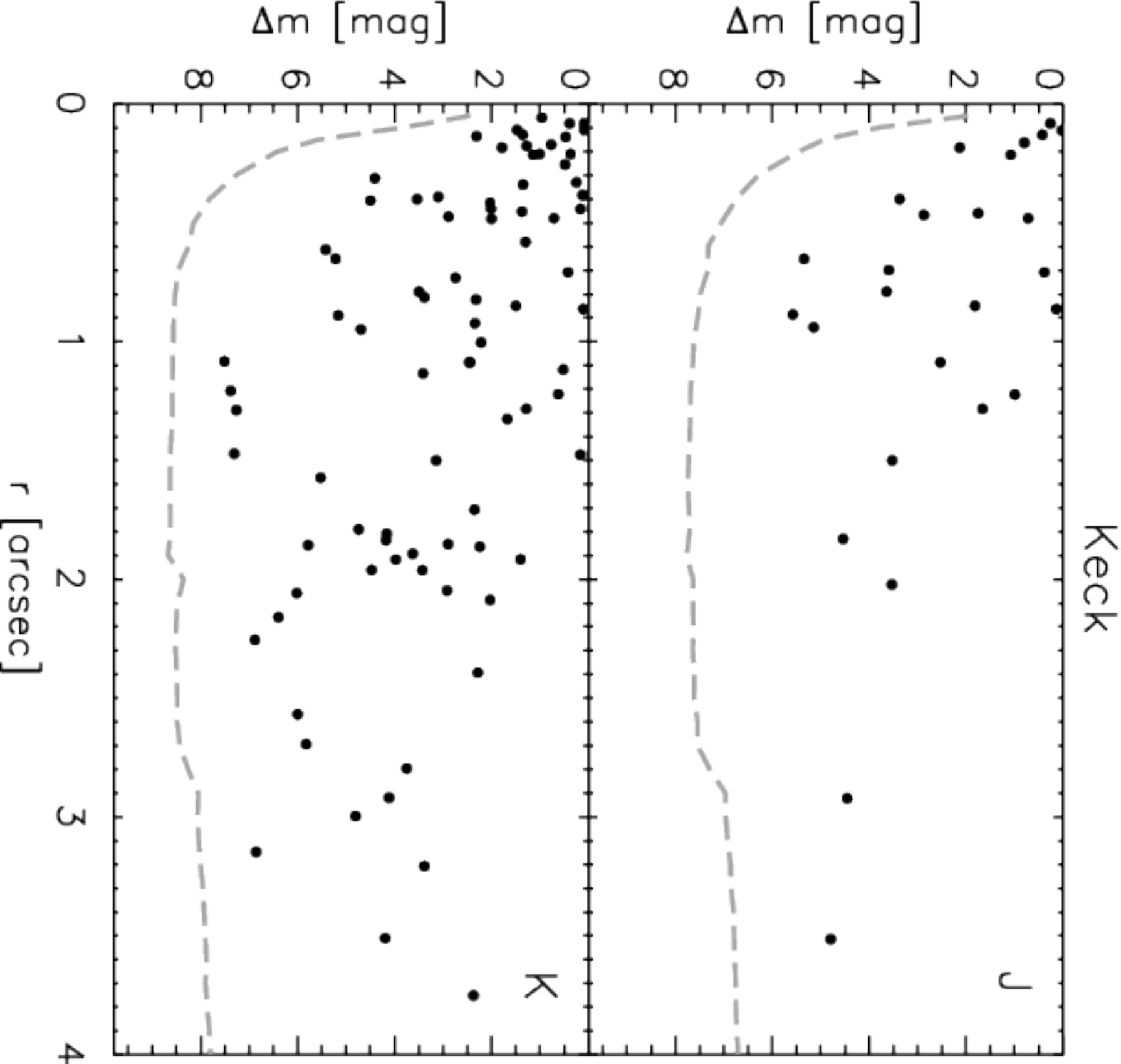}
\caption{Magnitude difference versus radial separation for all companions detected
around {\it Kepler} stars with the Keck 10-m telescope in the $J$-band ({\it top})
and $K$-band ({\it bottom}). The dashed lines are the median image sensitivities
from Figure \ref{sensitivites_Keck}.
\label{KOIs_Keck}}
\end{figure}

\begin{figure}[!]
\centering
\includegraphics[angle=90, scale=0.48]{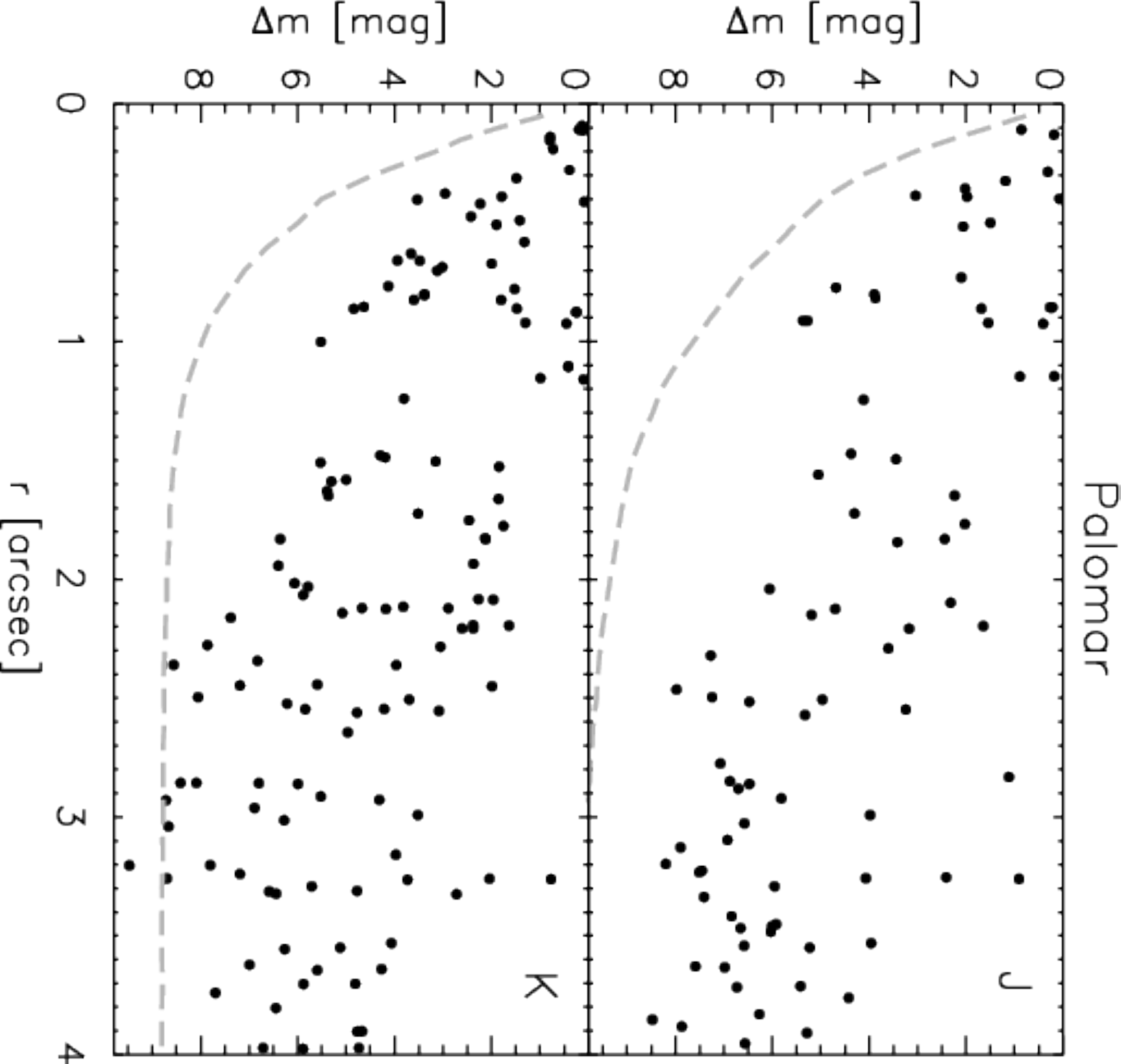}
\caption{Magnitude difference versus radial separation for all companions detected
around {\it Kepler} stars with the Palomar 5-m telescope in the $J$-band ({\it top})
and $K$-band ({\it bottom}). The dashed lines are the median image sensitivities
from Figure \ref{sensitivites_Palomar}.
\label{KOIs_Palomar}}
\end{figure}

\begin{figure}[!t]
\centering
\includegraphics[angle=90, scale=0.48]{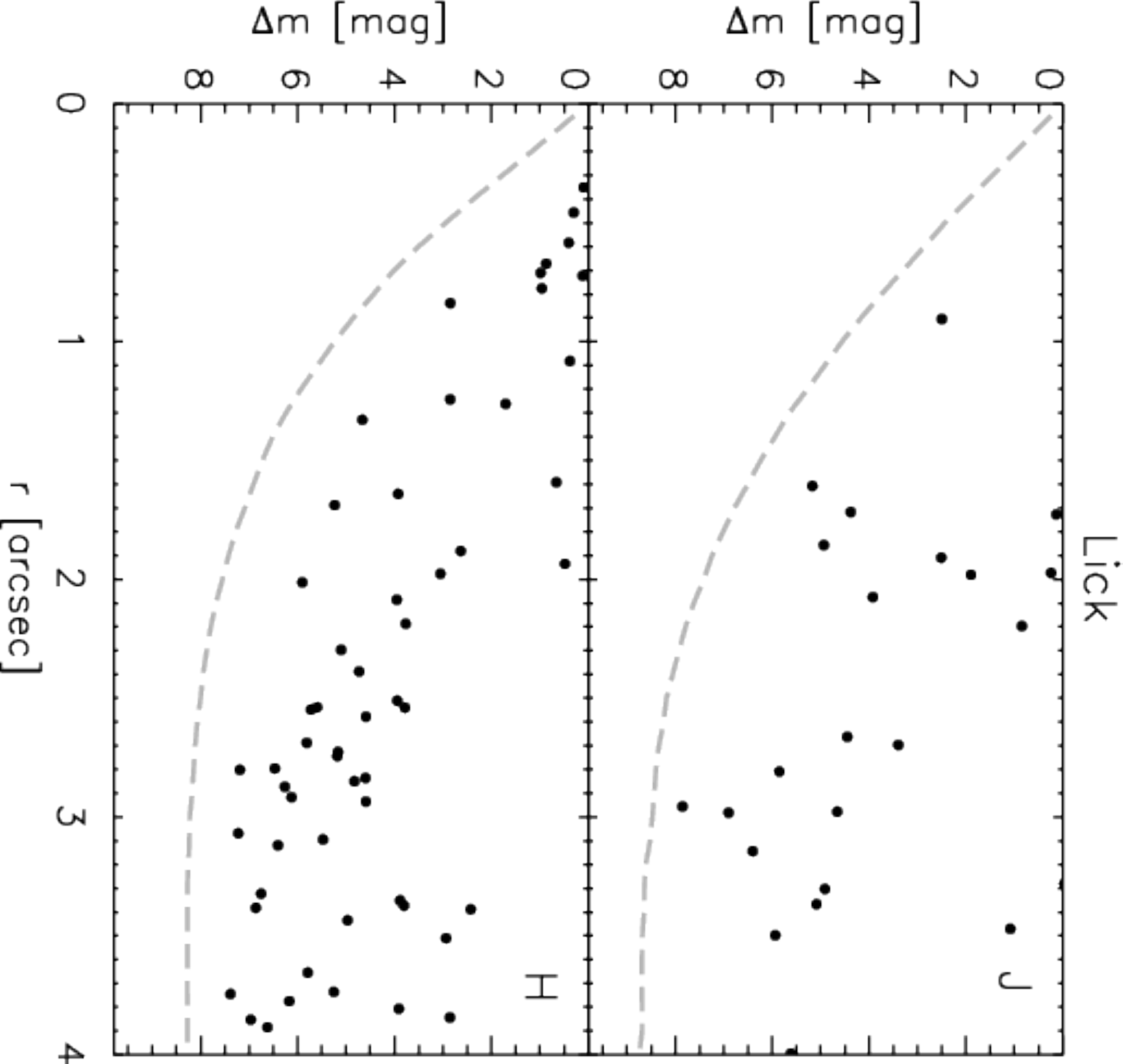}
\caption{Magnitude difference versus radial separation for all companions detected
around {\it Kepler} stars with the Lick 3-m telescope in the $J$-band ({\it top})
and $H$-band ({\it bottom}). The dashed lines are the median image sensitivities
from Figure \ref{sensitivites_Lick}.
\label{KOIs_Lick}}
\end{figure}

For those KOI host stars where a companion was detected, we measured the 
position and brightness of that companion relative to the primary. Figures 
\ref{KOIs_Keck} to \ref{KOIs_Lick} show the companions detected within 
4\arcsec\ in our Keck, Palomar, and Lick images; for each telescope, detections 
in two filter bands are shown ($J$ and $K$ for Keck and Palomar, $J$ and $H$ 
for Lick). Some companions have measurements in both filters. Of the 253 unique 
KOI stars observed at Keck, 75 have at least one companion detected within 
4\arcsec; for the 317 KOI stars observed at Palomar, this number is 116, and 
for the 310 KOI stars observed at Lick, 71 have such companions (see Table 
\ref{Companion_systems}). In Table \ref{Companion_systems}, we also list 
the number of KOI stars with one, two, three, and even four companions. 
These are just the companions we detected; there could be more companions 
that were too faint or too close to the primary stars to be found in our data.
Overall, at all three telescopes 770 unique KOI host stars were observed, 
and 242 of these stars have companions detected within 4\arcsec; thus, in our 
AO sample the observed fraction of systems consisting of at least two stars 
within 4\arcsec\ is 31\% ($\pm$2\%, assuming Poisson statistics).

\begin{deluxetable*}{lcccccccc}[!t]
\tablecaption{Number of KOI stars with Detected Companions and Fraction of 
Multiple Systems in AO and Speckle Images
\label{Companion_systems}}
\tablehead{
 \colhead{Telescope} & \colhead{N} & \colhead{N$_{\rm comp}$} & 
 \colhead{N$_{\rm comp=1}$} & \colhead{N$_{\rm comp=2}$} & 
 \colhead{N$_{\rm comp=3}$} & \colhead{N$_{\rm comp=4}$} &
 \colhead{f($<$1\arcsec)}  & \colhead{f($<$4\arcsec)} \\
\colhead{(1)} & \colhead{(2)} & \colhead{(3)} & \colhead{(4)} &
\colhead{(5)} & \colhead{(6)} & \colhead{(7)} & \colhead{(8)} & \colhead{(9)} }
\startdata
Keck & 253 & 75 & 61 & 11 & 3 & 0 & 17\% & 30\% \\
Palomar & 317 & 116 & 93 & 18 & 4 & 1 & 10\% & 37\% \\
Lick & 310 & 71 & 63 & 8 & 0 & 0 & 3\% & 23\% \\
Gemini North & 158 & 39\tablenotemark{a} & 37 & 2 & 0 & 0 & 21\% & \nodata \\
WIYN & 681 & 49\tablenotemark{a} & 49 & 0 & 0 & 0 & 6\% & \nodata \\
DCT & 75 & 7\tablenotemark{a} & 7 & 0 & 0 & 0 & 8\% & \nodata \\
\enddata
\tablecomments{Column (1) lists the telescope where the data were
obtained, column (2) the total number of unique KOI host stars observed, 
column (3) the number of KOI stars with at least one companion,
columns (4) to (7) the number of KOI stars with 1, 2, 3, and 4 companions,
respectively, and columns (8) and (9) give the fraction of multiple systems
with stars within 1\arcsec\ and 4\arcsec, respectively.}
\tablenotetext{a}{Of the stars with companions detected with Gemini North, WIYN,
and DCT, 7, 10, and 1, respectively, have companions that lie at a separation 
larger than 1\arcsec\ from the primary (one star observed at Gemini North has
both one companion within 1\arcsec\ and one companion at $>$ 1\arcsec). 
Thus, to calculate f($<$1\arcsec) in column (8), N$_{\rm comp}$ of 33, 
39, and 6, respectively, is used.}
\end{deluxetable*}

\subsubsection{Gemini North, WIYN, and DCT}

\begin{figure}[!]
\centering
\includegraphics[angle=90, scale=0.47]{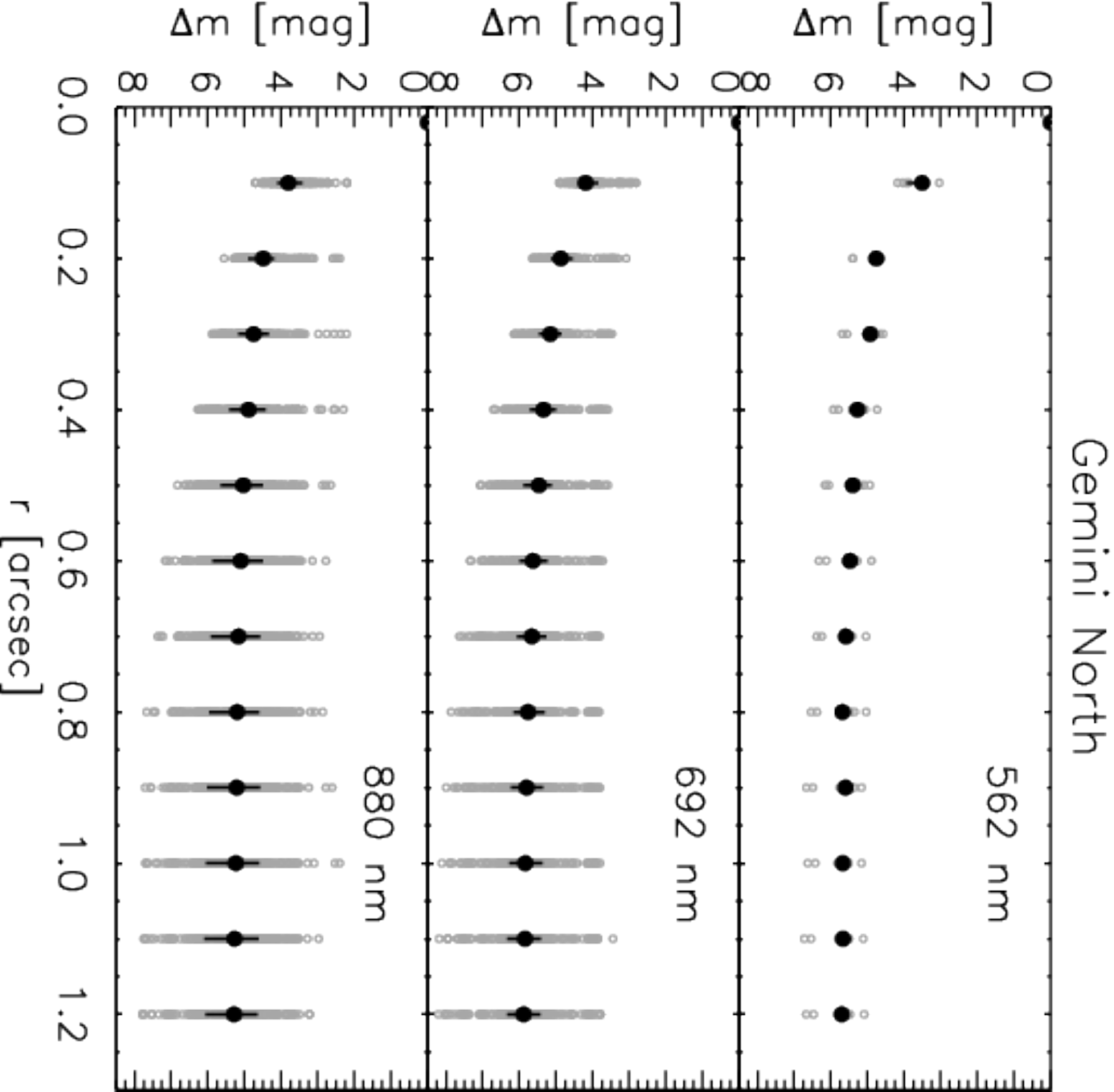}
\caption{Image sensitivities for observations of {\it Kepler} stars with the Gemini 8-m 
telescope. The median sensitivities and quartiles are plotted with black symbols.
The median FWHM of the stellar images was 0.02\arcsec.
\label{Sensitivites_Gemini}}
\end{figure}

\begin{figure}[!]
\centering
\includegraphics[angle=90, scale=0.47]{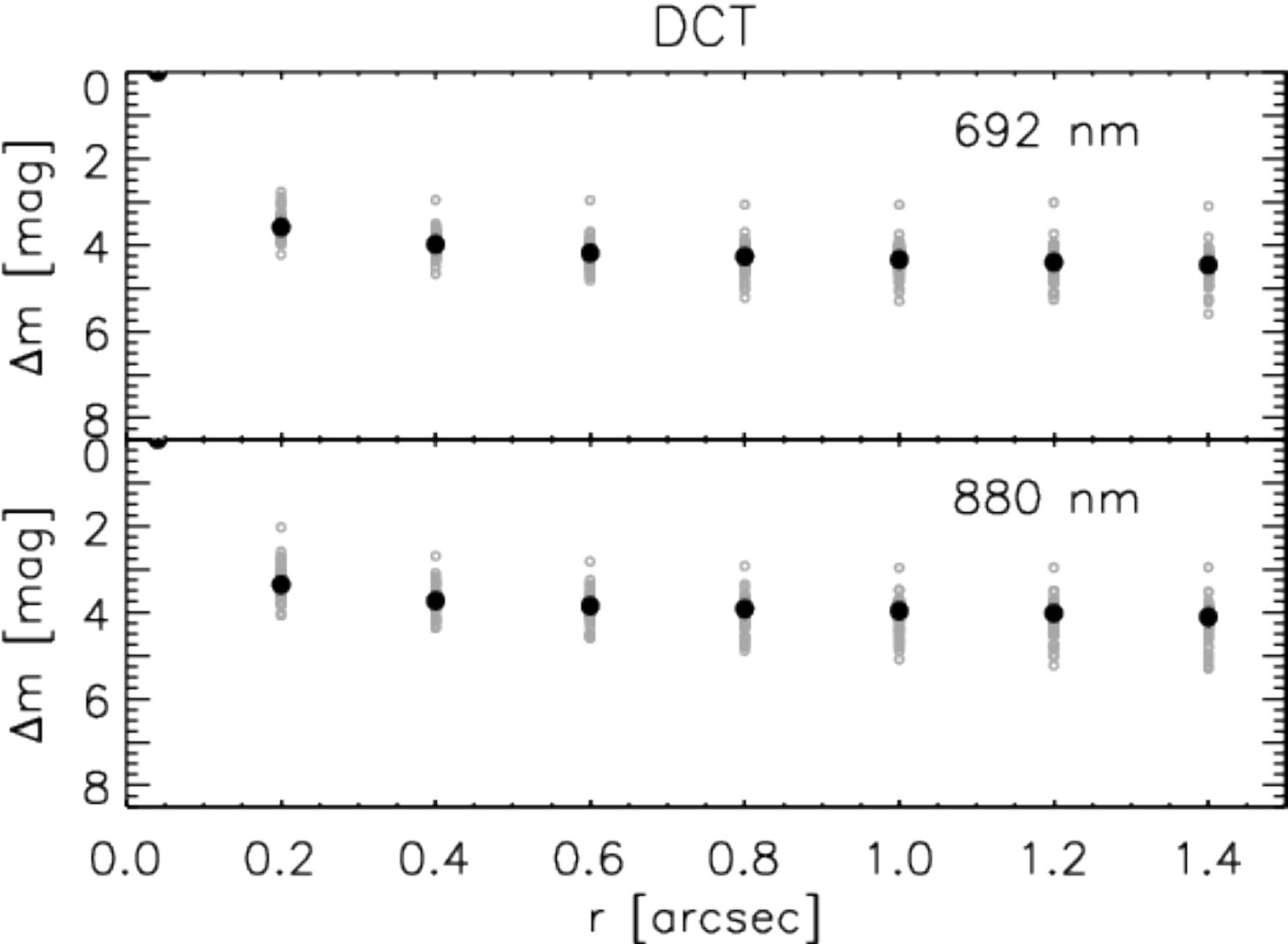}
\caption{Image sensitivities for observations of {\it Kepler} stars with the DCT 4-m 
telescope. The median sensitivities and quartiles are plotted with black symbols.
The median FWHM of the stellar images was 0.04\arcsec.
\label{Sensitivites_DCT}}
\end{figure}

\begin{figure}[!]
\centering
\includegraphics[angle=90, scale=0.47]{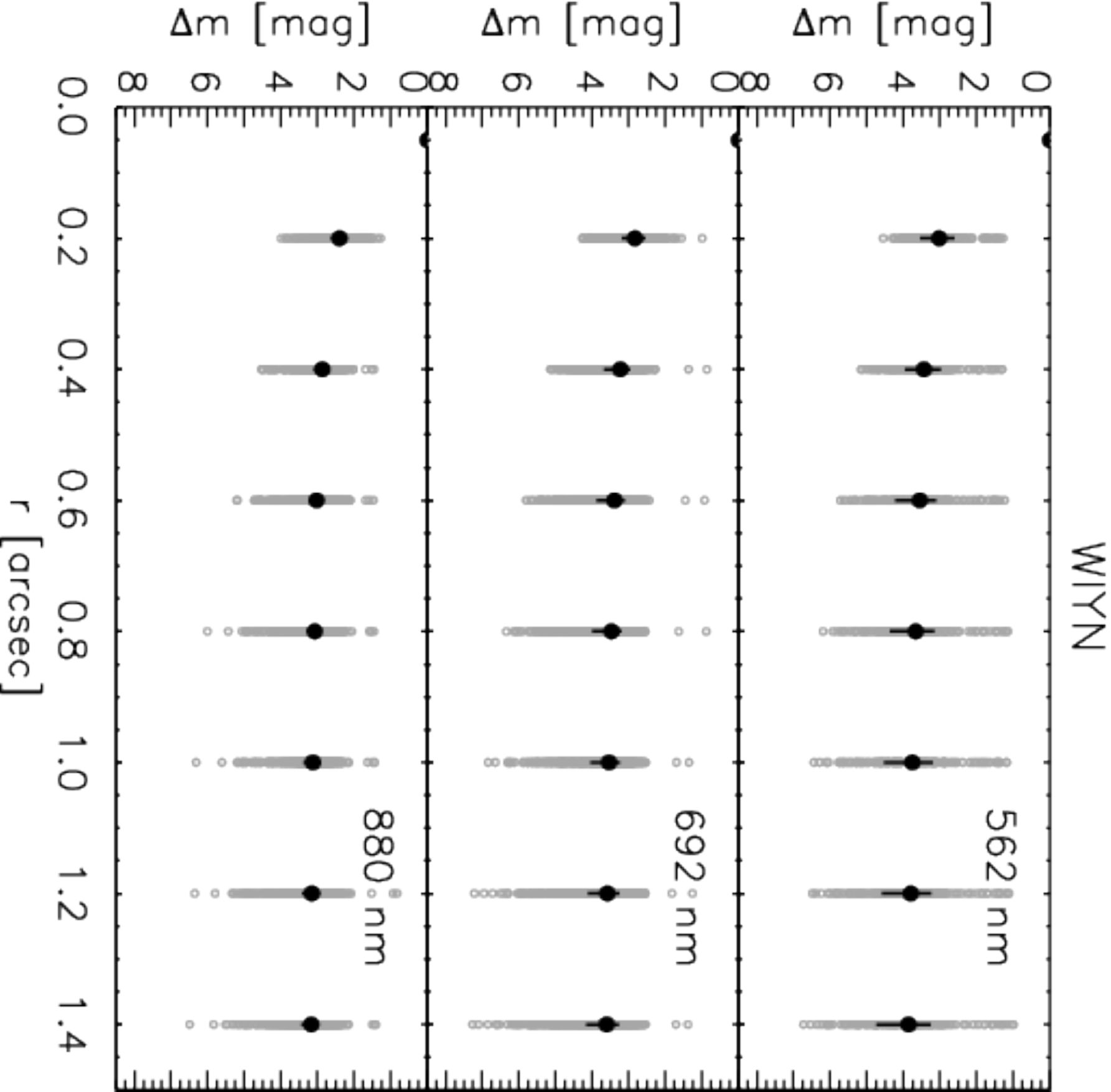}
\caption{Image sensitivities for observations of {\it Kepler} stars with the WIYN 3.5-m 
telescope. The median sensitivities and quartiles are plotted with black symbols.
The median FWHM of the stellar images was 0.05\arcsec.
\label{Sensitivites_WIYN}}
\end{figure}

\begin{figure}[!]
\centering
\includegraphics[angle=90, scale=0.53]{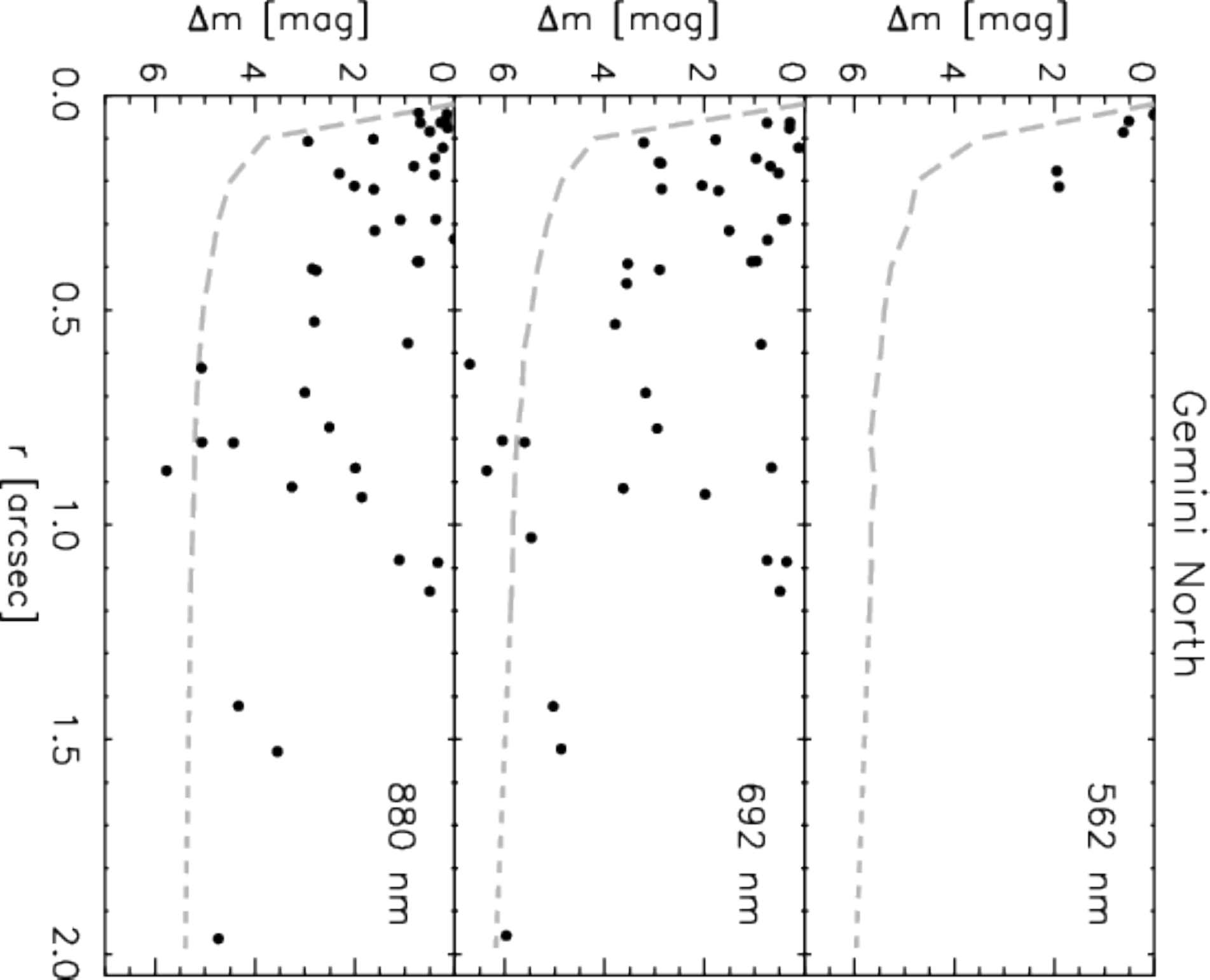}
\caption{Magnitude difference versus radial separation for all companions detected
around {\it Kepler} stars with the Gemini 8-m telescope at 562 nm ({\it top}),
692 nm ({\it middle}), and 880 nm ({\it bottom}). The dashed lines are the median
image sensitivities from Figure \ref {Sensitivites_Gemini}.
\label{KOIs_Gemini}}
\end{figure}

\begin{figure}[!]
\centering
\includegraphics[angle=90, scale=0.53]{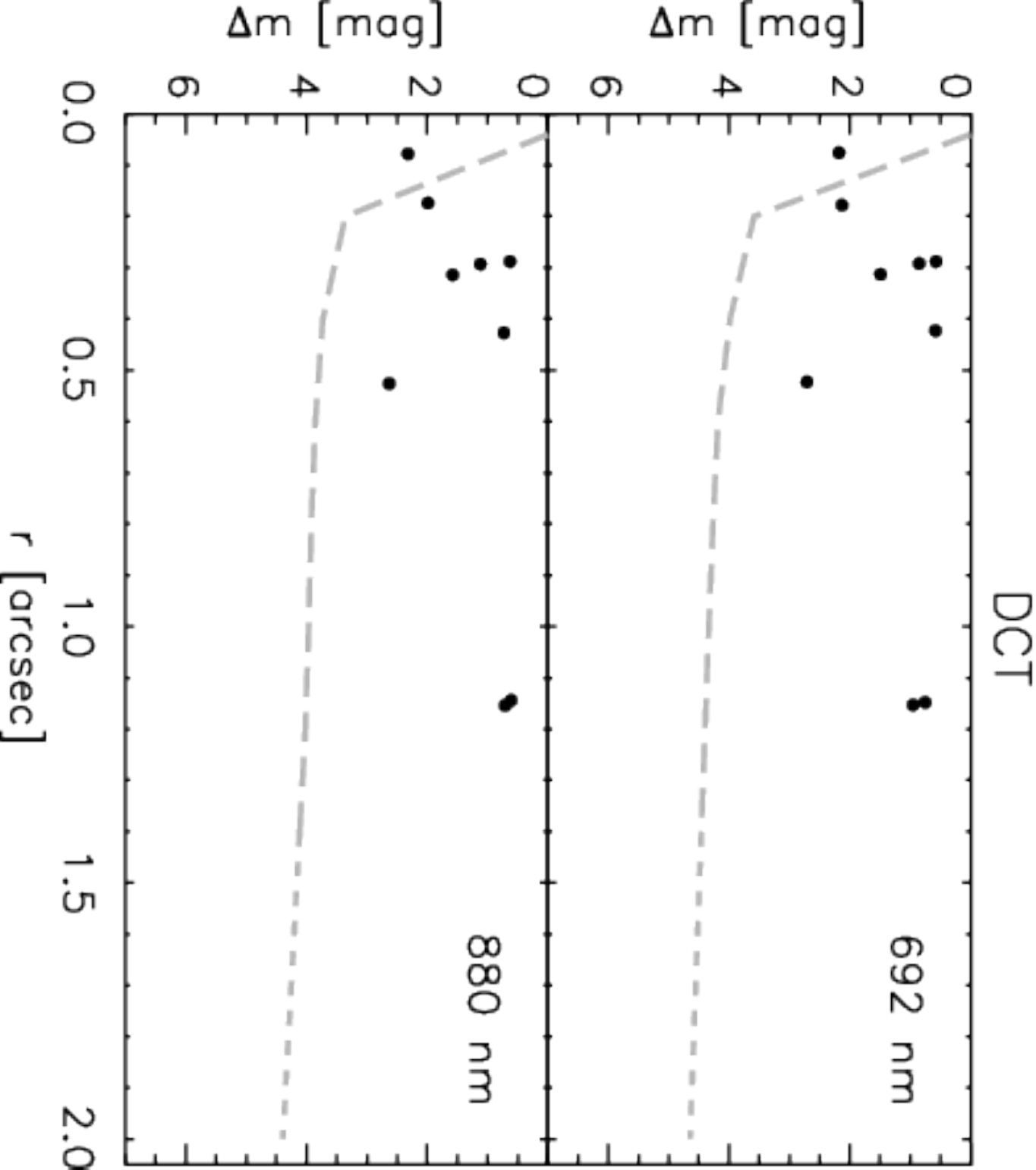}
\caption{Magnitude difference versus radial separation for all companions detected
around {\it Kepler} stars with the DCT 4-m telescope at 692 nm ({\it top}) and 
880 nm ({\it bottom}). The dashed lines are the median image sensitivities from 
Figure \ref {Sensitivites_DCT}.
\label{KOIs_DCT}}
\end{figure}

\begin{figure}[!]
\centering
\includegraphics[angle=90, scale=0.53]{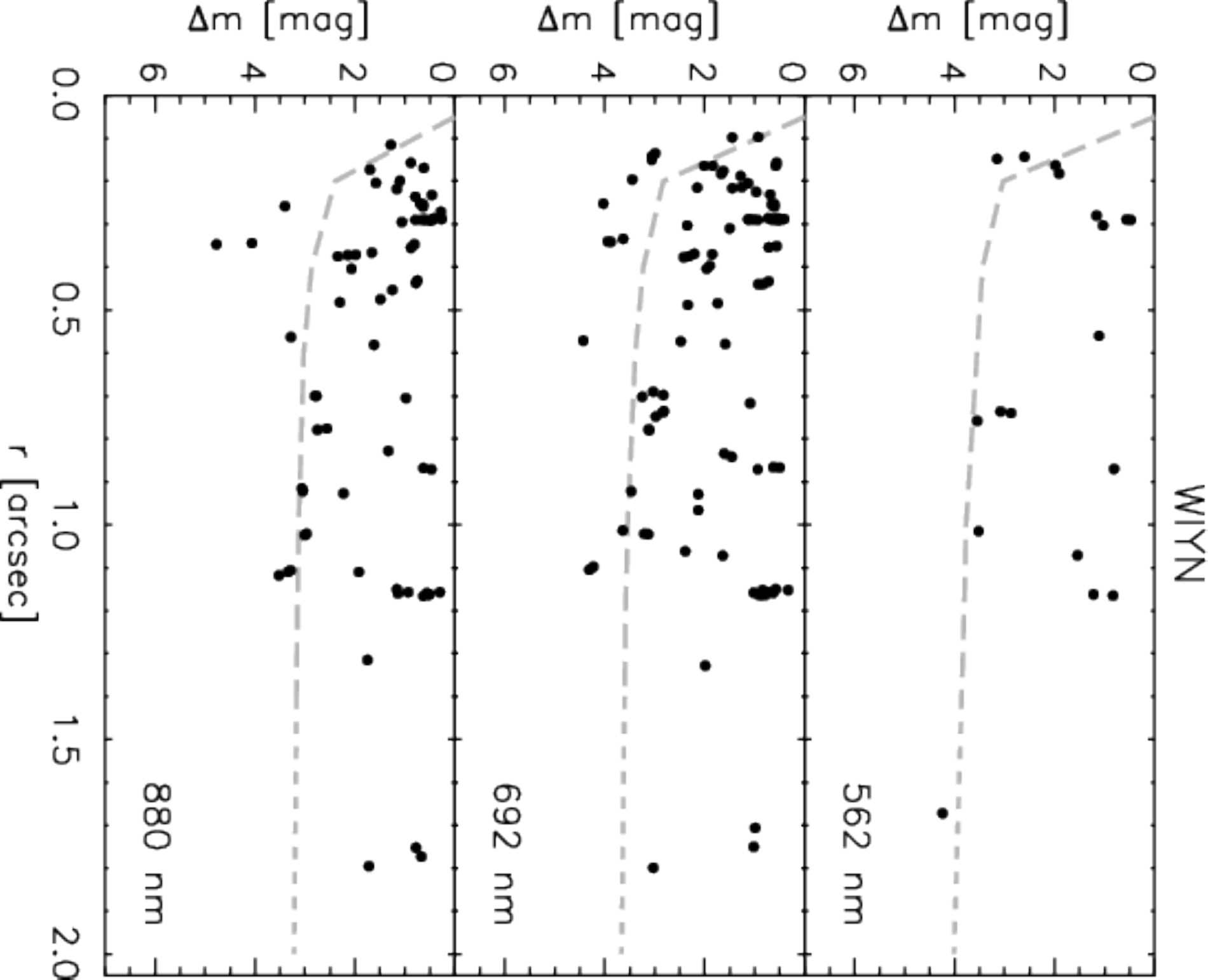}
\caption{Magnitude difference versus radial separation for all companions detected
around {\it Kepler} stars with the WIYN 3.5-m telescope at 562 nm ({\it top}),
692 nm ({\it middle}), and 880 nm ({\it bottom}). Some repeated observations
of the same star result in points stacked at the same position. The dashed 
lines are the median image sensitivities from Figure \ref {Sensitivites_WIYN}.
\label{KOIs_WIYN}}
\end{figure}

As with the AO data, we measured image sensitivities and the separations,
position angles, and brightness differences for any companions detected 
in the speckle images (note that the field of view of these images is much 
smaller than for the AO images). The 5$\sigma$ sensitivity limits are shown 
in Figures \ref{Sensitivites_Gemini} to \ref{Sensitivites_WIYN}. The FWHM 
of the stellar PSFs  were 0.02\arcsec\ at Gemini North, 0.04\arcsec\
at the DCT, and 0.05\arcsec\ at WIYN (see Table \ref{FWHM_table}).
The sensitivity to companions is relatively flat from about 0.3\arcsec\ to 
the edge of the field of view (about 1.4\arcsec\ from the central star), 
and it is lower than the sensitivity of the Keck and Palomar AO images.
However, within $\sim$~0.2\arcsec\ speckle interferometry is more
sensitive to companions than adaptive optics (median $\Delta m \sim$ 
4-5 in all three bands at Gemini North). 
Compared to the image sensitivities from \citet{horch14}, who used a 
subsample of the speckle data from Gemini North and WIYN presented 
in this work, our values for the WIYN 692 nm data are very similar, 
while our values for the Gemini North 692 nm data are somewhat
different. For the latter, our sensitivities are about 1 magnitude worse
below 0.4\arcsec\ and between 1.5 and 2.5 magnitudes less sensitive
in the 0.4\arcsec-1.2\arcsec\ range. This is likely a result of the larger 
sample studied here (158 vs.\ 35 stars in \citealt{horch14}) and thus 
a wider range of observing conditions.

Companions detected in speckle images are shown in Figures \ref{KOIs_Gemini} 
to \ref{KOIs_WIYN}; each individual detection is shown. In some cases 
a target was observed at the same facility with the same filter multiple times, 
resulting in more than one measurement for a certain companion; these multiple
measurements disagree in a few cases by up to $\sim$ 0.5 mag (see Fig.\
\ref{KOIs_WIYN}), likely a result of different observing conditions. At both Gemini 
North and WIYN, targets were typically observed at 692 and 880 nm, with some
targets also observed at 562 nm, while at DCT only the 692 and 880 nm filters
were used. We find at least one companion within the field of view ($\sim$~2\arcsec) 
around 39 of the 158 unique KOI host stars observed at Gemini North; this fraction 
is 7 out of 75 for the KOI stars observed at DCT and 49 out of 681 for the KOI 
stars observed at WIYN (see Table \ref{Companion_systems}). 
Except for two KOI host stars, multiple systems discovered in speckle images are 
binaries; only KOI 2626 and 2032 have two companions detected in Gemini North 
speckle images, and thus, if bound, they would form triple stellar systems.

Overall, at the three telescopes where DSSI was used, 828 unique KOI host stars 
were observed; of these, 85 have at least one companion detected within 
$\sim$~2\arcsec. This translates to an observed  fraction of multiple stellar 
systems in our speckle sample of 10$\pm$1\%. If we consider only companions 
within 1\arcsec\ of the primary star, we find that the observed fraction of multiple 
stellar systems is 8$\pm$1\%. These fractions are smaller than what we found from 
our AO data for companions within 4\arcsec, which is a result of the smaller field of 
view of the speckle images. If we only consider companions detected at separations 
$\leq$~1\arcsec, 10$\pm$1\% of KOI host stars observed with AO have companions 
(79 out of 770 stars), which is in agreement with the results from speckle imaging. 
The same fraction, 10$\pm$1\%, also results when combining the samples of stars 
we targeted with AO and speckle imaging (116 out of a total of 1189 unique KOI 
host stars have at least one companion within 1\arcsec).

\begin{figure}[!t]
\centering
\includegraphics[angle=90, scale=0.47]{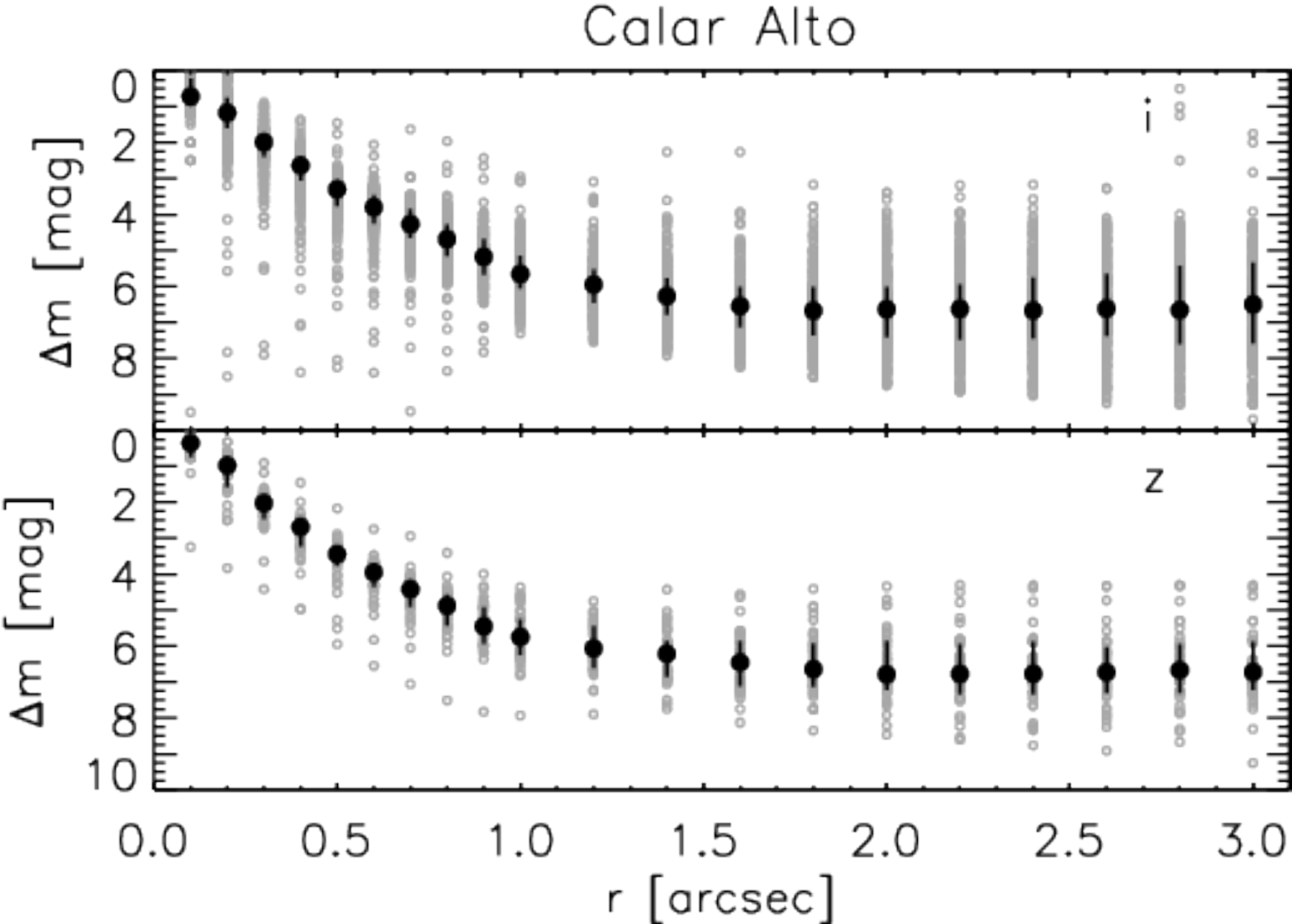}
\caption{Image sensitivities for observations of {\it Kepler} stars with the 
Calar Alto 2.2-m telescope. The median sensitivities and quartiles 
are plotted with black symbols.
\label{Sensitivites_CalarAlto}}
\end{figure}

\begin{figure}[!t]
\centering
\includegraphics[angle=90, scale=0.48]{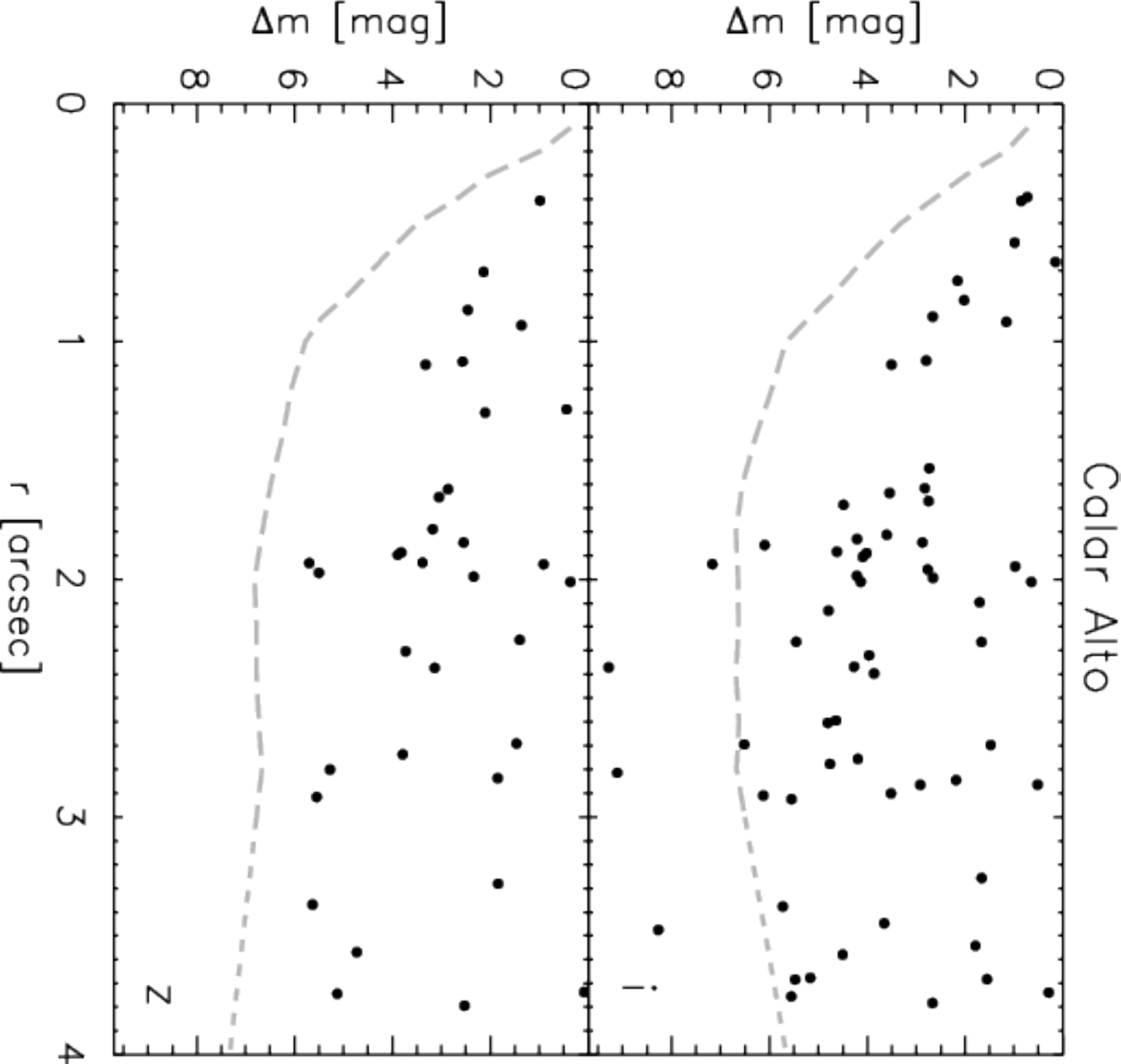}
\caption{Magnitude difference versus radial separation for all companions detected
around {\it Kepler} stars with the Calar Alto 2.2-m telescope in the $i$-band ({\it top}) 
and $z$-band ({\it bottom}). The dashed lines are the median image sensitivities from 
Figure \ref {Sensitivites_CalarAlto}. 
\label{KOIs_CalarAlto}}
\end{figure}

\subsubsection{Calar Alto}

\citet{lillo-box12,lillo-box14} used the lucky imaging technique to obtain 
high-resolution images for a sample of 234 KOI host stars. Given that 
this imaging method is different from the ones described above, we 
include in this section image sensitivity plots for the images taken with
the SDSS $i$ and $z$ filters and the AstraLux instrument (the data can 
be found on the CFOP site). 
The 5$\sigma$ sensitivities are shown in Figure \ref{Sensitivites_CalarAlto}; 
beyond about 1\arcsec, they are somewhat  lower than the sensitivities of 
most of our AO images (median $\Delta m \sim$ 6-7 versus median
$\Delta m >$ 8 for the AO $K$-band data), but they are substantially lower in the 
inner 0.5\arcsec. This is not surprising, given the different imaging techniques.

The companions detected within 4\arcsec\ by \citet{lillo-box12, lillo-box14} 
are shown in Figure \ref{KOIs_CalarAlto}. Only a few bright companions are 
found within $\sim$~1\arcsec, but several fainter companions ($\Delta m 
\sim$ 4-6) are revealed at separations larger than about 2\arcsec. In this 
separation range, there are a few companions with $\Delta m$ of 7-9.5 in 
the $i$-band (these detections lie above the median image sensitivity since 
the images used to extract them were likely obtained under exceptionally 
good observing conditions).
Of the 234 KOI host stars observed, 53 have at least one companion
at projected separations $<$~4\arcsec, which translates to an observed 
fraction of 23$\pm$3\%, a value just somewhat lower than the one derived
from our AO data. If only companions within 1\arcsec\ are considered, 
the observed fraction of stars with companions decreases to 3$\pm$1\%.
This value is much smaller than what we obtained from our AO and 
speckle images and can be understood in terms of the lower image
sensitivity of the lucky images within $\sim$ 1\arcsec\ from the
primary stars.

\newpage

\subsection{Compilation of all KOI Host Star Companions}

We combined the results on detected KOI companions from this work
and the literature (see Table \ref{obs_list} for references) to create 
a list with the separations and position angles of the companions and 
any $\Delta m$ values that were measured (including their derived
uncertainties, but with a floor of 0.01 mag). We limit this list to companions 
within 4\arcsec\ of each KOI host star. It is important to note that the 
companions are not necessarily bound; they could be background stars 
or galaxies that are just by chance aligned with a KOI host star, and more 
analysis is needed to determine whether they and their primary stars form 
bound systems \citep[see][]{teske15,hirsch16}. On the other hand, from 
simulations of stars in the {\it Kepler} field, \citet{horch14} found that most 
of the companions within $\sim$~1\arcsec\ are expected to be bound. 

When combining the measurements of detected companions, we used
results from both high-resolution imaging and seeing-limited imaging
(mainly the UKIRT survey). Some companions were detected in only one 
band, while others have detections in multiple bands. For the separation and
position angle of each companion, we averaged the results from different
measurements; they usually agreed fairly well, but in a few cases the 
position angles were discrepant (usually related to an orientation problem
in the image). When companions were measured multiple times in
the same band, we averaged their $\Delta m$ values in that band, weighted
by the inverse squared of the uncertainty in $\Delta m$ of each measurement.
If the standard deviation of the individual measurements was larger than 
the formal value of the combined $\Delta m$ uncertainties, we used it as 
the uncertainty of the average $\Delta m$ value. Thus, in a few cases where
the measurements were discrepant, the uncertainty of the combined $\Delta m$ 
value is fairly large.

The results of our KOI companion compilation are shown in Table 
\ref{photometry_tab}. It contains 2297 companions around 1903
primary stars; 330 KOI host stars have two or more companions stars.
We assign an identifier to each companion, choosing letters ``B'' to ``H'' 
for the first to seventh companion star. This nomenclature does not imply 
that the companions are actually bound (see note above); it is used to 
uniquely identify each companion star. 
There are eight KOI host stars with more than three companions:
KOI 113, 908, 1019, 1397, 1884, 3049, 3444, and 4399; most of these 
companions lie at separations $>$1\arcsec\ and may therefore be unbound.
We also list the KIC ID for each primary and companion star in Table 
\ref{photometry_tab}; in most cases, the two KIC IDs are the same, since 
the stars are not resolved in the KIC, but for 78 wide companions 
($\geq$ 2\arcsec\ from the primary), both objects can be found 
in the KIC.

\begin{deluxetable*}{cccccccccccccc}[h]\scriptsize
\rotate
\tablewidth{1.0\linewidth}
\tablecaption{Relative Photometry (${\Delta}m$), Separations and Position Angles 
for Companions to KOI Host Stars 
\label{photometry_tab}}
\tablehead{
  &  &  &  &  &  & \multicolumn{8}{c}{$\Delta m = m_{\mathrm{secondary}}-m_{\mathrm{primary}}$
 for photometric bands:} \\
\colhead{KOI} &  \colhead{ID} & \colhead{KICID$_{prim}$} & \colhead{KICID$_{sec}$} &  \colhead{d [$\arcsec$]} & \colhead{PA [\degr]} & 
\colhead{$F555W$} & \colhead{$F775W$} & \colhead{$i$} & \colhead{$z$} & \colhead{$LP600$} & \colhead{$562$} & 
\colhead{$692$} & \colhead{$880$} \\
\colhead{(1)} & \colhead{(2)} & \colhead{(3)} & \colhead{(4)} & \colhead{(5)} & \colhead{(6)} & 
\colhead{(7)} & \colhead{(8)} & \colhead{(9)} & \colhead{(10)} & \colhead{(11)} & \colhead{(12)} & \colhead{(13)} & \colhead{(14)}}
\startdata
    1 &  B &  11446443 &  (11446443) &  1.112 $\pm$ 0.051 &  136.2 $\pm$  1.1 &  \nodata &  \nodata &   3.950 $\pm$ 0.330 &  \nodata &  \nodata &  \nodata &   4.269 $\pm$ 0.150 &   3.379 $\pm$ 0.150 \\
    2 &  B &  10666592 &  (10666592) &  3.093 $\pm$ 0.050 &  266.4 $\pm$  1.0 &  \nodata &  \nodata &  \nodata &  \nodata &  \nodata &  \nodata &  \nodata &  \nodata \\
    2 &  C &  10666592 &  (10666592) &  3.849 $\pm$ 0.060 &   90.1 $\pm$  1.1 &  \nodata &  \nodata &  \nodata &  \nodata &  \nodata &  \nodata &  \nodata &  \nodata \\
    4 &  B &   3861595 &   (3861595) &  3.394 $\pm$ 0.062 &   74.8 $\pm$  1.0 &  \nodata &  \nodata &   4.460 $\pm$ 0.050 &  \nodata &  \nodata &  \nodata &  \nodata &  \nodata \\
    5 &  B &   8554498 &   (8554498) &  0.029 $\pm$ 0.050 &  142.1 $\pm$  1.0 &  \nodata &  \nodata &  \nodata &  \nodata &  \nodata &  \nodata &  \nodata &  \nodata \\
    5 &  C &   8554498 &   (8554498) &  0.141 $\pm$ 0.050 &  304.3 $\pm$  2.2 &  \nodata &  \nodata &  \nodata &  \nodata &  \nodata &   2.841 $\pm$ 0.389 &   3.036 $\pm$ 0.150 &  \nodata \\
    6 &  B &   3248033 &   (3248033) &  3.381 $\pm$ 0.050 &  307.8 $\pm$  1.0 &  \nodata &  \nodata &  \nodata &  \nodata &  \nodata &  \nodata &  \nodata &  \nodata \\
   10 &  B &   6922244 &   (6922244) &  3.128 $\pm$ 0.050 &  265.8 $\pm$  1.0 &  \nodata &  \nodata &  \nodata &  \nodata &  \nodata &  \nodata &  \nodata &  \nodata \\
   10 &  C &   6922244 &   (6922244) &  3.830 $\pm$ 0.050 &   89.3 $\pm$  1.0 &  \nodata &  \nodata &  \nodata &  \nodata &  \nodata &  \nodata &  \nodata &  \nodata \\
   12 &  B &   5812701 &   (5812701) &  0.603 $\pm$ 0.050 &  345.7 $\pm$  1.0 &  \nodata &  \nodata &  \nodata &  \nodata &  \nodata &  \nodata &  \nodata &  \nodata \\
   12 &  C &   5812701 &   (5812701) &  1.903 $\pm$ 0.050 &  320.3 $\pm$  1.0 &  \nodata &  \nodata &  \nodata &  \nodata &  \nodata &  \nodata &  \nodata &  \nodata \\
   13 &  B &   9941662 &   (9941662) &  1.144 $\pm$ 0.083 &  279.7 $\pm$  4.3 &  \nodata &  \nodata &   0.190 $\pm$ 0.060 &  \nodata &  \nodata &   1.008 $\pm$ 0.276 &   0.715 $\pm$ 0.210 &   0.619 $\pm$ 0.239 \\
   14 &  B &   7684873 &   (7684873) &  1.724 $\pm$ 0.050 &  273.5 $\pm$  1.0 &  \nodata &  \nodata &  \nodata &  \nodata &  \nodata &  \nodata &  \nodata &  \nodata \\
   18 &  B &   8191672 &   (8191672) &  0.912 $\pm$ 0.050 &  167.3 $\pm$  1.0 &  \nodata &  \nodata &  \nodata &  \nodata &  \nodata &  \nodata &  \nodata &  \nodata \\
   18 &  C &   8191672 &   (8191672) &  3.464 $\pm$ 0.050 &  110.5 $\pm$  1.1 &  \nodata &  \nodata &  \nodata &  \nodata &  \nodata &  \nodata &  \nodata &  \nodata \\
   21 &  B &  10125352 &    10125357 &  2.074 $\pm$ 0.050 &   59.8 $\pm$  1.0 &  \nodata &  \nodata &  \nodata &  \nodata &  \nodata &  \nodata &  \nodata &  \nodata \\
   28 &  B &   4247791 &   (4247791) &  0.560 $\pm$ 0.050 &   23.3 $\pm$  1.0 &  \nodata &  \nodata &  \nodata &  \nodata &  \nodata &   1.110 $\pm$ 0.150 &  \nodata &  \nodata \\
   41 &  B &   6521045 &   (6521045) &  1.832 $\pm$ 0.050 &  242.1 $\pm$  1.0 &  \nodata &  \nodata &   4.206 $\pm$ 0.097 &  \nodata &  \nodata &  \nodata &  \nodata &  \nodata \\
   41 &  C &   6521045 &   (6521045) &  3.434 $\pm$ 0.050 &  195.8 $\pm$  1.0 &  \nodata &  \nodata &  \nodata &  \nodata &  \nodata &  \nodata &  \nodata &  \nodata \\
   42 &  B &   8866102 &   (8866102) &  1.667 $\pm$ 0.061 &   35.7 $\pm$  2.1 &  \nodata &  \nodata &  \nodata &  \nodata &   3.040 $\pm$ 0.170 &   4.240 $\pm$ 0.150 &  \nodata &  \nodata \\
   43 &  B &   9025922 &   (9025922) &  3.341 $\pm$ 0.050 &   83.6 $\pm$  1.0 &  \nodata &  \nodata &  \nodata &  \nodata &  \nodata &  \nodata &  \nodata &  \nodata \\
   44 &  B &   8845026 &   (8845026) &  3.356 $\pm$ 0.068 &  124.7 $\pm$  1.3 &  \nodata &  \nodata &  \nodata &  \nodata &  \nodata &  \nodata &  \nodata &  \nodata \\
   45 &  B &   3742855 &   (3742855) &  3.140 $\pm$ 0.050 &   36.8 $\pm$  1.0 &  \nodata &  \nodata &  \nodata &  \nodata &  \nodata &  \nodata &  \nodata &  \nodata \\
   45 &  C &   3742855 &   (3742855) &  3.964 $\pm$ 0.051 &   72.9 $\pm$  2.3 &  \nodata &  \nodata &  \nodata &  \nodata &  \nodata &  \nodata &  \nodata &  \nodata \\
   51 &  B &   6056992 &   (6056992) &  3.510 $\pm$ 0.050 &  161.0 $\pm$  1.0 &  \nodata &  \nodata &  \nodata &  \nodata &   2.630 $\pm$ 0.070 &  \nodata &  \nodata &  \nodata \\
   53 &  B &   2445975 &     2445980 &  3.315 $\pm$ 0.050 &   95.5 $\pm$  1.0 &  \nodata &  \nodata &  \nodata &  \nodata &  \nodata &  \nodata &  \nodata &  \nodata \\
   53 &  C &   2445975 &     2445972 &  3.381 $\pm$ 0.050 &  210.9 $\pm$  1.0 &  \nodata &  \nodata &  \nodata &  \nodata &  \nodata &  \nodata &  \nodata &  \nodata \\
   68 &  B &   8669092 &   (8669092) &  0.735 $\pm$ 0.052 &  256.5 $\pm$  2.1 &  \nodata &  \nodata &  \nodata &  \nodata &  \nodata &   3.131 $\pm$ 0.348 &   2.874 $\pm$ 0.150 &  \nodata \\
   68 &  C &   8669092 &   (8669092) &  2.738 $\pm$ 0.060 &  256.7 $\pm$  4.0 &  \nodata &  \nodata &  \nodata &  \nodata &  \nodata &  \nodata &  \nodata &  \nodata \\
   68 &  D &   8669092 &   (8669092) &  3.406 $\pm$ 0.053 &  352.4 $\pm$  3.6 &  \nodata &  \nodata &  \nodata &  \nodata &  \nodata &  \nodata &  \nodata &  \nodata \\
\enddata
\tablecomments{Column (1) lists the KOI number of the host star, column (2) the identifier
we assigned to each companion star (``B'' for the first companion, ``C'' for the second companion,
etc), columns (3) and (4) contain the KIC ID of the primary and companion (``secondary'') star, 
respectively (a value in parentheses for the companion star means that it is not a distinct source 
in the KIC), columns (5) and (6) list the separation and position angle (from north through east), 
respectively, of the companion relative to the primary, and columns (7) to (20) list the difference 
in magnitudes between the primary and the companion star in different bands.}
\end{deluxetable*}

\begin{deluxetable*}{cccccccc}[!]\scriptsize
%\rotate
\tablewidth{1.0\linewidth}
\tablenum{8}
\tablecaption{{\bf Continued.} Relative Photometry (${\Delta}m$), Separations and Position Angles 
for Companions to KOI Host Stars}
\tablehead{
  &  & \multicolumn{6}{c}{$\Delta m = m_{\mathrm{secondary}}-m_{\mathrm{primary}}$
  for photometric bands:} \\
\colhead{KOI} &  \colhead{ID} & \colhead{$U$}  & \colhead{$B$} & \colhead{$V$} & 
\colhead{$J$}  & \colhead{$H$} & \colhead{$K$}  \\
\colhead{(1)} & \colhead{(2)} & \colhead{(15)} & \colhead{(16)} & \colhead{(17)} & \colhead{(18)} & 
\colhead{(19)} & \colhead{(20)}}
\startdata
    1 &  B &  \nodata &  \nodata &  \nodata &   2.800 $\pm$ 0.100 &   2.500 $\pm$ 0.100 &   2.359 $\pm$ 0.029 \\
    2 &  B &  \nodata &  \nodata &  \nodata &  \nodata &  \nodata &   7.525 $\pm$ 0.114 \\
    2 &  C &  \nodata &  \nodata &  \nodata &   5.745 $\pm$ 0.018 &  \nodata &   5.965 $\pm$ 0.045 \\
    4 &  B &  \nodata &  \nodata &  \nodata &   4.233 $\pm$ 0.010 &  \nodata &  \nodata \\
    5 &  B &  \nodata &  \nodata &  \nodata &  \nodata &  \nodata &   0.400 $\pm$ 0.062 \\
    5 &  C &  \nodata &  \nodata &  \nodata &  \nodata &  \nodata &   2.310 $\pm$ 0.199 \\
    6 &  B &  \nodata &  \nodata &  \nodata &   7.393 $\pm$ 0.126 &  \nodata &  \nodata \\
   10 &  B &  \nodata &  \nodata &  \nodata &   7.895 $\pm$ 0.032 &  \nodata &  \nodata \\
   10 &  C &  \nodata &  \nodata &  \nodata &   6.266 $\pm$ 0.032 &  \nodata &  \nodata \\
   12 &  B &  \nodata &  \nodata &  \nodata &  \nodata &  \nodata &   3.835 $\pm$ 0.010 \\
   12 &  C &  \nodata &  \nodata &  \nodata &  \nodata &  \nodata &   7.539 $\pm$ 0.043 \\
   13 &  B &  \nodata &  \nodata &  \nodata &   0.180 $\pm$ 0.031 &  \nodata &   0.145 $\pm$ 0.042 \\
   14 &  B &  \nodata &  \nodata &  \nodata &   4.304 $\pm$ 0.150 &  \nodata &   3.514 $\pm$ 0.150 \\
   18 &  B &  \nodata &  \nodata &  \nodata &   5.365 $\pm$ 0.041 &  \nodata &  \nodata \\
   18 &  C &  \nodata &  \nodata &  \nodata &   6.014 $\pm$ 0.122 &  \nodata &  \nodata \\
   21 &  B &  \nodata &  \nodata &  \nodata &   2.402 $\pm$ 0.010 &  \nodata &  \nodata \\
   28 &  B &  \nodata &  \nodata &  \nodata &  \nodata &  \nodata &  \nodata \\
   41 &  B &  \nodata &  \nodata &  \nodata &  \nodata &  \nodata &  \nodata \\
   41 &  C &  \nodata &  \nodata &  \nodata &  \nodata &  \nodata &  10.049 $\pm$ 0.162 \\
   42 &  B &  \nodata &  \nodata &  \nodata &   2.212 $\pm$ 0.026 &  \nodata &   1.873 $\pm$ 0.024 \\
   43 &  B &   1.199 $\pm$ 0.065 &   1.135 $\pm$ 0.049 &   1.098 $\pm$ 0.034 &   1.110 $\pm$ 0.016 &  \nodata &  \nodata \\
   44 &  B &  \nodata &  \nodata &  \nodata &   3.983 $\pm$ 0.021 &   3.803 $\pm$ 0.032 &   3.825 $\pm$ 0.010 \\
   45 &  B &  \nodata &  \nodata &  \nodata &   1.407 $\pm$ 0.087 &  \nodata &  \nodata \\
   45 &  C &  \nodata &  \nodata &  \nodata &  -2.503 $\pm$ 0.021 &  -2.092 $\pm$ 0.151 &  \nodata \\
   51 &  B &  \nodata &  \nodata &  \nodata &  \nodata &  \nodata &  \nodata \\
   53 &  B &  -0.952 $\pm$ 0.035 &  -0.579 $\pm$ 0.036 &  -0.378 $\pm$ 0.027 &  -0.099 $\pm$ 0.010 &  \nodata &  \nodata \\
   53 &  C &   1.170 $\pm$ 0.047 &   0.695 $\pm$ 0.037 &   0.500 $\pm$ 0.028 &  -0.295 $\pm$ 0.010 &  \nodata &  \nodata \\
   68 &  B &  \nodata &  \nodata &  \nodata &   2.025 $\pm$ 0.070 &  \nodata &   1.800 $\pm$ 0.020 \\
   68 &  C &  \nodata &  \nodata &  \nodata &   7.166 $\pm$ 0.090 &  \nodata &   6.200 $\pm$ 0.020 \\
   68 &  D &  \nodata &  \nodata &  \nodata &   6.498 $\pm$ 0.229 &  \nodata &   5.800 $\pm$ 0.020 \\
\enddata
\tablecomments{The full table is available in a machine-readable form in the online
journal. A portion is shown here for guidance regarding content and form.}
\end{deluxetable*}

\begin{figure*}[!t]
\centering
\includegraphics[angle=90, scale=0.54]{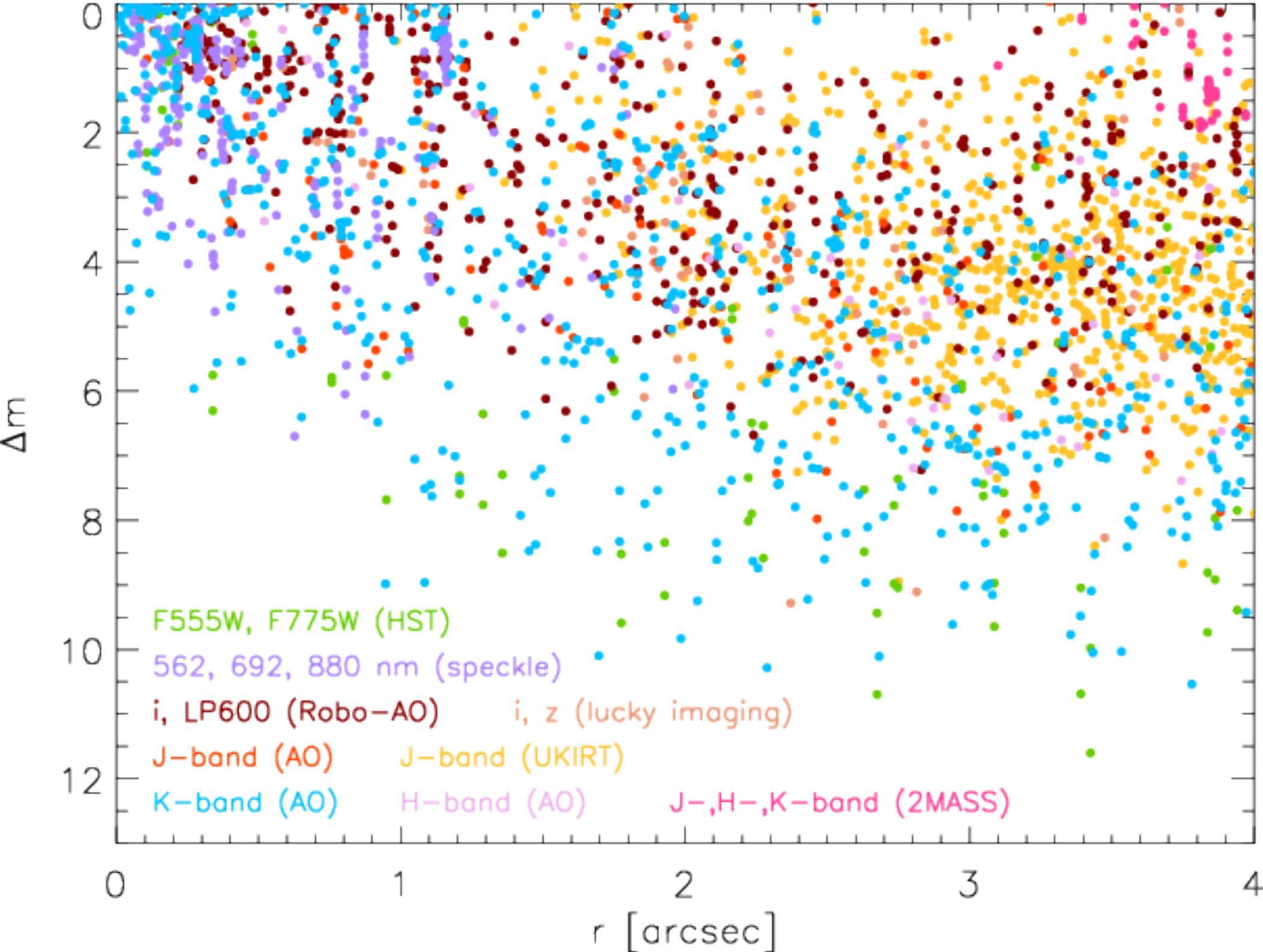}
\caption{Magnitude difference versus radial projected separation relative to the
primary star for all companions detected around {\it Kepler} planet host stars 
(i.e., stars that host at least one candidate or confirmed planet). The 
magnitude differences in the various photometric bands are color-coded. 
\label{KOI_planets_delta_m_radius}}
\end{figure*}

\begin{figure*}[!t]
\centering
\includegraphics[angle=90, scale=0.54]{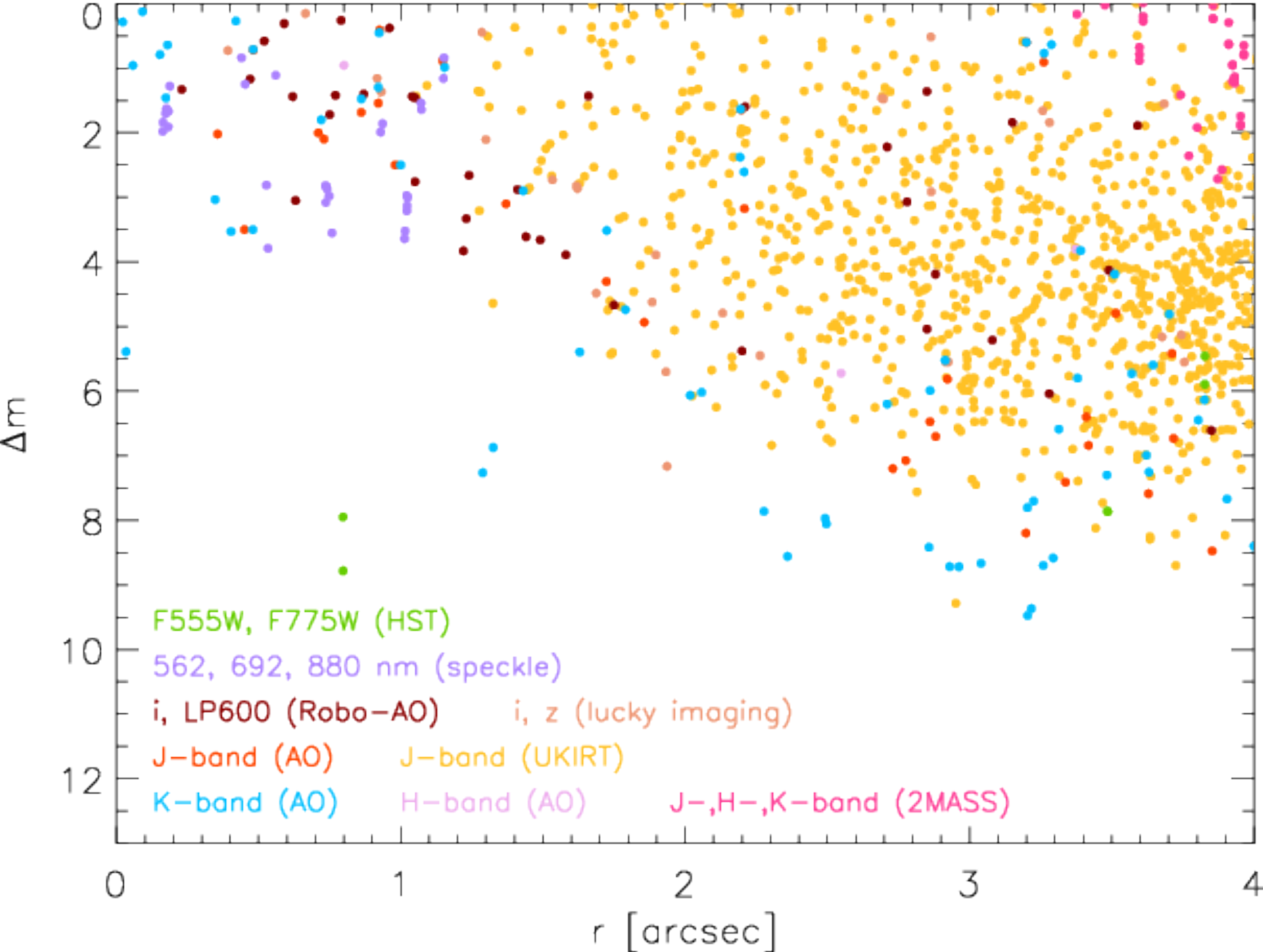}
\caption{Magnitude difference versus radial projected separation relative to the
primary star for all companions detected around {\it Kepler} stars with one or
more transit events that are all classified as false positive. 
 \label{KOI_FPs_delta_m_radius}}
\end{figure*}

The companions from Table \ref{photometry_tab} are plotted in Figures
\ref{KOI_planets_delta_m_radius} and \ref{KOI_FPs_delta_m_radius}. 
The two figures separate the KOIs identified as planet candidates or
confirmed planets from those identified as false positives (if a KOI host
star has both candidate or confirmed planets and one or more false
positive transit signals, its companions are shown in Figure 
\ref{KOI_planets_delta_m_radius}). Both Figures \ref{KOI_planets_delta_m_radius} 
and \ref{KOI_FPs_delta_m_radius} show the difference in magnitudes 
between primary and companion versus their projected separations, 
color-coded by the different photometric bands (some are grouped into 
the same color). Thus, if a companion has been detected in more than 
one photometric band, it will appear more than once in the figure, but 
usually with a different color and also likely different $\Delta m$ value.
Speckle and AO find the closest companions, while AO, lucky imaging, 
and in particular HST imaging find the faintest companions. Robo-AO
imaging detects companions down to 0.2\arcsec, with typical $\Delta m$ 
values (mostly in the $LP600$ band) between 0 and 5; just $\sim$~8\%
of companions found with Robo-AO are fainter than the primary by 
$\geq$ 5 mag. 
The UKIRT survey detects most companions at separations between 
2\arcsec\ and 4\arcsec\ (and beyond); typical $\Delta J$ values are 
below 7 mag.  

\begin{figure*}[!]
\centering
\includegraphics[scale=0.5]{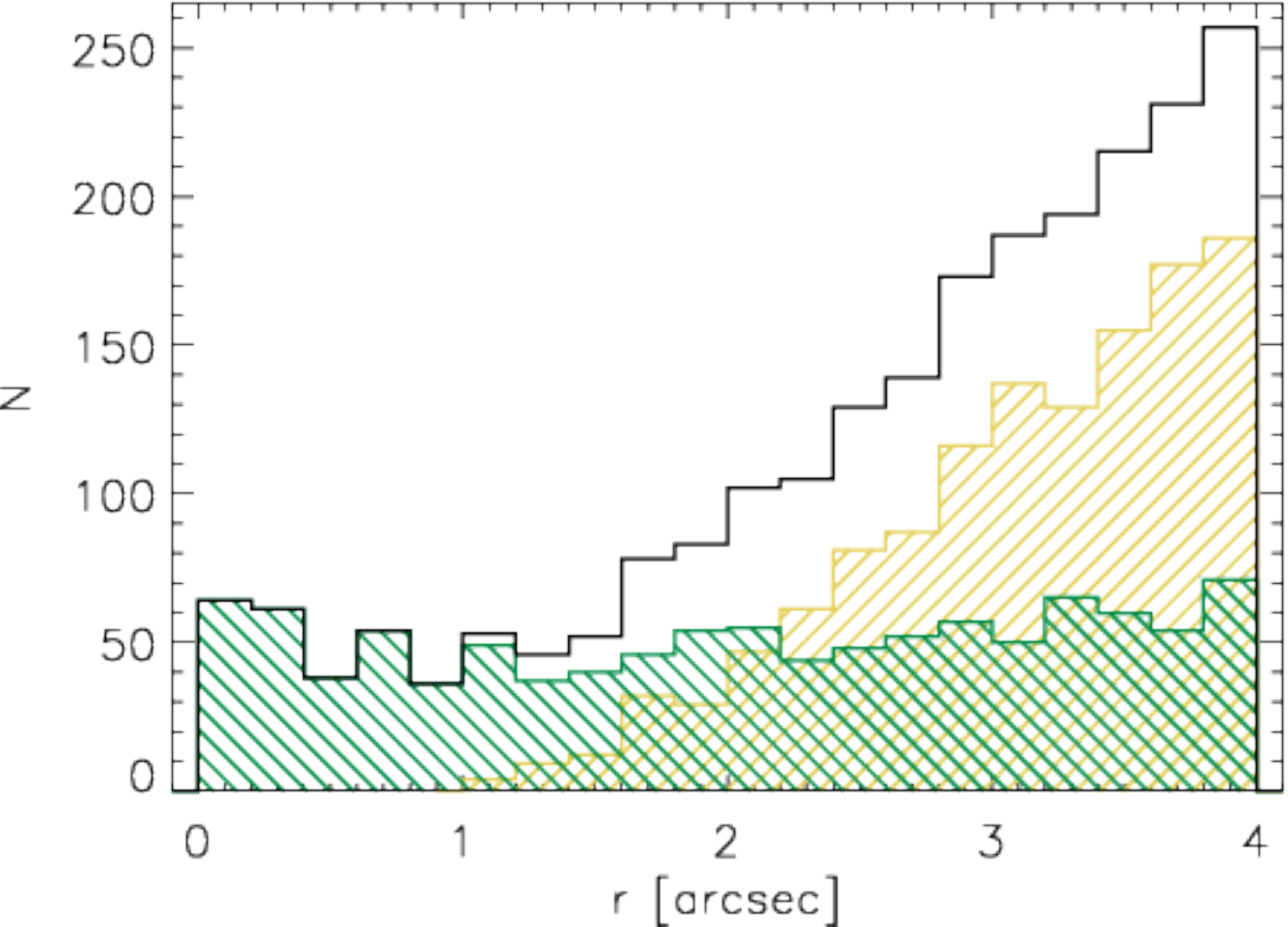}
\caption{Histogram of the distribution of radial projected separations for all detected
companions to KOI host stars ({\it black line}). The yellow shaded histogram 
is the distribution of companion separations for companions detected only in
UKIRT (and sometimes also 2MASS and UBV survey images), while the 
green shaded histogram is for the remaining companions detected in 
high-resolution images.
\label{KOI_dist_histo}}
\end{figure*}

Figure \ref{KOI_dist_histo} shows the distribution of projected separations 
between primary stars and companions for all KOI host stars. Also
shown are histograms for the separations of companions detected
only in UKIRT data (some of these stars are also detected in 2MASS
and UBV survey data) and of companions detected also (or only) in
high-resolution imaging data. The increase in numbers for separations 
larger than about 1.6\arcsec\ is clearly due to the detections of 
companions in UKIRT images. It is likely that many of these stars are
not actually bound companions, but background stars or galaxies.
Of the companions detected in high-resolution imaging data (thus,
excluding companions detected only in UKIRT, 2MASS, and UBV 
survey data), we find that 46$\pm$2\% are found at separations 
$<$ 2\arcsec; of these close companions, 53$\pm$3\% are within 
1\arcsec\ from the primary.

\begin{figure*}[!]
\centering
\includegraphics[angle=90,scale=0.7]{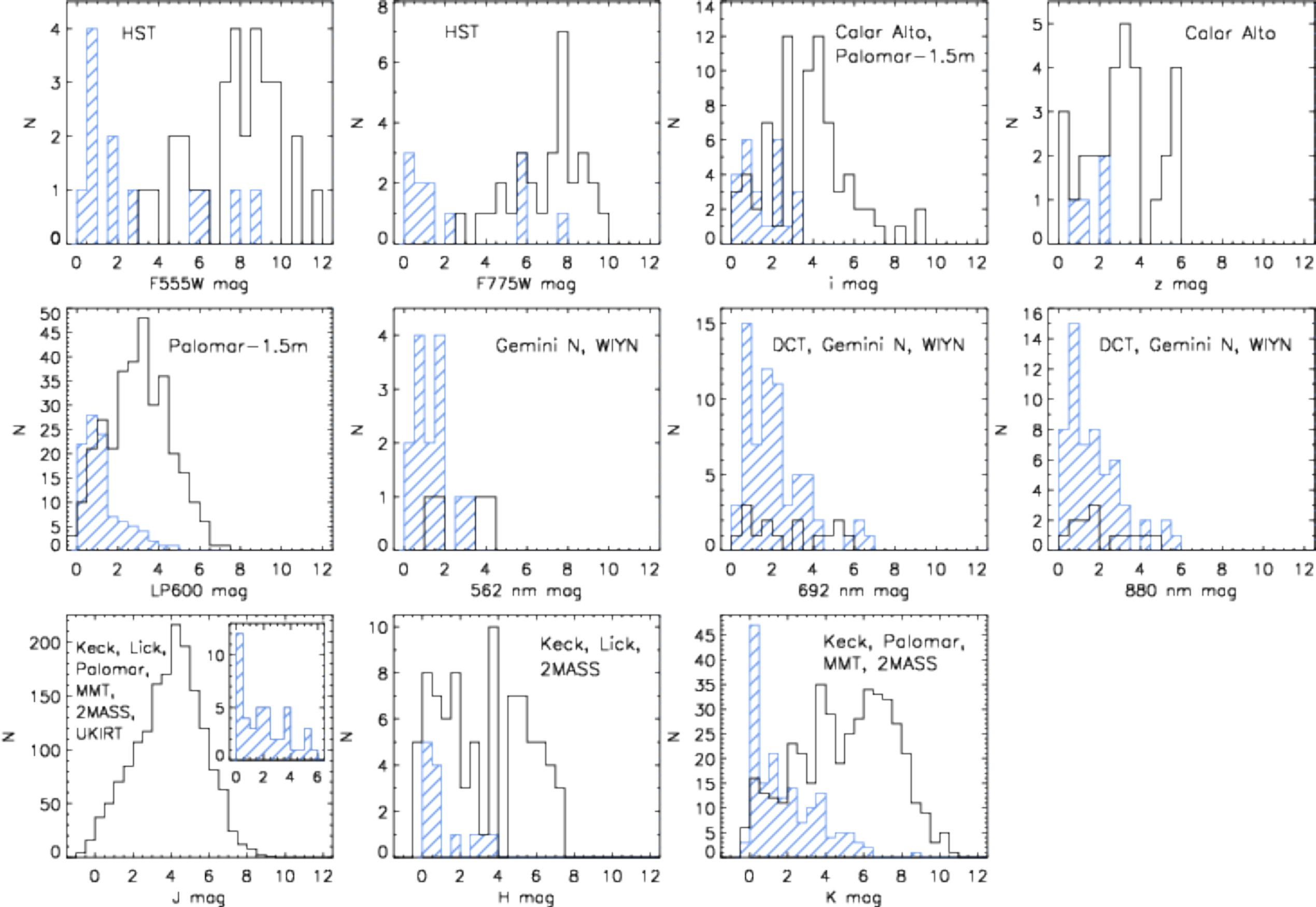}
\caption{Histograms of the distributions of $\Delta m$ values in various 
bands for all detected companions to KOI host stars. The blue shaded 
histogram is the distribution for companions separated by $\leq$ 
1\arcsec\ from the primary star, while the black histograms are for 
companions separated by $>$ 1\arcsec.
\label{KOI_mag_histo}}
\end{figure*}

The distributions of $\Delta m$ values in various bands for all detected 
companions to KOI host stars are shown in Figure \ref{KOI_mag_histo}.
The most sensitive observations are those obtained by HST; a few
companions have $\Delta m > 10$ in the $F555W$ band. Of the 
ground-based observations at optical wavelengths, the faintest
companions are detected in the $i$-band with lucky imaging at
Calar Alto. 
The three bands with the largest number of observations and thus
companion detections, $LP600$, $J$, and $K$ display different
distributions of $\Delta m$ values for companions at $>$ 1\arcsec\
from the primary stars. The $LP600$ band shows a broad peak
around 3.5 mag, while the $K$-band distribution is very wide,
spanning values up to 11 mag, with two peaks at 4 and 7 mag
and smaller peaks at 0 and 2.5 mag. The $J$ band, which is
dominated by UKIRT measurements, displays a broad peak
centered at 4-4.5 mag. 

Of particular interest are companions detected within 1\arcsec\ from 
the primary. Most of these companions are not much fainter than the 
primary ($\Delta m < 2$); this is partly an observational effect, as very 
faint companions next to brighter stars can be difficult to detect. 
At near-infrared wavelengths ($J$, $H$, $K$ bands), most close 
companions have $\Delta m$=0-0.5. In the $K$ band, almost all 
companions at $\leq$~1\arcsec\ and with $\Delta K$ $<$ 0.5 were
detected at Keck with adaptive optics.
It is these bright, close companions that will have the largest effect
on derived planet radii, as is described in the next section.

\subsection{Revised Transit Depths and Planet Radii}
\label{revised_radii}

\subsubsection{Background}

The observed transit depth $\delta_{\rm obs}$ in the {\it Kepler} bandpass 
($Kp$) is used to derive planet radii:
\begin{equation}
\delta_{\rm obs} = \frac{F_{\rm tot}-F_{\rm transit}}{F_{\rm tot}} = 
\left( \frac{R_p}{R_{\ast}} \right)^2, 
\end{equation}
where $F_{\rm tot}$ is the total out-of-transit flux, $F_{\rm transit}$
is the in-transit flux, and $R_p$ and $R_{\ast}$ are the planet and stellar
radius, respectively. However, if there is more than one star in a
system, the total flux is the sum of the stellar fluxes, but the in-transit flux
depends on which star the planet transits. Thus, the observed transit
depth becomes
\begin{equation}
\delta_{\rm obs} = \left( \frac{F_{\ast}}{F_{\rm tot}} \right)
\left( \frac{R_p}{R_{\ast}} \right)^2, 
\end{equation}
where the star symbol denotes the star with the transiting planet. Thus,
the transit depth is shallower, and the derived planet radius is smaller
when the additional stars in the system are not taken into account. 
Given that the radii of {\it Kepler} planets are derived assuming a single
star (with a correction factor applied to account only for flux 
dilution by nearby stars resolved in the KIC; \citealt{mullally15,coughlin16}), 
the presence of close companions results in an upward revision 
of the planet radius (see \citealt{ciardi15} for more details). 
If we assume that the planet orbits the primary star in a multiple system, the 
``corrected'' planet radius relative to the one derived assuming a single star 
($R_p$) is
\begin{equation}
R_{p, \rm corr} = R_p \sqrt{\frac{F_{\rm tot}}{F_{\rm prim}}},
\end{equation}
where $F_{\rm tot}$ is the combined out-of-transit flux of all stars in the 
system and $F_{\rm prim}$ is the flux of the primary star. For two stars, 
$F_{\rm tot}=F_{\rm prim}+F_{\rm sec}$ and, in magnitudes, 
$\Delta m = m_{\rm sec} - m_{\rm prim} = 
- 2.5 \log (F_{\rm sec}/F_{\rm prim})$; then the previous equation 
becomes:
\begin{equation}
R_{p, \rm corr} = R_p \sqrt{1+10^{-0.4 \Delta m}}
\end{equation}
The expression under the square root can be considered a ``planet 
radius correction factor''.
For more than one companion star, the previous equation converts to
\begin{equation}
R_{p, \rm corr} = R_p \sqrt{1+\sum_{i=1}^{N}{10^{-0.4 \Delta m_i}}},
\label{Rcorr}
\end{equation}
where the sum is for N companion stars with magnitude differences
$\Delta m_i$ relative to the primary star. 

These equations for calculating revised planet radii assume that
planets orbit the primary star; if they orbit one of the companion stars,
there is an additional dependence on stellar radii:
\begin{equation}
R_{p, \rm corr} = R_p \frac{R_{\rm sec}}{R_{\ast}} 
\sqrt{\frac{F_{\rm tot}}{F_{\rm sec}}},
\end{equation} 
where $R_{sec}$ and $F_{sec}$ are the radius and flux, respectively, of
the secondary star around which the planet orbits, and $R_{\ast}$ is 
the radius of the star when assumed to be single (i.e., the radius of the
primary star). In the case of two stars, the flux ratio $F_{\rm tot}/F_{\rm sec}$ 
is equal to $1+10^{0.4 \Delta m}$, where $\Delta m = 
m_{\rm sec} - m_{\rm prim}$.
If there is more than one companion star and the planet obits the 
secondary whose magnitude difference with respect to the primary is
$\Delta m_c$, the equation to derive revised planet radii becomes
\begin{equation}
R_{p, \rm corr} = R_p \frac{R_{\rm sec}}{R_{\ast}}
\sqrt{10^{0.4 \Delta m_c} (1+ \sum_{i=1}^{N}{10^{-0.4\Delta m_i}})}   
\label{Rcorr2}
\end{equation} 
Thus, as shown by \citet{ciardi15}, the radius correction factor can
become much larger if planets orbit a typically fainter (and smaller) 
companion star.
 
The $\Delta m_i$ values in Equations \ref{Rcorr} and \ref{Rcorr2} 
are for the {\it Kepler} bandpass. Thus, to revise the transit depths 
and thus planet radii for the systems in Table \ref{photometry_tab}, 
the magnitude differences between primary stars and companions 
have to be converted from the bandpass in which they were measured 
to the equivalent magnitude difference in the {\it Kepler} bandpass. 

For the HST F555W and F775W bandpasses, \citet{cartier15} derived the
following relation:
\begin{equation}
Kp = 0.236 + 0.406 \times F555W + 0.594 \times F775W,
\end{equation} 
with an uncertainty in the derived $Kp$ value of $\sigma(Kp) = 
\sqrt{0.019^2(F555W-F775W)^2+0.027^2}$.
 
\citet{lillo-box14} found a linear correlation between the {\it Kepler} magnitudes
(for $13 < Kp < 16$) and the SDSS $i$ ($i_{SDSS}$) magnitudes:
\begin{equation}
i_{SDSS} = 0.947 Kp + 0.510
\end{equation}

\citet{everett15} used a library of model spectra to derive relationships between 
stellar properties and the conversion from magnitudes in the speckle filters to 
those in $Kp$. They then applied the conversion to each KOI host star based 
on the stellar properties reported for the star. As an approximation, we assume
that magnitudes measured in the 692 nm DSSI band are the same as for the 
$Kp$ bandpass.
Also the $LP600$ filter used by Robo-AO at the Palomar 1.5-m telescope
is similar to the {\it Kepler} bandpass, and thus we can assume $\Delta LP600 =
\Delta Kp$ \citep{law14, baranec16, ziegler16}.

\citet{howell12} used photometry from the KIC and 2MASS magnitudes to derive
relations between the infrared color $(J-K_s)$ and the {\it Kepler} magnitude.
They inferred
\begin{eqnarray}
Kp - K_s & = & 0.314377 + 3.85667 \Delta + 3.176111 \Delta^2 \nonumber \\
& - & 25.3126 \Delta^3 + 40.7221 \Delta^4  \nonumber \\
& - & 19.2112 \Delta^5 \quad \mathrm{for \; dwarfs} \nonumber \\
Kp - K_s & = & 0.42443603 + 3.7937617 \Delta -2.3267277 \Delta^2 \nonumber \\
& + & 1.4602553 \Delta^3 
\mathrm{\quad for \;  giants,}
\end{eqnarray} 
where $\Delta=J-K_s$ (see \citealt{ciardi11} on how dwarfs and giants are
separated in the KIC). Typical uncertainties in the derived $Kp-K_s$ colors
are 0.083 mag for dwarfs and 0.065 mag for giants. Given that, as in
\citet{howell12}, we measured the $K$-band magnitude of our targets in
a band that is slightly different from the $K_s$ band, using the above
relationships to convert our measured $K$-band magnitude to a $Kp$
magnitude adds an additional uncertainty of about 0.03 mag 
\citep[see][]{howell12}.

For cases where only a $J$- or only a $K$-band magnitude is known, but
not both, \citet{howell12} derived the following relations:
\begin{eqnarray*}
Kp - J & = & -398.04666 + 149.08127 J - 21.952130 J^2 \nonumber \\
& + & 1.5968619 J^3 - 0.057478947 J^4 \nonumber \\
& + & 0.00082033223 J^5 \nonumber \\
& \mathrm{for} & \mathrm{10\, mag < J < 16.7\, mag} \nonumber \\
Kp - J & = & 0.1918 + 0.08156 J \nonumber \\
& \mathrm{for} & \mathrm{J > 16.7\, mag}
\end{eqnarray*} 
and
\begin{eqnarray}
Kp - K_s & = & -643.05169 + 246.00603 K_s - 37.136501 K_s^2 \nonumber \\
& + & 2.7802622 K_s^3 - 0.10349091 K_s^4 \nonumber \\
& + & 0.0015364343 K_s^5 \nonumber \\
& \mathrm{for} & \mathrm{10\, mag < K_s < 15.4\, mag} \nonumber \\
Kp - K_s & = & -2.7284 + 0.3311 K_s \nonumber \\
& \mathrm{for} & \mathrm{K_s > 15.4\, mag}
\end{eqnarray} 
$Kp$ magnitude estimates using these equations have an uncertainty
of about 0.6-0.8 mag. 

\newpage

\subsubsection{Radius Correction Factors for {\it Kepler} Planets}

Using the relations from Equations 8-11, we converted the measured $\Delta m$ 
values to $\Delta Kp$ values for observations in the $F555W$, $F775W$, $i$, 
$LP600$, 692 nm, $J$, and $K$ bands; of the 1903 KOI host stars that
have nearby stars, just 12 do not have observations in any of these seven
bands (and therefore do not have $\Delta Kp$ values derived for them).
With $\Delta Kp$ for the companion stars, Equations \ref{Rcorr} and
\ref{Rcorr2} can be used to derive correction factors for the planet radii.

However, Equation \ref{Rcorr2} also requires the ratio of the stellar radii
of the secondary and primary star. To derive an estimate of this ratio, we 
used the table with colors and effective temperatures for dwarf stars from 
\citet{pecaut13} and assumed that primary and secondary stars are bound. 
We also assumed that $Kp$ magnitudes correspond to $R$ magnitudes, 
and we adopted effective temperatures ($T_{eff}$) for the primary stars from 
the stellar parameters of \citet{huber14}. Using the tabulated $T_{eff}$ 
values, we derived $V-R$ ($=V-Kp$) colors and absolute $V$ magnitudes 
($M_{V}$) for the primary stars. Assuming primary and secondary stars are 
bound and therefore at the same distance from the Sun implies 
$M_{V, \rm sec} = m_{V, \rm sec}-m_{V, \rm prim}+ M_{V, \rm prim}$ or 
$M_{V, \rm sec} = Kp_{\rm sec}+(V-R)_{\rm sec}-Kp_{\rm prim}-
(V-R)_{\rm prim}+M_{V, \rm prim}$.
We found the $(V-R)_{\rm sec}$ color from the \citet{pecaut13} table 
that yielded a self-consistent $M_{V, \rm sec}$ value. With $M_{V, \rm sec}$
determined, the effective temperature and luminosity of the secondary
star are also known. Then, $\frac{R_{\rm sec}}{R_{\rm prim}} =
\sqrt{\frac{L_{\rm sec}}{L_{\rm prim}}}\left(\frac{T_{eff, \rm prim}}
{T_{eff, \rm sec}}\right)^2$. If a star had more than one companion
within 4\arcsec, we adopted the $\frac{R_{\rm sec}}{R_{\rm prim}}$
ratio and $\Delta m_c$ value of the brightest companion star (highest
luminosity as derived from its $M_V$ value) in Equation \ref{Rcorr2}.
We also checked that the brightness of the companion star, assumed to
be bound to the primary star, was still consistent with the transit depth; for 
example, a transit depth of 0.1\% is consistent with a companion star that 
is up to 7.5 mag fainter than the primary (in this case, the planet would fully 
obscure the companion star during transit). As a result, 249 companion 
stars were excluded from being the planet host star (roughly half of them are 
hosts to only false positive transit events).

For the 1891 KOI host stars with companions for which we derived $\Delta Kp$ 
values, we calculated factors to revise the planet radii  to take the flux dilution 
into account. We derived such factors assuming the planets orbit the primary 
star (see Table \ref{Rcorr1_tab}) and assuming the planets orbit the brightest 
companion star (see Table \ref{Rcorr2_tab}). 
We included the small number of stars with companions resolved in the KIC; 
even though flux dilution by these companions should already be accounted for 
in the planet radii listed in the latest KOI table, we did not attempt to evaluate 
this correction term and decided to treat all companions uniformly.
Both Tables \ref{Rcorr1_tab} and \ref{Rcorr2_tab} list correction factors derived 
from from measurements of $\Delta m$ in different bands; since the measurements 
were done in different filters at different telescopes, and there are uncertainties in 
converting them to $\Delta Kp$, the derived correction factors are expected to differ 
somewhat. 
Moreover, there are some cases in which a star has more than one companion,
and not all companions are detected in all bands (for example, a faint companion
close to the primary star is only detected in a Keck AO image, while a brighter 
companion at a larger distance is only measured in a UKIRT image). Therefore,
radius correction factors, which depend on the sum of the $\Delta m_i$ values
of the companion stars, are different for different bands for these stars. 

We also computed a weighted average of the radius correction factors for 
each star by using the inverse of the square of the uncertainty as weight. 
Given that we derived up to four correction factors from the $J$- and 
$K$-band measurements, we used the individual correction factors 
derived from $J$- or $K$-band measurements if companions were 
only measured in one of these two bands. If measurements in both 
the $J$- and $K$-band were available, we instead used the average 
of the correction factors derived from the relationships between $J-K_s$ 
color and $Kp-K_s$ for dwarfs and giants. However, in Tables \ref{Rcorr1_tab}
and \ref{Rcorr2_tab} there are 20 and 7 stars, respectively, for which the 
latter two correction factors differed by more than 25\%; for these we used 
the correction factors derived from the $J-$ and $K$-band in our calculation 
of the weighted average. Also, for one star in Table \ref{Rcorr1_tab} (KOI 2971),
the radius correction factor derived from the $J$-band band was very close to 1
and discrepant with the values derived from the other bands (since only a more
distant, faint companions was detected in $J$, but closer, brighter companions 
were detected in the other bands); the discrepant value was not included in 
the weighted average. 

\begin{deluxetable*}{cccccccccc}[!h]\scriptsize
\rotate
\tablewidth{1.0\linewidth}
\tablenum{9}
\tablecaption{Planet Radius Correction Factors Assuming Planets Orbit the Primary Stars, Derived from $\Delta m$
Measurements in Various Bands, and Weighted Average
\label{Rcorr1_tab}}
\tablehead{ 
\colhead{KOI} & \colhead{F555W,F775W} & \colhead{i} &
\colhead{692} & \colhead{LP600} & \colhead{J} & \colhead{K} & 
\colhead{$J-K$(dwarf)} & \colhead{$J-K$(giant)} & \colhead{Weighted average}  \\
\colhead{(1)} & \colhead{(2)} & \colhead{(3)} & \colhead{(4)} & \colhead{(5)} & \colhead{(6)} & 
\colhead{(17)} & \colhead{(8)} & \colhead{(9)} & \colhead{(10)}}
\startdata
  1 &  \nodata &  1.0107 $\pm$ 0.0034 &  1.0098 $\pm$ 0.0013 &  \nodata &  1.0517 $\pm$ 0.0465 &  1.0810 $\pm$ 0.0650 &  1.0165 $\pm$ 0.0065 &  1.0193 $\pm$ 0.0057 &  1.0102 $\pm$ 0.0018 \\
  2 &  \nodata &  \nodata &  \nodata &  \nodata &  1.0029 $\pm$ 0.0024 &  1.0017 $\pm$ 0.0015 &  1.0033 $\pm$ 0.0002 &  1.0036 $\pm$ 0.0002 &  1.0034 $\pm$ 0.0002 \\
  4 &  \nodata &  1.0065 $\pm$ 0.0003 &  \nodata &  \nodata &  1.0133 $\pm$ 0.0108 &  \nodata &  \nodata &  \nodata &  1.0065 $\pm$ 0.0003 \\
  5 &  \nodata &  \nodata &  1.0301 $\pm$ 0.0041 &  \nodata &  \nodata &  1.3946 $\pm$ 0.3033 &  \nodata &  \nodata &  1.0301 $\pm$ 0.0041 \\
  6 &  \nodata &  \nodata &  \nodata &  \nodata &  1.0006 $\pm$ 0.0005 &  \nodata &  \nodata &  \nodata &  1.0006 $\pm$ 0.0005 \\
 10 &  \nodata &  \nodata &  \nodata &  \nodata &  1.0013 $\pm$ 0.0011 &  \nodata &  \nodata &  \nodata &  1.0013 $\pm$ 0.0011 \\
 12 &  \nodata &  \nodata &  \nodata &  \nodata &  \nodata &  1.0215 $\pm$ 0.0175 &  \nodata &  \nodata &  1.0215 $\pm$ 0.0175 \\
 13 &  \nodata &  1.3532 $\pm$ 0.0179 &  1.2319 $\pm$ 0.0407 &  \nodata &  1.3555 $\pm$ 0.2605 &  1.3665 $\pm$ 0.2719 &  1.3301 $\pm$ 0.0233 &  1.3338 $\pm$ 0.0249 &  1.3314 $\pm$ 0.0226 \\
 14 &  \nodata &  \nodata &  \nodata &  \nodata &  1.0123 $\pm$ 0.0123 &  1.0295 $\pm$ 0.0291 &  1.0019 $\pm$ 0.0004 &  1.0026 $\pm$ 0.0006 &  1.0022 $\pm$ 0.0005 \\
 18 &  \nodata &  \nodata &  \nodata &  \nodata &  1.0047 $\pm$ 0.0042 &  \nodata &  \nodata &  \nodata &  1.0047 $\pm$ 0.0042 \\
 21 &  \nodata &  \nodata &  \nodata &  \nodata &  1.0539 $\pm$ 0.0430 &  \nodata &  \nodata &  \nodata &  1.0539 $\pm$ 0.0430 \\
 28 &  \nodata &  \nodata &  \nodata &  \nodata &  \nodata &  \nodata &  \nodata &  \nodata &  \nodata \\
 41 &  \nodata &  1.0083 $\pm$ 0.0008 &  \nodata &  \nodata &  \nodata &  1.0000 $\pm$ 0.0000 &  \nodata &  \nodata &  1.0083 $\pm$ 0.0008 \\
 42 &  \nodata &  \nodata &  \nodata &  1.0300 $\pm$ 0.0046 &  1.0174 $\pm$ 0.0144 &  1.0070 $\pm$ 0.0058 &  1.0405 $\pm$ 0.0044 &  1.0379 $\pm$ 0.0041 &  1.0349 $\pm$ 0.0044 \\
 43 &  \nodata &  \nodata &  \nodata &  \nodata &  1.1544 $\pm$ 0.1189 &  \nodata &  \nodata &  \nodata &  1.1544 $\pm$ 0.1189 \\
 44 &  \nodata &  \nodata &  \nodata &  \nodata &  1.0125 $\pm$ 0.0104 &  1.0106 $\pm$ 0.0087 &  1.0093 $\pm$ 0.0011 &  1.0095 $\pm$ 0.0011 &  1.0094 $\pm$ 0.0011 \\
 45 &  \nodata &  \nodata &  \nodata &  \nodata &  3.6294 $\pm$ 1.3971 &  \nodata &  \nodata &  \nodata &  3.6294 $\pm$ 1.3971 \\
 51 &  \nodata &  \nodata &  \nodata &  1.0434 $\pm$ 0.0027 &  \nodata &  \nodata &  \nodata &  \nodata &  1.0434 $\pm$ 0.0027 \\
 53 &  \nodata &  \nodata &  \nodata &  \nodata &  1.8535 $\pm$ 0.5374 &  \nodata &  \nodata &  \nodata &  1.8535 $\pm$ 0.5374 \\
 68 &  \nodata &  \nodata &  1.0348 $\pm$ 0.0047 &  \nodata &  1.0854 $\pm$ 0.0736 &  1.1121 $\pm$ 0.0883 &  1.0567 $\pm$ 0.0108 &  1.0532 $\pm$ 0.0118 &  1.0378 $\pm$ 0.0057 \\
 69 &  \nodata &  \nodata &  \nodata &  \nodata &  \nodata &  1.0000 $\pm$ 0.0000 &  \nodata &  \nodata &  1.0000 $\pm$ 0.0000 \\
 70 &  \nodata &  \nodata &  \nodata &  \nodata &  1.0093 $\pm$ 0.0076 &  1.0116 $\pm$ 0.0098 &  1.0055 $\pm$ 0.0005 &  1.0053 $\pm$ 0.0005 &  1.0054 $\pm$ 0.0005 \\
 72 &  \nodata &  \nodata &  \nodata &  \nodata &  \nodata &  1.0005 $\pm$ 0.0005 &  \nodata &  \nodata &  1.0005 $\pm$ 0.0005 \\
 75 &  \nodata &  \nodata &  \nodata &  \nodata &  1.0012 $\pm$ 0.0011 &  1.0008 $\pm$ 0.0006 &  1.0008 $\pm$ 0.0003 &  1.0009 $\pm$ 0.0003 &  1.0008 $\pm$ 0.0003 \\
 84 &  \nodata &  \nodata &  \nodata &  \nodata &  \nodata &  1.0000 $\pm$ 0.0000 &  \nodata &  \nodata &  1.0000 $\pm$ 0.0000 \\
 85 &  \nodata &  \nodata &  \nodata &  \nodata &  \nodata &  1.0001 $\pm$ 0.0001 &  \nodata &  \nodata &  1.0001 $\pm$ 0.0001 \\
 97 &  \nodata &  1.0056 $\pm$ 0.0028 &  \nodata &  \nodata &  1.0123 $\pm$ 0.0101 &  1.0093 $\pm$ 0.0077 &  1.0125 $\pm$ 0.0013 &  1.0125 $\pm$ 0.0013 &  1.0113 $\pm$ 0.0015 \\
 98 &  \nodata &  1.3811 $\pm$ 0.0511 &  1.2436 $\pm$ 0.0512 &  \nodata &  1.3233 $\pm$ 0.2461 &  1.3226 $\pm$ 0.2414 &  1.3359 $\pm$ 0.0468 &  1.3337 $\pm$ 0.0459 &  1.3174 $\pm$ 0.0493 \\
 99 &  \nodata &  1.0002 $\pm$ 0.0001 &  \nodata &  \nodata &  1.0041 $\pm$ 0.0041 &  1.0032 $\pm$ 0.0027 &  1.0039 $\pm$ 0.0021 &  1.0040 $\pm$ 0.0018 &  1.0002 $\pm$ 0.0001 \\
100 &  \nodata &  \nodata &  \nodata &  \nodata &  \nodata &  \nodata &  \nodata &  \nodata &  \nodata \\
102 &  \nodata &  \nodata &  \nodata &  \nodata &  1.1844 $\pm$ 0.1390 &  \nodata &  \nodata &  \nodata &  1.1844 $\pm$ 0.1390 \\
103 &  \nodata &  \nodata &  \nodata &  \nodata &  1.0003 $\pm$ 0.0005 &  1.0002 $\pm$ 0.0002 &  1.0001 $\pm$ 0.5044 &  1.0002 $\pm$ 0.0003 &  1.0002 $\pm$ 0.0002 \\
105 &  \nodata &  \nodata &  \nodata &  \nodata &  1.0007 $\pm$ 0.0008 &  1.0162 $\pm$ 0.0132 &  1.0004 $\pm$ 0.0003 &  1.0004 $\pm$ 0.0002 &  1.0004 $\pm$ 0.0002 \\
\enddata
\tablecomments{Column (1) lists the KOI number of the host star, columns 
(2)-(9) the radius correction factors calculated as shown in Equation \ref{Rcorr}, 
derived from ${\Delta}m$ measurements in different bands converted to 
$\Delta Kp$ values (see text for details), and column (10) the weighted
average of the correction factors from the previous columns. \\
The full table is available in a machine-readable form in the online
journal. A portion is shown here for guidance regarding content and form.}
\end{deluxetable*}

\begin{deluxetable*}{ccccccccccc}[!h]\scriptsize
\rotate
\tablewidth{1.0\linewidth}
\tablenum{10}
\tablecaption{Planet Radius Correction Factors Assuming Planets Orbit the Brightest Companion Stars, 
Derived from $\Delta m$ Measurements in Various Bands, and Weighted Average
\label{Rcorr2_tab}}
\tablehead{ 
\colhead{KOI} & \colhead{ID} & \colhead{F555W,F775W} & \colhead{i} &
\colhead{692} & \colhead{LP600} & \colhead{J} & \colhead{K} & 
\colhead{$J-K$(dwarf)} & \colhead{$J-K$(giant)} & \colhead{Weighted average}  \\
\colhead{(1)} & \colhead{(2)} & \colhead{(3)} & \colhead{(4)} & \colhead{(5)} & \colhead{(6)} & 
\colhead{(17)} & \colhead{(8)} & \colhead{(9)} & \colhead{(10)} & \colhead{(11)}}
\startdata
  1 & B &   \nodata &    3.305 $\pm$   0.982 &    3.333 $\pm$   0.864 &    \nodata &    2.027 $\pm$   0.911 &    1.682 $\pm$   0.756 &    2.532 $\pm$   0.751 &    2.386 $\pm$   0.656 &    2.926 $\pm$   0.819 \\
  2 & C &    \nodata &    \nodata &    \nodata &    \nodata &    \nodata &    \nodata &    \nodata &    2.206 $\pm$   0.560 &    2.206 $\pm$   0.560 \\
  4 & B &    \nodata &    3.622 $\pm$   0.910 &    \nodata &    \nodata &    2.834 $\pm$   1.262 &    \nodata &    \nodata &    \nodata &    3.353 $\pm$   1.030 \\
  5 & B &    \nodata &    \nodata &    \nodata &    \nodata &    \nodata &    1.496 $\pm$   0.742 &    \nodata &    \nodata &    1.496 $\pm$   0.742 \\
  6 & B &    \nodata &    \nodata &    \nodata &    \nodata &    6.880 $\pm$   3.094 &    \nodata &    \nodata &    \nodata &    6.880 $\pm$   3.094 \\
 10 & C &    \nodata &    \nodata &    \nodata &    \nodata &    \nodata &    \nodata &    \nodata &    \nodata &    \nodata \\
 12 & B &    \nodata &    \nodata &    \nodata &    \nodata &    \nodata &    2.288 $\pm$   1.020 &    \nodata &    \nodata &    2.288 $\pm$   1.020 \\
 13 & B &    \nodata &    1.487 $\pm$   0.375 &    1.574 $\pm$   0.425 &    \nodata &    1.487 $\pm$   0.721 &    1.484 $\pm$   0.724 &    1.487 $\pm$   0.378 &    1.484 $\pm$   0.377 &    1.511 $\pm$   0.390 \\
 14 & B &    \nodata &    \nodata &    \nodata &    \nodata &    2.467 $\pm$   1.114 &    1.938 $\pm$   0.876 &    4.129 $\pm$   1.084 &    3.688 $\pm$   0.969 &    3.909 $\pm$   1.028 \\
 18 & B &    \nodata &    \nodata &    \nodata &    \nodata &    \nodata &    \nodata &    \nodata &    \nodata &    \nodata \\
 21 & B &    \nodata &    \nodata &    \nodata &    \nodata &    1.734 $\pm$   0.775 &    \nodata &    \nodata &    \nodata &    1.734 $\pm$   0.775 \\
 28 & B &    \nodata &    \nodata &    \nodata &    \nodata &    \nodata &    \nodata &    \nodata &    \nodata &    \nodata \\
 41 & B &    \nodata &    3.604 $\pm$   0.917 &    \nodata &    \nodata &    \nodata &    \nodata &    \nodata &    \nodata &    3.604 $\pm$   0.917 \\
 42 & B &    \nodata &    \nodata &    \nodata &    2.162 $\pm$   0.566 &    2.811 $\pm$   1.252 &    4.267 $\pm$   1.900 &    1.860 $\pm$   0.472 &    1.905 $\pm$   0.484 &    1.999 $\pm$   0.515 \\
 43 & B &    \nodata &    \nodata &    \nodata &    \nodata &    1.551 $\pm$   0.709 &    \nodata &    \nodata &    \nodata &    1.551 $\pm$   0.709 \\
 44 & B &    \nodata &    \nodata &    \nodata &    \nodata &    3.109 $\pm$   1.385 &    3.355 $\pm$   1.494 &    2.464 $\pm$   0.626 &    2.451 $\pm$   0.623 &    2.457 $\pm$   0.624 \\
 45 & C &    \nodata &    \nodata &    \nodata &    \nodata &    1.887 $\pm$   1.111 &    \nodata &    \nodata &    \nodata &    1.887 $\pm$   1.111 \\
 51 & B &    \nodata &    \nodata &    \nodata &    2.271 $\pm$   0.572 &    \nodata &    \nodata &    \nodata &    \nodata &    2.271 $\pm$   0.572 \\
 53 & C &    \nodata &    \nodata &    \nodata &    \nodata &    1.779 $\pm$   0.945 &    \nodata &    \nodata &    \nodata &    1.779 $\pm$   0.945 \\
 68 & B &    \nodata &    \nodata &    2.030 $\pm$   0.526 &    \nodata &    1.591 $\pm$   0.719 &    1.489 $\pm$   0.673 &    1.640 $\pm$   0.428 &    1.695 $\pm$   0.445 &    1.815 $\pm$   0.473 \\
 69 & B &    \nodata &    \nodata &    \nodata &    \nodata &    \nodata &    \nodata &    \nodata &    \nodata &    \nodata \\
 70 & B &    \nodata &    \nodata &    \nodata &    \nodata &    3.522 $\pm$   1.569 &    3.328 $\pm$   1.483 &    2.778 $\pm$   0.704 &    2.805 $\pm$   0.711 &    2.791 $\pm$   0.707 \\
 72 & B &    \nodata &    \nodata &    \nodata &    \nodata &    \nodata &    7.718 $\pm$   3.444 &    \nodata &    \nodata &    7.718 $\pm$   3.444 \\
 75 & B &    \nodata &    \nodata &    \nodata &    \nodata &    5.705 $\pm$   2.556 &    6.789 $\pm$   3.024 &    3.664 $\pm$   1.084 &    3.467 $\pm$   0.951 &    3.566 $\pm$   1.020 \\
 84 & B &    \nodata &    \nodata &    \nodata &    \nodata &    \nodata &    \nodata &    \nodata &    \nodata &    \nodata \\
 85 & B &    \nodata &    \nodata &    \nodata &    \nodata &    \nodata &    \nodata &    \nodata &    \nodata &    \nodata \\
 97 & B &    \nodata &    3.816 $\pm$   1.357 &    \nodata &    \nodata &    2.981 $\pm$   1.328 &    3.302 $\pm$   1.471 &    2.043 $\pm$   0.518 &    2.043 $\pm$   0.518 &    2.268 $\pm$   0.625 \\
 98 & B &    \nodata &    1.408 $\pm$   0.372 &    1.424 $\pm$   0.397 &    \nodata &    1.393 $\pm$   0.673 &    1.387 $\pm$   0.668 &    1.331 $\pm$   0.344 &    1.332 $\pm$   0.344 &    1.383 $\pm$   0.369 \\
 99 & B &    \nodata &    \nodata &    \nodata &    \nodata &    4.348 $\pm$   1.963 &    4.748 $\pm$   2.114 &    2.722 $\pm$   0.927 &    2.693 $\pm$   0.827 &    2.708 $\pm$   0.878 \\
100 & B &    \nodata &    \nodata &    \nodata &    \nodata &    \nodata &    \nodata &    \nodata &    \nodata &    \nodata \\
102 & B &    \nodata &    \nodata &    \nodata &    \nodata &    1.482 $\pm$   0.682 &    \nodata &    \nodata &    \nodata &    1.482 $\pm$   0.682 \\
103 & B &    \nodata &    \nodata &    \nodata &    \nodata &    \nodata &    \nodata &    \nodata &    \nodata &    \nodata \\
105 & B &    \nodata &    \nodata &    \nodata &    \nodata &    \nodata &    3.053 $\pm$   1.360 &    \nodata &    \nodata &    3.053 $\pm$   1.360 \\
\enddata
\tablecomments{Column (1) lists the KOI number of the host star, column (2)
the identifier of the companion star (see Table \ref{photometry_tab}) assumed 
to host the planet(s), columns (3)-(10) the radius correction factors calculated 
as shown in Equation \ref{Rcorr2}, derived from ${\Delta}m$ measurements in 
different bands converted to $\Delta Kp$ values (see text for details), and 
column (10) the weighted average of the correction factors from the previous 
columns. \\
The full table is available in a machine-readable form in the online
journal. A portion is shown here for guidance regarding content and form.}
\end{deluxetable*}

When assuming planets orbit the primary stars (Table \ref{Rcorr1_tab}),
we find overall satisfactory agreement between the different radius correction 
factors for each KOI host star, especially considering the approximations 
involved in converting magnitude differences into the {\it Kepler} bandpass. 
The mean correction factors derived from the HST, $i$, $LP600$, and 692 nm
bands are very similar, 1.072, 1.082, 1.102, and 1.114, respectively. 
Using just the $J$- and $K$-band magnitude differences individually, the mean 
correction factors derived from them are 1.053 and 1.101, respectively. 
When the $J-K$ colors are considered, assuming dwarf stars the average factor 
is 1.131, while it is 1.100 if giant stars are assumed. The somewhat larger 
correction factors for the $LP600$, 692 nm, and $K$-bands are related 
to the larger number of bright ($\Delta m <  2$), close ($\leq$ 1\arcsec) companions 
detected in these bands (see Figure \ref{KOI_mag_histo}). Also, the correction
factor for the $J$-band is somewhat lower since most companions in the 
$J$-band were detected by UKIRT, which mostly found more distant, fainter
companions.
As can be seen from Equation \ref{Rcorr}, an equal-brightness companion 
in a binary system results in a correction factor of 1.41, while the factor 
decreases to, e.g., 1.18, 1.08, and 1.03 for $\Delta m$ values of 1, 2, 
and 3 mag, respectively.

Under the assumption that planets orbit the brightest companion star
(Table \ref{Rcorr2_tab}), the radius correction factors become larger. 
In this case, the mean correction factors derived from the HST, 
$i$, $LP600$, and 692 nm bands are 4.18, 2.75, 2.23, and 2.12, 
respectively. The average factors derived from the $J$- and $K$-band 
individually, and from the $J-K$-color relationship for dwarf and
giant stars, are 3.21, 3.44, 2.31, and 2.25, respectively. 
Larger factors are due to fainter companions.

\begin{figure}[!]
\centering
\includegraphics[angle=90,scale=0.4]{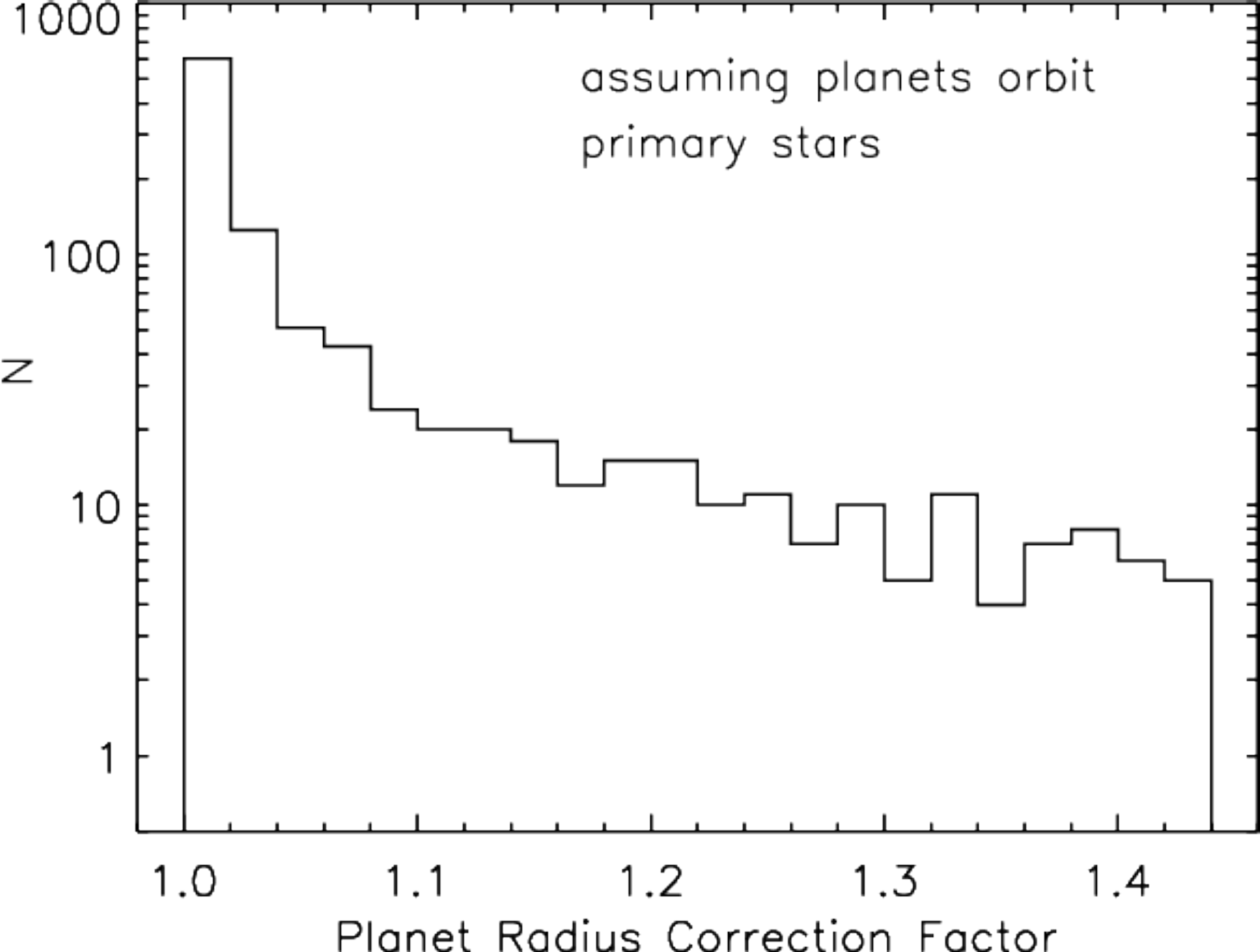}
\caption{Histogram of the average planet radius correction factors derived
from the measurements of detected companions to host stars of KOI
planets in different bands, assuming that planets orbit the primary stars
(one correction factor per star). There are six additional values between 
1.54 and 1.86 that are not shown.
\label{KOI_Prad_correction1}}
\end{figure}

\begin{figure}[!]
\centering
\includegraphics[angle=90,scale=0.4]{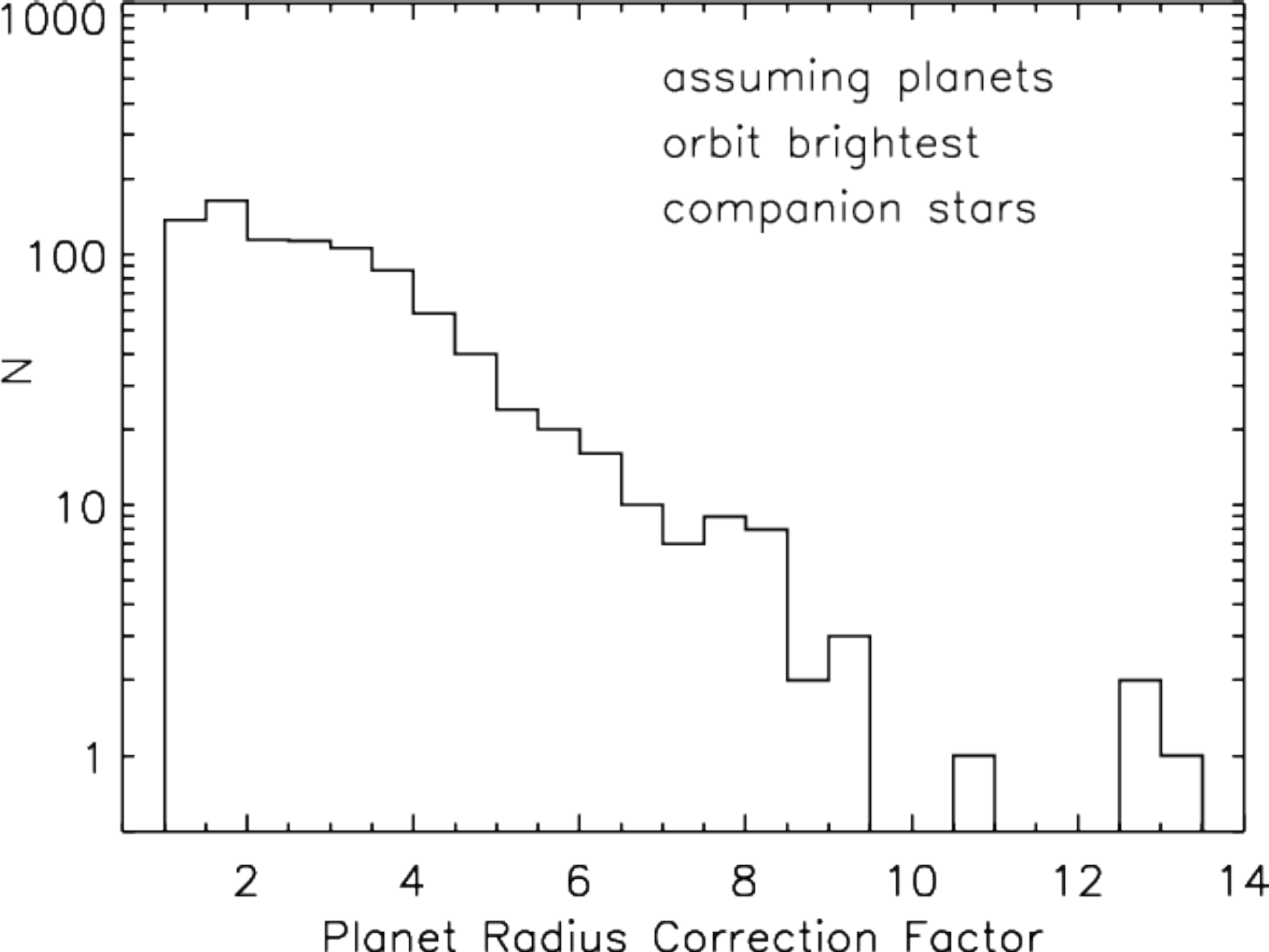}
\caption{Histogram of the average planet radius correction factors derived
from the measurements of detected companions to host stars of KOI
planets in different bands, assuming that planets orbit the brightest
companion stars (one correction factor per star). One additional value at
16.4 is not shown.
\label{KOI_Prad_correction2}}
\end{figure}

In Figures \ref{KOI_Prad_correction1} and \ref{KOI_Prad_correction2},
we show the histograms of average planet radius correction factors for 
planets transiting KOI host stars that have at least one companion within 
4\arcsec\ detected in imaging observations (both high-resolution and 
seeing-limited). The correction factors in Figure \ref{KOI_Prad_correction1} 
assume planets orbit the primary stars, while the factors in Figure 
\ref{KOI_Prad_correction2} assume planets orbit the brightest companion
stars. For all planets orbiting a certain KOI host star (primary or companion), 
the radius correction factor is the same, since it just depends on stellar 
properties (brightness ratios between primary and companion stars, and, 
for planets orbiting companion stars, ratios of stellar radii). Therefore, there 
is one planet radius correction factor per KOI host star. In Figures 
\ref{KOI_Prad_correction1} and \ref{KOI_Prad_correction2} we only include 
{\it Kepler} stars with companions that are hosts to planets, thus excluding 
those stars with only false positive events; this leaves 1036 stars in Fig.\ 
\ref{KOI_Prad_correction1} (out of the 1891 stars for which we derived 
$\Delta Kp$ values), and 922 stars in Fig.\ \ref{KOI_Prad_correction2} 
(out of the 1642 stars for which measured transit depths are still 
consistent with the brightness of the companion).

Assuming planets orbit the primary star, the mean and median planet radius
correction factors for planet host stars are 1.06 and 1.01, respectively. 
There is a monotonic decrease in the frequency of correction factor values as 
the value increases. When assuming planets orbit the brightest companion
star, the mean and median planet radius correction factors are 3.09 and
2.69, respectively; 90\% of values are between 1.12 and 5.22, and just 
18 stars have correction factors between 8.0 and the largest value,
16.4.

\begin{figure}[!]
\centering
\includegraphics[angle=90,scale=0.4]{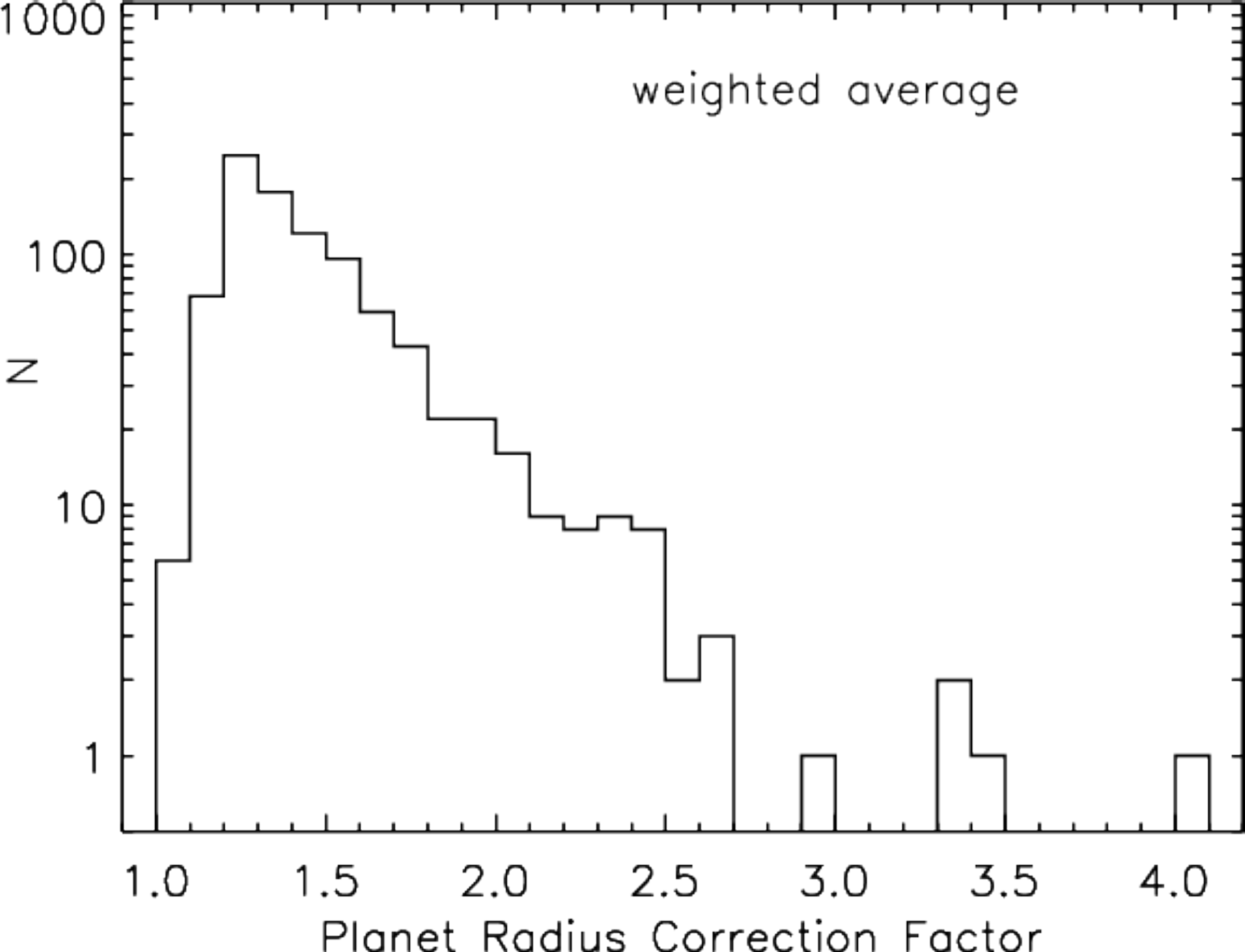}
\caption{Histogram of the weighted average of planet radius correction factors 
shown in Figures \ref{KOI_Prad_correction1} and \ref{KOI_Prad_correction2}
(see text for details).
\label{KOI_Prad_correction3}}
\end{figure}

If we calculate a weighted average of the planet radius correction factors,
using weights of, e.g., 0.8 and 0.2 for the factors that assume planets orbit 
the primary stars and brightest companion stars, respectively, we derive a
mean value of 1.47 and and a median value of 1.38. These weights
exemplify the assumption that planets are more likely to orbit primary rather 
than secondary stars, since the former typically have more massive 
protoplanetary disks \citep[e.g.,][]{akeson14} which may lead to more 
efficient planet formation. Assuming that planets are equally likely to 
orbit the primary and secondary star, the mean and median correction factors 
increase somewhat to 2.08 and 1.85, respectively.
The histogram of our weighted average correction factors is shown in Figure 
\ref{KOI_Prad_correction3}; three quarters of values are below 1.58. 
Therefore, in a more realistic scenario where most planets orbit the primary stars, 
but some orbit secondary stars, the average planet radius correction factors are 
typically $\lesssim$ 2. Our mean weighted average radius correction
factor is very similar to the result from \citet{ciardi15}, who modeled multiple
stellar systems and derived that, on average, planet radii are underestimated 
by a factor of 1.5. However, we note that these average or median correction 
factors should not be applied to individual planets in multiple stellar systems 
to correct their radii; each system is unique, and the increase in radius is either 
relatively small, of the order of several percent, if a planet orbits the primary star 
(which is expected for many, if not most, planets), or of the order of a few if it
is determined that the planet orbits a fainter companion star.

\section{Discussion}
\label{disc}

\subsection{Effect of Companions on Planet Radii}

\begin{figure}[!t]
\centering
\includegraphics[angle=90, scale=0.39]{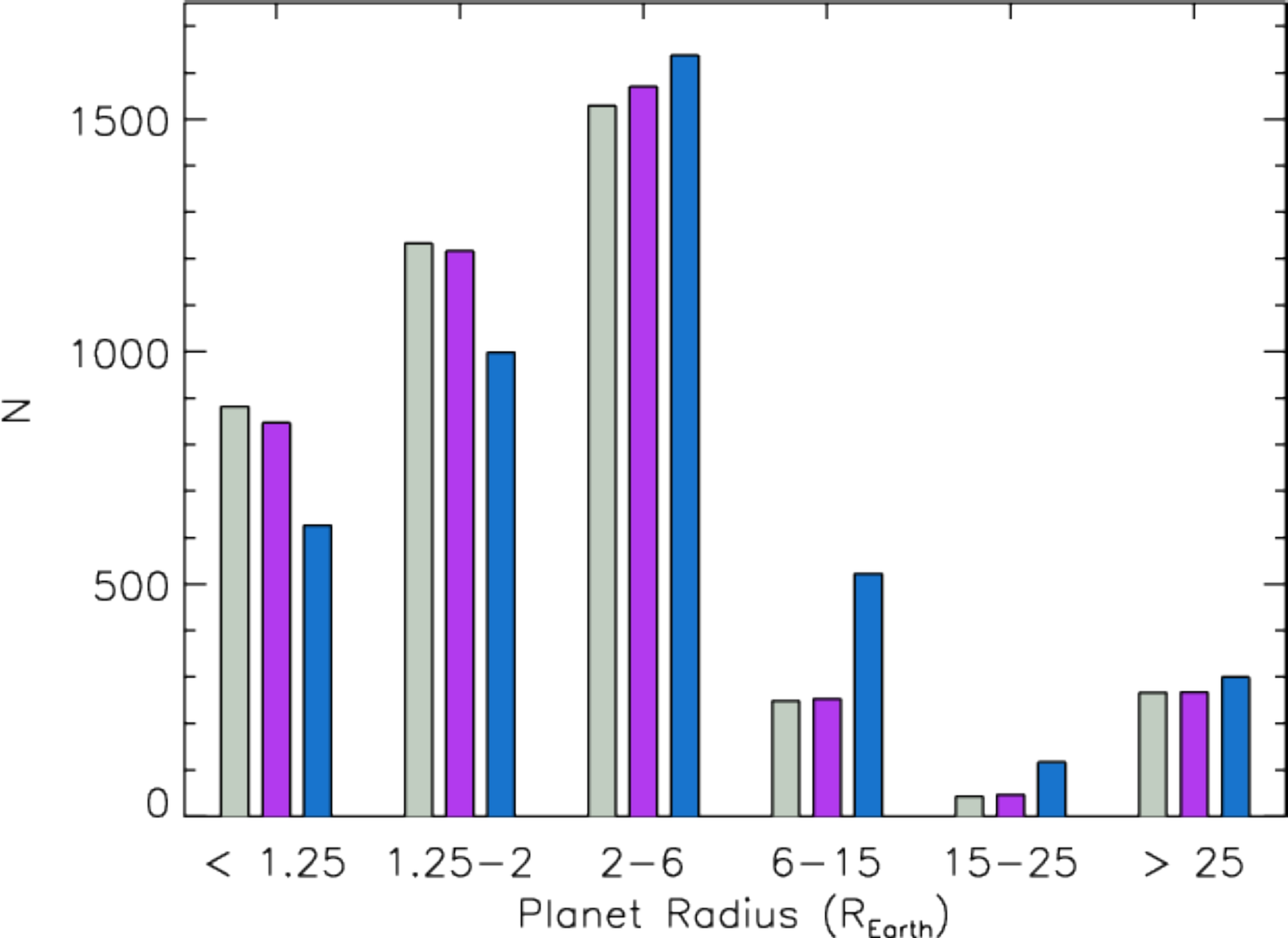}
\caption{Histograms of planet radii of all KOI planet candidates and confirmed 
planets from the latest KOI cumulative table targeted by high-resolution
imaging ({\it gray}) and the same planet radii, but corrected using the average 
radius correction factors for each star derived in section \ref{revised_radii} 
assuming planets orbit the primary stars ({\it purple}) and assuming planets 
orbit the brightest companion stars ({\it blue}).
\label{KOIs_corr_Rp_histo}}
\end{figure}

We applied the average radius correction factors for each star derived in the 
previous section to the radii of all those KOI planets whose host stars were 
targeted by high-resolution imaging and had a companion detected in the 
$F555W$, $F775W$, $i$, $LP600$, 692 nm, $J$, or $K$ bands (for stars
without detected companions, the correction factor is 1). In Figure 
\ref{KOIs_corr_Rp_histo} we show the distributions of planet radii before
and after applying correction factors; the planet radii are binned in ranges
often used to group different classes of planets (e.g., super-Earths have
$\sim$1.25-2 \RE, Neptunes have $\sim$ 2-6 \RE). 
When assuming that planets orbit the primary stars (purple bars in Fig.\
\ref{KOIs_corr_Rp_histo}), there are only slight changes in the various 
histogram bins; the number of planets with radii from 0.5 to 2 \RE\ 
decreases somewhat, while there are more planets in the 2-6 \RE\ range. 
Under the assumption that planets orbit the brightest companion stars
(blue bars in Fig.\ \ref{KOIs_corr_Rp_histo}), the number of planets with 
radii from 0.5 to 2 \RE\ decreases noticeably, while the number of planets 
with radii larger than 2 \RE\ increases. Table \ref{Prad_corrections} lists 
the number of planets in certain radius bins before and after the correction 
factors were applied to the planet radii. 

\begin{deluxetable}{ccccc}
\tablewidth{1.0\linewidth}
\tablenum{11}
%\tabletypesize{\scriptsize}     %{\tiny}   %{\footnotesize}
\tablecaption{Number of KOI Planets in Different Radius Bins, Orbiting Stars Targeted
by High-resolution Imaging
\label{Prad_corrections}}
\tablehead{ 
\colhead{$R_p$ range (\RE)} & \colhead{N$_{\rm all}$} & \colhead{N$_{\rm all,obs.}$} 
& \colhead{N$_{\rm obs.,corr.}^{\rm prim}$} & \colhead{N$_{\rm obs.,corr.}^{\rm sec}$} \\
\colhead{(1)} & \colhead{(2)} & \colhead{(3)} & \colhead{(4)} & \colhead{(5)}}
\startdata
   0.25 -- 1.25  & 983  &  882 & 847 & 627 \\
   1.25 -- 2.0  &  1337 &  1233 & 1217 & 999  \\
   2.0 -- 6.0  &  1622  &  1529 & 1571 & 1638 \\
   6.0 -- 15  &  297 & 249 & 253 & 522 \\
   15 -- 25  &  73 &  43 & 47 & 117 \\
   $>$ 25\tablenotemark{a} &  394 & 267 & 268 & 300 \\
\enddata
\tablecomments{
Column (1) lists the range for the planet radius, column (2) the number 
of KOI planets from the latest KOI cumulative table with radii within that 
radius range, columns (3), (4), and (5) the number of KOI planets from 
column (2) that were targeted by high-resolution imaging, split into the 
same radius ranges, but for the planets in column (3) no correction was 
applied to the radius, while for the planets in columns (4) and (5) the 
radii have been corrected with the average radius correction factors 
for each star assuming the planets orbit the primary or the brightest 
companion star, respectively (see text for details).
\tablenotetext{a}{These very large ``planets'' are mostly planet candidates;
some will likely not be confirmed, and others will likely be confirmed as
planets with much smaller radii.}}
\end{deluxetable}

\begin{figure}[!t]
\centering
\includegraphics[angle=90, scale=0.39]{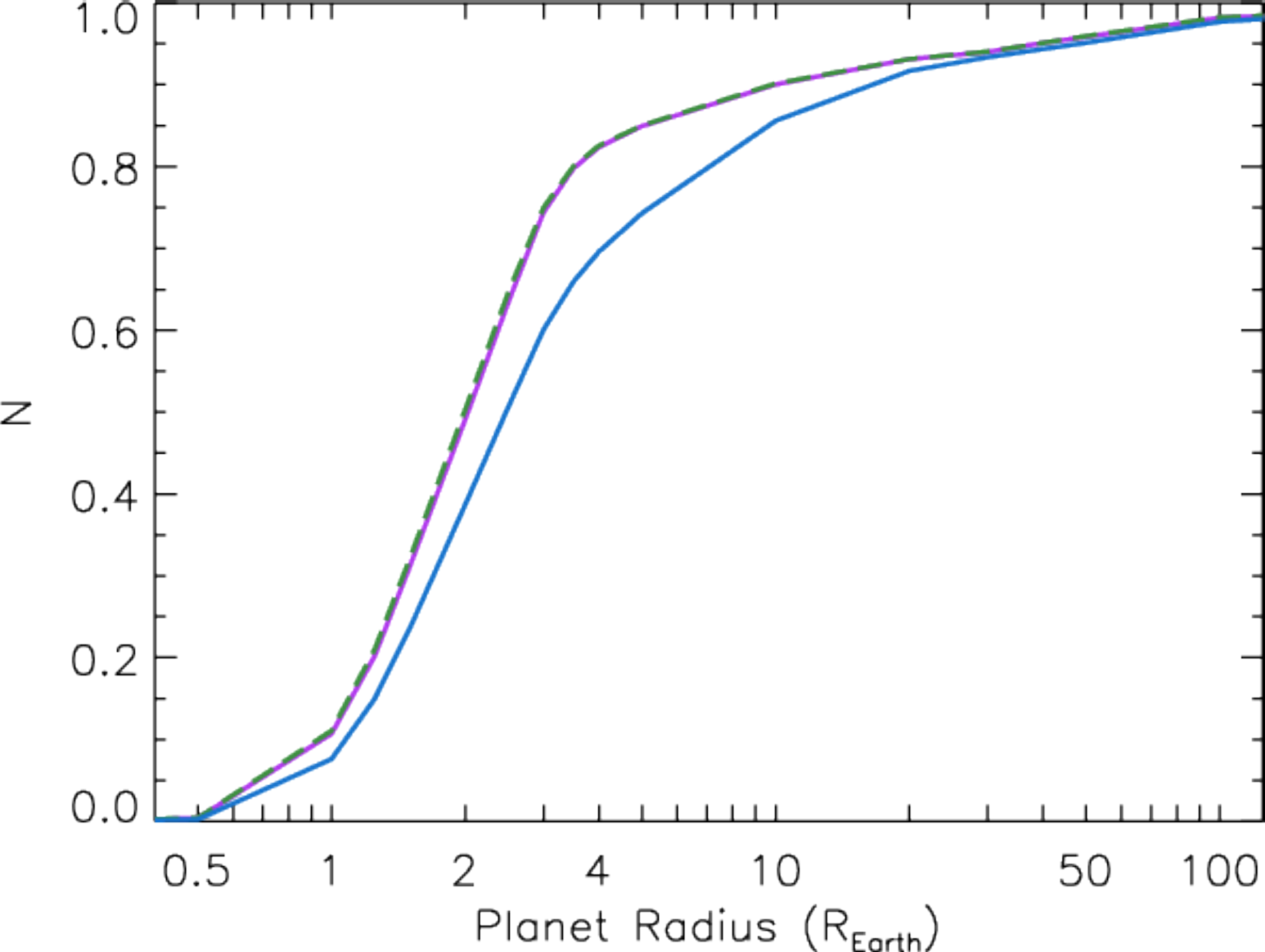}
\caption{Cumulative distribution of planet radii of all KOI planet candidates and 
confirmed planets from the latest KOI cumulative table targeted by high-resolution
imaging ({\it green, dashed line}) and the same planet radii, but corrected using 
the average radius correction factors for each star derived in section \ref{revised_radii} 
assuming planets orbit the primary stars ({\it purple}) and assuming planets orbit 
the brightest companion stars ({\it blue}).
\label{KOIs_corr_Rp_cumul}}
\end{figure}

In Figure \ref{KOIs_corr_Rp_cumul} we show the effect of correcting
planet radii due to the transit dilution by a companion star as 
cumulative distributions of planet radii, before and after the correction
was applied. As in the previous figure (Fig.\ \ref{KOIs_corr_Rp_histo})
there is hardly any effect on the cumulative distribution if we assume 
that planets orbit the primary stars, but the distribution of planet radii 
clearly shifts to larger radii if planets are assumed to orbit the brightest 
companion stars.

Since we do not know if planets orbit the primary star or one of the
companion stars, the true effect of transit dilution by companions lies
in between the two extreme cases shown in Figures \ref{KOIs_corr_Rp_histo}
and \ref{KOIs_corr_Rp_cumul}. Overall, we can state that, without accounting 
for the presence of companions, the number of small planets ($R_p \lesssim$ 
4 \RE) will be overestimated, while the number of planets with radii larger
than 4 \RE\ range will be underestimated. 
Moreover, the search sensitivity to small planets will be lower if target stars are
actually part of multiple systems, since transits of small planets will be diluted
by the light from the other stars in the system and thus more difficult to detect.
On the other hand, after applying radius correction factors, the decrease 
in the number of planets with radii up to 25\% of the Earth's radius ranges 
between 4 and 29\% (and those with radii up to two times the Earth's radius 
between 2 and 23\%; see Table \ref{Prad_corrections}), depending on whether
planets are assumed to orbit the primary or brightest secondary star. Thus,
the occurrence rate of Earth-sized and smaller (all presumably rocky) planets 
would have to be revised, but it would decrease by no more than $\sim$~25\%.

\begin{deluxetable*}{llcccc}[!t]
\tablewidth{1.0\linewidth}
\tablenum{12}
%\tabletypesize{\scriptsize}     %{\tiny}   %{\footnotesize}
\tablecaption{Observed Companion Star Fractions from the Literature
and This Work
\label{comp_fractions}}
\tablehead{ 
\colhead{Reference} & \colhead{Obs. Technique} & \colhead{f($<1\arcsec$)} 
& \colhead{f($<2\arcsec$)} & \colhead{f($<3\arcsec$)} & \colhead{f($<4\arcsec$)} \\
\colhead{(1)} & \colhead{(2)} & \colhead{(3)} & \colhead{(4)} & \colhead{(5)} &
\colhead{(6)}}
\startdata
\citet{adams12} & AO & \nodata & 20$\pm$5\% & \nodata & \nodata \\
\citet{adams13} & AO & \nodata & 17$\pm$12\% & \nodata & 33$\pm$17\% \\
\citet{dressing14} & AO & \nodata & 14$\pm$4\% & \nodata & 31$\pm$6\% \\
\citet{wang14} & AO & \nodata & 5$\pm$3\% & 12$\pm$5\% & 25\tablenotemark{a}$\pm$7\% \\
\citet{law14} & Robo-AO & \nodata & \nodata & 7.4\tablenotemark{b}$\pm$1\% & \nodata \\
\citet{baranec16} & Robo-AO & \nodata & \nodata & 10.6\tablenotemark{b}$\pm$1.1\% & 17.6$\pm$1.5\% \\
\citet{ziegler16} & Robo-AO & \nodata & \nodata & \nodata & 12.6$\pm$0.9\% \\
this work & AO & 10$\pm$1\% & 18$\pm$2\% & 26$\pm$2\% & 31$\pm$2\% \\
\citet{howell11} & speckle & \nodata & 6$\pm$2\% & \nodata & \nodata \\
\citet{horch14} & speckle, WIYN & 7$\pm$1\% & \nodata & \nodata & \nodata \\
\citet{horch14} & speckle, Gemini N & 23$\pm$8\% & \nodata & \nodata & \nodata \\
this work & speckle & 8$\pm$1\% & 10\tablenotemark{c}$\pm$1\% & \nodata & \nodata \\
\citet{lillo-box12,lillo-box14} & lucky imaging & \nodata & \nodata & 17$\pm$3\% & \nodata \\
\citet{gilliland15,cartier15} & HST imaging & \nodata & \nodata & 26\tablenotemark{b}$\pm$11\% & \nodata \\
this work & UKIRT imaging & \nodata & \nodata & \nodata & 19.1\tablenotemark{d}$\pm$0.5\% \\
\enddata
\tablecomments{
Column (1) lists the reference, column (2) the observing technique, 
and columns (3) to (6) the fractions of stars with companions within 
1\arcsec, 2\arcsec, 3\arcsec, and 4\arcsec, respectively, from the 
primary star.
\tablenotetext{a}{This value is for a companion star fraction within 5\arcsec\
of the primary star.}
\tablenotetext{b}{This value is for a companion star fraction within 2.5\arcsec\ 
of the primary star}
\tablenotetext{c}{This number is likely slightly underestimated, since in several
cases the field of view of the speckle images did not extend out to 2\arcsec,
but slightly below this limit.}
\tablenotetext{d}{This value is for the companion star fraction in UKIRT images
for projected separations between $\sim$ 1\arcsec\ and 4\arcsec\ from the 
primary star.}}
\end{deluxetable*}

As mentioned before, relevant to this discussion is the question whether 
companions are bound. In this work we refer to nearby stars as 
``companions'', even though they may not form a bound system with the 
primary star. An unbound object will still dilute the transit depth, but the 
planet radius correction factor depends on which star the planet orbits.
Determining the fraction of gravitationally bound companions is challenging,
but different methods can be used to derive it. The most direct way to determine
whether a companion is bound is to obtain observations at multiple epochs to 
detect common proper motion (or even orbital motion); while this is not
feasible to carry out for all detected companions to KOI host stars, the {\it Gaia}
mission \citep{prusti16} will be able to determine whether many of these 
systems are bound based on their proper motion or parallax. 
\citet{horch14} used simulations of star counts in the {\it Kepler} field, adding
companions following the known distributions of binaries in the solar
neighborhood \citep{duquennoy91, raghavan10}, and compared them to
the detected companions and sensitivity limits of their DSSI observations 
at WIYN and Gemini North. They concluded that companions within 
1\arcsec\ are likely to be bound.

If a star with companions has been observed in more than one filter,
the color of the stars, combined with isochrone fits, can also be used to 
estimate the probability that the companions are bound to their primaries 
\citep[e.g.,][] {lillo-box12, everett15, teske15}. If a common isochrone exists for the 
primary and its companion(s), it is possible that they form a bound system. 
\citet{hirsch16} use the sample of stars presented in this work that have 
observations in at least two different bands to perform isochrone fits 
and thus derive probabilities that systems are bound. We refer to their 
work for the identification of bound systems among the KOI planet host 
stars and their implications.

\subsection{Fraction of KOI Host Stars with Companions}

The fraction of KOI host stars with companions depends on the technique 
used to obtain observations and detect companions, as well as the range 
in projected separations chosen to measure this fraction. Also, differences 
between various studies are expected due to different selection criteria for 
the sample of KOI stars targeted by observations; for example, one study 
might have favored brighter stars or stars with a certain type of planets.
In Table \ref{comp_fractions} we list the observed companion star fractions 
from this work and from the literature. No corrections were applied due to 
sensitivity and completeness limits.
Overall, the AO and speckle results presented in this work agree broadly 
with previous studies, which typically used smaller samples to derive the 
fraction of KOI host stars with companions. The observed fraction of KOI 
host stars with companions is about 10\% for companions within 1\arcsec\ 
and increases to about 30\% for companions at separations up to 4\arcsec\ 
from the primary. 
 
The most complete sample of KOI host stars targeted by high-resolution
imaging is the one from Robo-AO \citep{law14, baranec16, ziegler16}; they 
observed 3320 unique stars. The observed fraction of stars with companions 
varies somewhat between the three studies; when we combine their 
observations, we derive observed companion star fractions of 8.8$\pm$0.5\% 
and 12.9$\pm$0.6\% for companions within 2.5\arcsec\ and 4\arcsec,
respectively. Compared to AO observations, these fractions are about
a factor of two smaller. One likely explanation is the lower sensitivity to
companions in the Robo-AO data; at projected separations between 
about 1\arcsec\ and 2\arcsec\ from the primary star, Robo-AO can detect 
companions at the 5-$\sigma$ level if their $\Delta m$ values are smaller 
than $\sim$ 3.5, 5, and 6.5 for low, medium, and high performance, 
respectively \citep{law14, baranec16, ziegler16}, while, e.g., our AO 
data can detect companions down to $\Delta m$ $\sim$ 8. AO imaging is 
also more sensitive within 0.5\arcsec\ from the primary star.

\begin{deluxetable*}{lccc}
\tablewidth{1.0\linewidth}
\tablenum{13}
\tablecaption{KOI Host Stars with Companions at 1\arcsec--4\arcsec\ in UKIRT Data
\label{UKIRT_tab}}
\tablehead{ 
\colhead{KOI Host Stars} & \colhead{N} & \colhead{N$_{\mathrm comp}$} & \colhead{f} \\
\colhead{(1)} & \colhead{(2)} & \colhead{(3)} & \colhead{(4)}}
\startdata
All & 7557 & 1446 & 19.1\% $\pm$ 0.5\% \\
Planets (confirmed or candidate) & 3665 & 645 & 17.6\% $\pm$ 0.7\% \\
False Positives (and no planets) & 3892 & 801 & 20.6\% $\pm$ 0.7\% \\
False Positives (and possibly also planets) & 4014 & 822 & 20.5\% $\pm$ 0.7\% \\
\enddata
\tablecomments{
Column (1) lists the KOI type (all, planets, false positives), column (2) the number of 
KOI host stars, column (3) the number of KOI stars with at least one companion at 
$\leq$ 4\arcsec\ detected in UKIRT data, and column (4) the observed fraction of 
multiple systems.}
\end{deluxetable*}

While the UKIRT survey did not obtain high-resolution imaging observations,
it yields uniform information on companions for all KOIs, albeit not at very close
projected separations.  
Of the 7557 KOI host stars in the latest KOI cumulative table, 1446 (or
19\%) have one or more companions detected in the UKIRT $J$-band 
images (at separations between $\sim$ 1\arcsec\ and 4\arcsec). 
When considering only planet host stars, the observed fraction of stars
with companions is $\sim$~17\%, while it is $\sim$~20\% for stars with
transit events classified only as false positives (see Table \ref{UKIRT_tab}).
The latter fraction remains unchanged if we consider host stars that have 
at least one false positive signal, but may also have planets. It is clear that 
the presence of companions contributes to false positive transit signals; 
in fact, the {\it Kepler} pipeline flags some false positives as having a centroid 
offset, i.e., the centroid of the image during the transit and outside of the transit 
is offset, indicating that the transits occur on a nearby star. Nonetheless, the
sample of KOI host stars with companions detected within 4\arcsec\ in UKIRT
images is not dominated by false positives; the observed fraction of 17\% for
planet host stars is somewhat larger than that obtained by Robo-AO, and, as
expected, lower than that from AO imaging.

Given that some companions are missed due to their large brightness 
difference or separation from the primary (too close or too far), it is expected 
that the observed companion star fraction as detailed in this paper is lower 
than the true companion star fraction. For solar-type stars in the solar 
neighborhood (at $<$~25 pc), the fraction of stars with bound companions 
out to projected separations of $\sim$ 10,000 AU is measured to be
about 44\%; the peak in the distribution of companion separations is at about 
50 AU, with just 11.5\% of stars having companions at $>$~1000 AU 
\citep{raghavan10}. Furthermore, the mass-ratio distribution for multiple systems 
is roughly flat between $\sim$ 0.2 and 0.95, with few low-mass companions and 
a large number of close-to-equal-mass pairs of stars \citep{raghavan10}. 
At the average distance of about 830 pc to the {\it Kepler} stars \citep{huber14}, 
a separation of 2\arcsec\ corresponds to 1660 AU (and 4\arcsec\ to 3320 AU); 
thus, among the {\it Kepler} stars surveyed, it is likely that some fainter companions 
within 4\arcsec\ and brighter companions very close to the primary 
($\lesssim$~0.1\arcsec) were not detected.

Simulations of stars in the {\it Kepler} field, including multiple systems,
can be used to assess the true companion star fractions. For example,
\citet{horch14} reproduced their observed companion star fractions 
from speckle observations with companion star fractions from simulations
where a companion star fraction of 40\%-50\% was adopted. They 
concluded that the binary fraction of KOI planet host stars is consistent 
with that of field stars. Other work \citep{wang15a} found that KOI
host stars of giant planet candidates have no companions for stellar
separations smaller than 20 AU; at larger separations, the multiplicity
rate increases to the value expected from field stars. Also, stars with 
multiple transiting planets seem to have a lower multiplicity rate in 
the 1-100 AU stellar separation range \citep{wang15b}. Thus, when
deriving true companion star fractions from observed ones, it is 
important to consider selection effects regarding planet properties.
Such an analysis is beyond the scope of this paper, but the data
presented here offer an ideal starting point for future work.

\section{Summary and Conclusions}
\label{conclude}

In this work we have summarized results from seven years of follow-up 
imaging observations of KOI host stars, including work done by teams 
from the {\it Kepler} Follow-Up Observation Program and by other groups. 
Overall, 3557 stars that host KOIs, mostly planet candidates, were targeted 
with high-resolution imaging from the optical to the near-IR. Of the stars that 
host at least one KOI planet candidate or confirmed planet, 87\% have been 
covered by high-resolution imaging. In addition to these observations, the 
{\it Kepler} field has been surveyed with the UKIRT telescope in the $J$-band 
and in $UBV$ filters with the WIYN 0.9-m telescope.

We have presented in detail the results from our adaptive optics 
imaging at the Keck II, Palomar 5-m, and Lick 3-m telescopes, and 
from speckle imaging at the Gemini North, WIYN, and DCT telescopes. 
In the larger field of view of the AO images, we find that 31$\pm$2\% 
of KOI host stars are observed to have at least one companion within 
4\arcsec; within 1\arcsec, the observed companion fraction decreases 
to 10$\pm$1\%. The observed fraction of stars with companions at 
$<$ 1\arcsec\ in the speckle images is 8$\pm$1\%, very similar to the 
AO result.
 
We have combined results from our adaptive optics and speckle 
images with those published in the literature and on CFOP to create 
a catalog of companion stars to KOI host stars that lie within 4\arcsec. 
Our list contains 2297 companions around 1903 primary stars.
From high-resolution imaging, we find that the observed fraction of KOI host
stars with companions within 1\arcsec\ is $\sim$ 10\% (increasing to $\sim$
30\% for companions within 4\arcsec); using the complete sample
observed by UKIRT, this observed fraction amounts to 19\%, but for the 
separation range of 1\arcsec-4\arcsec. Given the sensitivity and completeness
limits of the observations, the actual fraction of {\it Kepler} stars with 
companions is higher than the observed one; in particular, companions that 
are faint or very close (in projected separation) to the primary ($\lesssim$ 
0.1\arcsec, $\Delta m \gtrsim$ 6-8) are not detected.

We have converted the measured magnitude differences between primary 
and companion stars to magnitude differences in the {\it Kepler} bandpass, 
and then we have calculated radius correction factors for the planet radii due
to transit depth dilution by the companion star. We have calculated such factors
first by assuming that planets orbit their primary stars, then by assuming 
planets orbit the brightest companion star (the latter also requires an 
estimate of the ratio of the stellar radii). Even though it is likely that only 
a fraction of the detected companions is actually bound to the primary, 
even a background source would contaminate the light curve as measured 
by {\it Kepler} and thus would have to be taken into account when deriving 
planet radii. We find mean and median correction factors of 1.06 and 1.01, 
respectively, for radii of KOI planets assuming that planets orbit the primary 
stars. Under the assumption that planets orbit the brightest companion star, 
we find mean and median radius correction factors of 3.09 and 2.69, 
respectively. In reality, the average planet radius correction factor lies in between 
these values, likely closer to the case when planets are assumed to orbit the
primary stars, so it can be expected to be of the order of $\sim$ 1.5-2.0.
We caution that these average correction factors do not apply to any
specific planet; each planet in a multiple stellar system has its own radius 
correction factor depending on the stars in the system and whether the planet
orbits the primary or a companion star.

When we apply the average planet radius correction factors for each star
to the radii of {\it Kepler} confirmed and candidate planets, we find that, by 
accounting for the dilution of the transit depths by companions, the number of 
small planets ($R_p \lesssim$ 2 \RE) decreases (by $\sim$ 2-23\%), 
while the number of planets with $R_p >$ 6 \RE\ increases (by up to 68\%). 
The exact numbers depend on the actual number of companions (some 
of which are missed by observations), on whether planets orbit primary or 
companion stars, and, related to this, whether companions are bound. 
We note that the decrease in the number of small planets due the 
effects of a companion star is noticeable, but it will not result in a large 
revision of the occurrence rate of small (and thus likely rocky planets); 
this rate would be lowered by at most  $\sim$ 25\%.

Work by \citet{hirsch16} that builds on the sample of companion stars 
presented in this paper addresses the question of whether companions 
are actually bound and thus gives more comprehensive and accurate 
estimates for corrected planet radii. 
Furthermore, spectroscopic follow-up of KOI host stars has yielded 
improved stellar parameters and therefore, in many cases, more precise 
planetary radii. High-resolution spectroscopy can also reveal companions
not resolved by high-resolution imaging \citep[see, e.g.,][]{marcy14,kolbl15}.
Thus, {\it Kepler} follow-up observations are essential not only for 
confirming transiting planet candidates, but also for determining planetary 
parameters as accurately as possible and therefore deriving the occurrence 
rate of planets of different sizes and compositions, including those planets 
that are most similar to our own Earth. This type of follow-up observations
will also be crucial for upcoming space missions like TESS and PLATO,
which will conduct large surveys for transiting exoplanets and likely yield
thousands of new planets.

\acknowledgments
We thank the Robo-AO team, in particular its leaders Christoph Baranec,
Nicholas Law, Reed Riddle, and Carl Ziegler for sharing their results on 
robotic laser adaptive optics imaging of KOI host stars in their publications 
and on CFOP.
We also thank Adam Kraus and his team for sharing their results on the 
multiplicity of KOI host stars obtained with adaptive optics imaging and 
non redundant aperture-mask interferometry in their recent publication.
The results from these publications provided substantial input for this work.
Support for this work was provided by NASA through awards issued by
JPL/Caltech.
This research has made use of the NASA Exoplanet Archive, which is operated 
by the California Institute of Technology, under contract with NASA under the 
Exoplanet Exploration Program.
It has also made use of data products from the Two Micron All Sky Survey, which 
is a joint project of the University of Massachusetts and the Infrared Processing 
and Analysis Center/Caltech, funded by NASA and the NSF. NASA's Astrophysics 
Data System Bibliographic Services were also used.
Some of the data presented in this work were obtained at the W.M. Keck Observatory, 
which is operated as a scientific partnership among the California Institute of 
Technology, the University of California and the National Aeronautics and Space 
Administration. The Observatory was made possible by the generous financial 
support of the W.M. Keck Foundation. The authors wish to recognize and 
acknowledge the very significant cultural role and reverence that the summit 
of Mauna Kea has always had within the indigenous Hawaiian community.  
We are most fortunate to have the opportunity to conduct observations from 
this mountain. 
This work is also based in part on observations at Kitt Peak National Observatory, 
National Optical Astronomy Observatory, which is operated by the Association 
of Universities for Research in Astronomy (AURA) under a cooperative agreement 
with the National Science Foundation. The WIYN Observatory is a joint facility of 
the University of Wisconsin-Madison, Indiana University, the National Optical 
Astronomy Observatory and the University of Missouri. 
Part of the observations were also obtained at the Gemini Observatory, which is 
operated by AURA under a cooperative agreement with the NSF on behalf of 
the Gemini partnership.
Some of the results in this work are based on observations with the NASA/ESA 
{\it Hubble Space Telescope}, obtained at the Space Telescope Science Institute, 
operated by AURA, Inc., under NASA contract NAS 5-26555.

\end{document}